\documentclass[a4paper,12pt]{article}
\pdfoutput=1
\usepackage{epsfig,graphicx,xcolor,amsbsy,amssymb,latexsym,amsfonts,amsmath,,tcolorbox,setspace,stackrel}
\usepackage{pstricks}
\usepackage{color}
\usepackage{mathrsfs}
\usepackage{soul}
\usepackage[numbers,square,comma, compress]{natbib}
\usepackage{placeins}
\usepackage{wrapfig,framed,caption}
\usepackage{tensor}
\usepackage[makeroom]{cancel}
\newcounter{relctr} 
\everydisplay\expandafter{\the\everydisplay\setcounter{relctr}{0}} 

\AtBeginDocument{} 

\usepackage{wasysym}

\usepackage{multirow}

\definecolor{shadecolor}{gray}{0.925}

\usepackage{eurosym}
\usepackage[a4paper]{geometry}
\usepackage{hyperref}
\usepackage[T1]{fontenc}
\usepackage[font=small,labelfont=bf,tableposition=top]{caption}

\DeclareCaptionLabelFormat{andtable}{#1~#2  \&  \tablename~\thetable}

\usepackage{graphicx}
\usepackage{tikz}
\usetikzlibrary{decorations.pathreplacing,decorations.markings,snakes}
\usetikzlibrary{decorations.pathmorphing}
\usetikzlibrary{patterns}
\usetikzlibrary{calc}
\usepackage{xcolor}
\usepackage{amsmath,amssymb,amsfonts,pstricks,setspace}
\usepackage{floatrow}
\newfloatcommand{capbtabbox}{table}[][\FBwidth]
\usepackage{blindtext}
\usepackage{caption}
\usepackage[enableskew]{youngtab}
\usepackage{derivative}

\numberwithin{equation}{section}

\usepackage{wrapfig}

\newcommand{\bea}{\begin{eqnarray}\displaystyle}
\newcommand{\eea}{\end{eqnarray}}

\newcommand{\hp}[1]{\widehat{p}^{#1}}

\newcommand{\av}{\langle d\rangle_p}
\newcommand{\efc}[1]{\mathfrak{d}^{#1}}

\newcommand{\met}[2]{\mathfrak{g}_{#1 #2}}

\newcounter{lp}
\title{
{\bf Fisher Information and Dynamical Sampling I}\\[10pt]

Part 1: Clustering\\[10pt]
}

\begingroup
\renewcommand{\thefootnote}{\fnsymbol{footnote}}

\author{
\large \textsc{Mattia Carrino\footnotemark[1]\;\,\footnotemark[2]\;\,\footnotemark[3]}
\,\, and \,\,
\textsc{Stefan~Hohenegger\footnotemark[4]}
}

\footnotetext[1]{\tt mattia.carrino01@universitadipavia.it}
\footnotetext[4]{\tt s.hohenegger@ip2i.in2p3.fr}

\endgroup

\begin{document}

\maketitle
\thispagestyle{empty}

\maketitle
\thispagestyle{empty}
\begin{center}
\renewcommand{\thefootnote}{\fnsymbol{footnote}}\vspace{-0.5cm}
${}^{\footnotemark[1]}$ National PhD Programme in One Health approaches to infectious diseases and life science research, Department of Public Health, Experimental and Forensic Medicine, University of Pavia, Pavia, 27100, Italy\\[0.3cm]
${}^{\footnotemark[2]}$Dept. of Physics E. Pancini, Universit\`a di Napoli Federico II, via Cintia, 80126 Napoli, Italy\\[0.3cm] 
${}^{\footnotemark[3]}$ Centre of Excellence in Respiratory Pathogens (CERP), Hospices Civils de Lyon (HCL) and Centre International de Recherche en Infectiologie (CIRI), Équipe Santé Publique, Épidémiologie et Écologie Évolutive des Maladies Infectieuses (PHE3ID), Inserm U1111, CNRS UMR5308, ENS de Lyon, Université Claude Bernard Lyon 1, Lyon, France\\[0.3cm]
${}^{\footnotemark[4]}$ Universit\'e Claude Bernard Lyon 1, CNRS/IN2P3, IP2I Lyon, UMR 5822, Villeurbanne, F-69100, France\\[1.2cm] 
\end{center}

\begin{abstract}
Information theory is a powerful framework to capture aspects of dynamical systems with multiple degrees of freedom. Mathematically, the dynamics can be represented as a continuous curve $\mathcal{C}$ on a suitable hyperplane in flat space and the Fisher information provides the norm of an infinitesimal displacement along this curve. In many applications, however,  we do not have direct access to $\mathcal{C}$. Instead, we have to reconstruct the latter from a time-series of measurements (obtained as samples of size $n$), which are represented by an ordered set of points $\widehat{\mathcal{C}}$ on the same hyperplane. In this work, we calculate the bias of the Fisher information for large $n$, which provides a quantitative estimation for how accurately the dynamics of a system can be reconstructed from a given set of sampled data. Based on this result, we show that a clustering of the degrees of freedom reduces the bias and thus improves the accuracy with which the new system can be described with the same data. Inspired by a concrete proposal for such a clustering in \cite{FILOCHE2025130647}, we provide a quantitive assessment of the loss of information, which allows to estimate how much information about the dynamics of a system can reliably be extracted based on a given set of data. We illustrate our findings in the case of a simple compartmental model. Although the latter is inspired by epidemiology, the results of this work are applicable to very general dynamical models with multiple degrees of freedom.
\end{abstract}

\newpage

\tableofcontents

\section{Introduction}
Analysing dynamical systems from an information theoretical perspective has a long standing history: notably in the context of communication theory \cite{Fisher,ShannonEntropy,WeaverShannon,Hotelling,Rao}, questions such as the maximal rate at which information can be transmitted by a channel \cite{Badii_Politi_1997,Castiglione_Falcioni_Lesne_Vulpiani_2008} or the optimal compression of data \cite{McMillian,Huffman} have been studied using a variety of mathematical tools \cite{Goldmann,LESNE_2014}. The latter, in fact, are applicable to a large number of different systems in electrical engineering, physics (\emph{e.g.} statistical mechanics \cite{PhysRev.106.620,Jaynes2,Touchette_2009}), computer science or economics. In most of these examples, the state of a system with a general number of degrees of freedom, is measured at periodic time intervals, from which a(n effective) model is inferred to describe the future dynamics. It is of particular practical interest to understand the influence of noise on the predictive power of this model, \emph{i.e.} the fact that the measurements generally only allow to reconstruct an approximate state of the system at any given moment in time. Information theoretical and statistical tools allow to make statements about the maximal information that can be obtained about a system from such measurements.

A particular branch of information theory that we shall rely on in this work is information geometry \cite{Lauritzen,Jeffreys,amari2000methods,AmariLoss,AmariRao}, \emph{i.e.} the application of elements of (Riemannian) geometry to describe and analyse probabilistic systems: indeed, the general object of study in this paper is a  \emph{statistical model} $\mathcal{C}$ as a system of $N$ dynamical degrees of freedom, which are encoded in a time-dependent probability\footnote{In certain applications, notably if the law governing the dynamics of $\mathcal{C}$ is of a deterministic nature, the $p^\mu$ are called \emph{frequencies}. In this work, in order to make contact with the literature (\emph{e.g.} \cite{CoverInformation}) we shall refer to them as \emph{probabilities}.} distribution, namely a set of functions $p^\mu(t)\in[0,1]$ (with $\mu=1,\ldots,N+1$), such that $\sum_{\mu=1}^{N+1}p^\mu(t)=1$ for all times $t\in\Xi\subset\mathbb{R}$. As we shall review (and extend) in this paper, there are two different geometric notions that can be studied in this framework: on the one hand, the \emph{Fisher information} \cite{FisherInfo} $g_{tt}$ associated with $\mathcal{C}$ can be understood as the metric on a Riemannian manifold (see \emph{e.g.} \cite{amari2000methods} for an excellent review). While in the current setup, this simply means that $g_{tt}$ is a smooth function of time, this notion becomes more sophisticated for probability distributions that depend on several parameters. On the other hand, $\mathcal{C}$ can be understood as a (continuous) curve on a hyperplane of $\mathbb{R}^{N+1}$, which we shall call the simplex $\Delta_{N+1}$. A notion of distance on $\Delta_{N+1}$ is provided by the \emph{Shahshahani metric} $\mathfrak{g}_{\mu\nu}$ \cite{Shahshahani} (see also \cite{Kimura}). These two geometric notions are, in fact, related: the norm (with respect to the Shahshahani metric) of an infinitesimal displacement along the curve $\mathcal{C}$ is given by the Fisher information $g_{tt}$ (see \emph{e.g.} \cite{Harper2009InformationGA,SANDHOLM2008666,SandholmBook,MERTIKOPOULOS2018315,Hofbauer_Sigmund_1998}). Thus, the Fisher information is an indispensable tool for studying the dynamics of the model $\mathcal{C}$ and describes 'how much' the system changes in an infinitesimal time interval. In fact, in the case of models with a single degree of freedom (\emph{i.e.} $N=1$), it was demonstrated in \cite{Filoche:2024xka} that the entire dynamics can be re-organised in terms~of~$g_{tt}$.

As mentioned above, in many applications, we try to determine the model $\mathcal{C}$ through \emph{time-series sampling} in the form of measuring the probability distribution at certain time intervals $dt$. These measurements produce a discrete statistical model $\widehat{\mathcal{C}}$ which consists of \emph{sampled} probabilities $\widehat{p}^\mu\in[0,1]$ at discrete points in time. These probabilities are generally subject to noise and therefore differ from the $p^\mu(t)$ that we wish to determine. This prevents us from accurately capturing the dynamics of the system: the difference between the (sampled) probabilities at two different times is now superimposed by the statistical noise. Therefore, in this work, we first quantify the impact of sampling on the Fisher information: we assume that $\widehat{p}(t)$ at any time $t$ has been obtained as a sequence of $n\in\mathbb{N}$ independent and identically distributed (i.i.d.) random data points (based on the probability distribution $p(t)$). In other words, we assume that the probability for a specific sampling to be returned is a multinomial distribution, which for large enough $n$ can be analysed using methods in information theory and statistics (such as the method of types \cite{Csiszar_Types,Csiszar_Korner_2011,Longo,CoverInformation} or large deviation theory \cite{Hoefding,DemboZeitouni,BenderOrszag,Touchette_2009}). Indeed, from a geometric perspective, the probability to find $\widehat{p}$ is strongly localised around $p(t)$ and depends on the distance to the latter (in the sense of the Shahshahani-metric). Following a proposal in \cite{FILOCHE2025130647} (see eq.(\ref{eq:metr_phat})), we extract $g_{tt}(t)$ from the discrete sampling data $\widehat{p}(t+dt/2)$ and $\widehat{p}(t-dt/2)$, with $dt$ a (small) time-interval. For large $n$ and small $dt$, we find the bias of the Fisher information to be $\sim \frac{2N}{n dt^2}$ (see eq~(\ref{ExpValueFisher})). We first remark that this result goes beyond the \emph{central limit theorem}, since $g_{tt}$ is not computed as the expectation value of a sequence of i.i.d. random variables, but instead from two such variables (namely $\widehat{p}^\mu(t\pm dt)$). This is necessary since, in order to capture the dynamics of the system, $g_{tt}$ compares $\mathcal{C}$ at two different times. For more details on this discussion, we refer the reader to Appendix~\ref{App:CentralLimit}. Secondly, to leading order, the result for the bias of the Fisher information is universal and only depends on the number of degrees of freedom $N$ as well as the quality of the samplig (\emph{i.e.} $n$ and $dt$). This indeed suggests two strategies to reduce the bias and therefore improve the information that can be extracted from a system from a finite amount of data:
\begin{itemize}
\item[\emph{(i)}] {\bf clustering}: since the bias of $g_{tt}$ depends on $N$, reducing the (effective) number of degrees of freedom improves the accuracy of the Fisher information obtained from sampling. However, at the same time, eliminating degrees of freedom from $\mathcal{C}$ generally diminishes the Fisher information and therefore reduces the accuracy for describing the dynamics. Thus a careful coarse graining of the degrees of freedom of the system is required to minimise the loss of information, while optimising the reduction in the bias of $g_{tt}$. In this work, we shall discuss \emph{clustering} (following a proposal in \cite{FILOCHE2025130647}) as a means to optimise this balance.
\item[\emph{(ii)}] {\bf time averaging and filtering}: the result of the bias also suggest that (effectively) increasing $dt$ also reduces the bias. More accurately, the latter can strongly be reduced by modifying the prescription on how to compute the Fisher information from the discrete $\widehat{\mathcal{C}}$ by taking into account $\widehat{p}$ at multiple times $t$. A first example of this idea was proposed in \cite{FILOCHE2025130647} by first Gaussian filtering the data. We shall explore this idea in more detail in the companion paper \cite{Companion}.
\end{itemize} 

In this paper, we shall exclusively focus on {\it (i)}: in \cite{FILOCHE2025130647} clustering has been analysed as a means to extract information about degrees of freedom that behave 'similarly'. Indeed, while mostly focused on epidemiological applications (namely the spread of a pathogen through a host population), a general proposal was made to minimise the loss of information (the reduction in the Fisher information) by clustering together degrees of freedom with a similar change of \emph{self-information} $\dot{\mathcal{I}}^\mu(t)=\frac{\dot{p}^\mu(t)}{p^\mu(t)}$ (where a dot denotes a time derivative).\footnote{In \cite{FILOCHE2025130647} a slightly different convention for the normalisation was chosen, which is less convenient for the applications in the current paper.} It was argued in \cite{FILOCHE2025130647} that such a clustering groups together degrees of freedom that are similarly adapted to the remaining dynamical system by linking $\dot{\mathcal{I}}^\mu(t)$ to an effective coupling $d^\mu(t)$: applying this type of clustering to the spread of SARS-CoV-2 in France (by analysing the data from over 500.000 patients during a 4 year period), it was demonstrated that such a regrouping gives insights into the genetic adaptation of the pathogen and highlights mutations that grant competitive advantages to new variants. 

In this work, we shall further formalise and generalise the approach proposed in \cite{FILOCHE2025130647}, by first expressing the loss of information $\Delta g_{tt}$ as the sum over the variances of the couplings $d$ across a cluster. This allows to quantify the change in the Fisher information in terms of the number of clusters and the composition of the individual clusters and makes contact to quantities defined within the dynamical system under consideration. Secondly, this $\Delta g_{tt}$ can be compared to the bias of the Fisher information due to sampling. Our computations allow to quantitatively gauge how much information can reliably be extracted from a given amount of data by describing the dynamics of only a coarse-grained set of effective degrees of freedom. Depending on how many of the latter are chosen and how they are combined together allows to minimise $\Delta g_{tt}$ while at the same time minimising the statistical bias.

In order to illustrate our theoretical computations, we demonstrate them at the hand of a simple dynamical system. Inspired by epidemiological applications, we consider a simple compartmental model \cite{McKendrick1926,Kermack:1927} that groups a fixed population into different classes with regards to their epidemiological status with respect to a number $N+1$ of pathogens. Their infectivity and pathogenicity is modelled by a set of numerical parameters, which determine together with the overall status of the population (concretely the number of healthy individuals) the couplings $d^\mu$ for each variant. Despite its simplicity, the model allows to numerically verify our theoretical computations and demonstrates that clustering is indeed an effective way to improve the accuracy of reproducing the Fisher information extracted from time-series samplings. The numerical model also allows us to address another question, namely how to choose the clustering itself directly from the sampled data. Indeed, we provide numerical indications that an optimal clustering can reliably be formed when combining sampled data from more points in time in the discrete model $\widehat{\mathcal{C}}$. This suggests that even more information can be extracted from the latter by making better use of the time series sampling of the data. We will discuss highly effective techniques of this type in the companion paper~\cite{Companion}.

This paper is organised as follows: Section~\ref{sec:info_theory} introduces and reviews important notions of information theory and in particular builds upon geometric notions that have been scattered in the literature to link distances between points of a statistical model $\mathcal{C}$ to the Fisher information, in particular in the presence of statistical noise. Section~\ref{Sect:Sampling} quantifies this notion by calculating the expectation value (and variance) of the Fisher information computed from the discrete, sampled statistical model $\widehat{\mathcal{C}}$. Section~\ref{Sect:Clustering} introduces the clustering of degrees of freedom as a means of reducing the bias for the Fisher information obtained from a time-series sampling procedure, while minimising the loss of information. It notably calculates the expectation value of the Fisher information for a clustered version of the statistical model $\widehat{\mathcal{C}}$ and provides further results related to the expectation value of the sampled change of information per degree of freedom. These results are relevant in gauging whether clusterings that minimise the loss of information can be obtained directly from the sampled data. Section~\ref{Sect:Examples} illustrates our finding for a simple compartmental model and provides first numerical indications that the clustering of sampled data can be improved by taking into account more than two sampling points of $\widehat{\mathcal{C}}$. Section~\ref{Sect:Conclusions} contains our conclusions and an outlook for future work, notably the companion paper~\cite{Companion}. This paper is supplemented by two appendices, which contain details on technical computations, as well as further background material on statistics.

\section{Elements of Information Theory}\label{sec:info_theory}
In this Section, we shall review a number of concepts, which have appeared in various places in the literature, and which shall lead us to a re-formulation of dynamical processes from an information-theoretical perspective. For more details, we refer the reader to a number of reviews~\cite{Lauritzen,Jeffreys,amari2000methods}.

\subsection{Probability Distributions and Dynamical Systems}\label{Sect:ProbabilityDistribution}

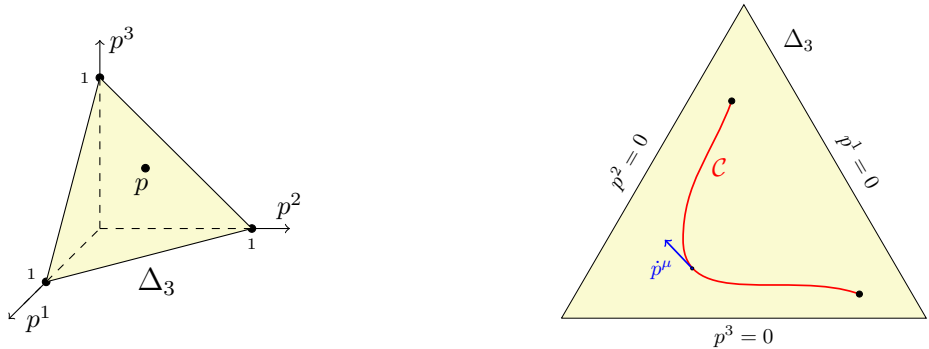
\begin{figure}[h]
\begin{center}
\scalebox{1}{\parbox{4cm}{
\begin{center}
\begin{tikzpicture}
\draw[->] (0,0) -- (-1.2,-1.2);
\node at (-0.8,-1.2) {\footnotesize $p^1$};
\draw[->] (0,0) -- (0,2.5);
\node at (2.5,0.3) {\footnotesize $p^2$};
\draw[->] (0,0) -- (2.5,0);
\node at (0.3,2.5) {\footnotesize $p^3$};
\draw[fill=black] (-0.707,-0.707) circle (0.05cm);
\node at (-0.9,-0.6) {\tiny $1$}; 
\draw[fill=black] (2,0) circle (0.05cm);
\node at (2,-0.2) {\tiny $1$}; 
\draw[fill=black] (0,2) circle (0.05cm);
\node at (-0.2,2) {\tiny $1$}; 
\draw[fill=blue!10!yellow!20!white] (2,0) -- (0,2) -- (-0.707,-0.707) -- (2,0);
\draw[dashed] (0,2) -- (0,0);
\draw[dashed] (2,0) -- (0,0);
\draw[dashed] (-1,-1) -- (0,0);
\draw[fill=black] (0.6,0.8) circle (0.05cm);
\node at (0.75,-0.7) {$\Delta_3$};
\node at (0.55,0.55) {\footnotesize $p$};
\end{tikzpicture}
\end{center}
}}
\hspace{3cm}
\scalebox{0.8}{\parbox{6cm}{\begin{tikzpicture}
\draw[fill=blue!10!yellow!20!white] (-3,0) -- (3,0) -- (0,5.19615) -- (-3,0);
\node at (0,-0.3) {\footnotesize $p^3=0$};
\node[rotate=60] at (-1.9,2.6) {\footnotesize $p^2=0$};
\node[rotate=-60] at (1.9,2.6) {\footnotesize $p^1=0$};
\draw[red,thick] (1.9,0.4) to [out=160,in=270] (-1,1.3)  to [out=90,in=250] (-0.2,3.6);
\draw[fill=black] (-0.2,3.6) circle (0.05cm);
\draw[fill=black] (1.9,0.4) circle (0.05cm);
\node[red] at (-0.4,2.5) {$\mathcal{C}$};
\node at (0.9,4.6) {$\Delta_3$};
\draw[blue,thick,->] (-0.85,0.825) -- (-1.3,1.3);
\node[blue] at (-1.35,0.8) {\footnotesize$\dot{p}^\mu$};
\draw[fill=blue] (-0.85,0.825) circle (0.025cm);
\end{tikzpicture}}}
\end{center}
\caption{Left panel: Schematic representation of a probability distribution $p$ on $\Delta_{3}$. Right panel: Schematic representation of a model $\mathcal{C}$ on $\Delta_3$. The blue vector represents the tangent vector $\dot{p}^\mu$ to $\mathcal{C}$.}
\label{Fig:Simplex}
\end{figure}

Let $\mathbb{V}\cong\{1,\ldots,N+1\}$ (with $N\in\mathbb{N}$) be a discrete set and define a \emph{probability distribution} as a map
\begin{align}
p:\,\mathbb{V}&\longrightarrow [0,1]&&\text{with} &&\sum_{\mu=1}^{N+1}p^\mu=1\,,\nonumber\\[-10pt]
\mu&\longmapsto p^\mu\,,\label{ProbDistr}
\end{align}
Geometrically, a probability distribution can be represented as a point on the following hyperplane of $\mathbb{R}^{N+1}$, which we shall refer to as simplex \cite{CoverInformation,MERTIKOPOULOS2018315}
\begin{align}
\Delta_{N+1}:=\left\{(p^1,\ldots,p^{N+1})\in [0,1]^{ N+1}\bigg|\sum_{\mu=1}^{N+1}p^\mu=1\right\}\subset\mathbb{R}^{N+1}\,.\label{DefSimplexInit}
\end{align}
An example for $N=2$ is schematically shown in the left panel of Figure~\ref{Fig:Simplex}. Furthermore, we define a \emph{statistical model} \cite{amari2000methods} (also called {\it parametric model} or simply \emph{model}) on $\mathbb{V}$ as a family of such probability distributions parametrised in the following way
\begin{align}
\mathcal{C}=\{p(t)|t\in\Xi\subset \mathbb{R}\}\,,\label{statisticalmodel}
\end{align}
where $\Xi \subset \mathbb{R}$ is a suitable interval.
Any member of the family $\mathcal{C}$ is a probability distribution in the sense of (\ref{ProbDistr}) (\emph{i.e.} notably $\sum_{\mu=1}^{N+1}p^\mu(t)=1$ $\forall t\in\Xi$). From a physical perspective, we shall interpret $t$ as a time variable, such that $\mathcal{C}$ describes (an aspect of) the dynamics of some underlying system with $N+1$ degrees of freedom (which are labelled by $\mu$).\footnote{We refer to \cite{Harper2009InformationGA,SANDHOLM2008666,SandholmBook,Filoche:2024xka,FILOCHE2025130647} for examples of systems in which parts of the dynamics have been reformulated as a statistical model~\ref{statisticalmodel}. We also remark that the $p^\mu(t)$ are sometimes called \emph{frequencies} rather than probabilities, to reflect the fact that they are determined by an underlying deterministic system. Since we shall use numerous concepts of information theory, we shall prefer to use the term probabilities.}

Geometrically, $\mathcal{C}$ corresponds to a curve on $\Delta_{N+1}$, as is schematically shown in the right panel of Figure~\ref{Fig:Simplex} for $N=2$. Throughout this work, we shall only work with models $\mathcal{C}$ which~are~represented by smooth curves, \emph{i.e.} concretely, we shall assume the $p^\mu(t)$ to be continuously differentiable functions of $t\in\Xi$. Furthermore, we shall limit ourselves to $p(t)$ in the interior of $\Delta_{N+1}$, \emph{i.e.} we shall assume $p^\mu(t)\neq 0$ $\forall \mu=1,\ldots,N+1$ and $t\in\Xi$.

At every point of $\mathcal{C}$, we can define the \emph{Fisher information}~\cite{FisherInfo} (see also \cite{amari2000methods})
\begin{align}
&g_{tt}(t):=\sum_{\mu=1}^{N+1}p^\mu(t)\left(\tfrac{d}{dt}\,\log p^\mu(t)\right)^2=\sum_{\mu=1}^{N+1}\frac{(\dot{p}^\mu(t))^2}{p^\mu(t)}\,,&&\text{with} &&\dot{p}^\mu(t):=\frac{dp^\mu}{dt}(t)\,.\label{FisherInformationDef}
\end{align}
We can further re-write this expression by introducing the amount of self-information (or \textit{information} for short)\footnote{Compared to \cite{FILOCHE2025130647}, we have defined the information with the logarithm to base $e$ and with a factor of $-1$. Notice that in the convention (\ref{eq:information}), $\mathcal{I}^\mu\leq 0$.} of the degree of freedom labelled by $\mu$ at time $t$
\begin{align}
&\mathcal{I}^\mu(t) := \log(p^\mu(t))\,,&&\text{such that} && \dot{\mathcal{I}}^\mu(t)= \frac{\dot{p}^\mu(t)}{p^\mu(t)}\,.\label{eq:information}
\end{align}
We then have the following presentation of the Fisher information
\begin{align}
g_{tt}(t)=\sum_{\mu=1}^{N+1}p^\mu(t)\,(\dot{\mathcal{I}}^\mu(t))^2\,,\label{FisherInformationIs}
\end{align}
as the expectation value of the squares of the change of information.

In \cite{FILOCHE2025130647}, the change of information has been related to so-called \emph{effective couplings} $d^\mu$ 
\begin{align}
&\dot{\mathcal{I}}^\mu(t)=d^\mu(t)-\av(t)\,,&&\text{with} &&\av(t)=\sum_{\mu=1}^{N+1}p^\mu(t)\,d^\mu(t)\,,\label{Couplings}
\end{align}
where $\av$ is interpreted as the average of the couplings $d^\mu$. The $d^\mu$ were interpreted as an effective way to describe the dynamics of $p^\mu$ in relation to the remaining degrees of freedom of the system. If the equations of motion of the entire system are known (see \emph{e.g.} Section~\ref{Sect:Examples} for an illustrative example), the $d^\mu$ can be calculated explicitly. In order to make the relation between the couplings $d^\mu$ and the dynamics of the system more tangible, we also write the Fisher information in the following form\footnote{Here and in the following, whenever there is no danger of confusion, we shall drop the explicit arguments~$t$.}
\begin{align}
g_{tt}(t)=\sum_{\mu=1}^{N+1}p^\mu\,\left(d^\mu\right)^2-\av^2=:\text{Var}(d)\,.
\end{align}
Thus, from an information theoretical perspective, the Fisher information calculates the {\it variance} of the couplings $d^\mu$, with respect to the probability distribution $p(t)$.


\subsection{Geometry of the Dynamics}\label{Sect:GeometryShash}
It is well established \cite{Rao,Jeffreys,amari2000methods,BURBEA1982575,Nielsen} that the Fisher information can also be interpreted as a Riemannian metric with coordinates $t\in\Xi$ (see \cite{amari2000methods} for an overview of generalisations to higher dimensional $\Xi$): for this reason, $g_{tt}(t)$ is sometimes also referred to as the Fisher information \emph{metric}. In the current context, however, this simply means that $g_{tt}(t)$ is a positive definite, smooth function of $t$. 

However, as discussed for example in \cite{Harper2009InformationGA,SANDHOLM2008666,SandholmBook,MERTIKOPOULOS2018315,Hofbauer_Sigmund_1998}, from its definition, $g_{tt}$ also furnishes a metric on the simplex $\Delta_{N+1}$: in order to see this, we re-write (\ref{FisherInformationDef}) in the following way
\begin{align}
ds^2=g_{tt}(t)\,dt^2= \sum_{\mu,\nu=1}^{N+1}\met{\mu}{\nu}(p(t))\,dp^\mu(t)\,dp^\nu(t)\,,&&\text{with} &&\met{\mu}{\nu}=\frac{\delta_{\mu\nu}}{p^\mu}\,,\label{NDimShashahani}
\end{align}
where $\met{\mu}{\nu}$ is the \emph{Shahshahani metric} \cite{Shahshahani} (see also \cite{Kimura} for earlier work and \cite{MERTIKOPOULOS2018315} for a modern

\begin{wrapfigure}{l}{0.3\textwidth}
\vspace{-0.5cm}
\centering
\includegraphics[width = 5cm]{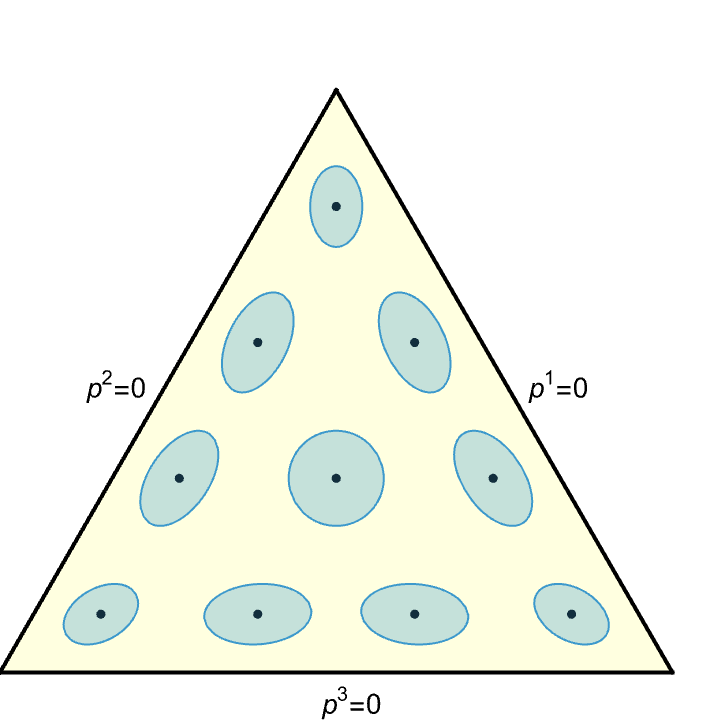}
\vspace{-0.55cm}
\caption{Disks of same norm at different points of the simplex $\Delta_3$: each blue disk represent points with a distance $\leq \sqrt{0.03}$ to the central point.}
\label{Fig:UnitCirclesShash}
\end{wrapfigure}

\noindent
review), which is well-defined in the interior of $\Delta_{N+1}$ (where~$p^\mu\neq 0$ $\forall\mu=1,\ldots,N+1$). Concretely, equation (\ref{NDimShashahani}) states that the Fisher information defines the norm $ds^2$ of the infinitesimal displacement along the curve $\mathcal{C}$ on the simplex $\Delta_{N+1}$. The tangent vector $\dot{p}^\mu$ is schematically shown in the right panel of Figure~\ref{Fig:Simplex} and its norm is determined by the Fisher information according to (\ref{NDimShashahani}): to obtain a better intuition about the latter, Figure~\ref{Fig:UnitCirclesShash} shows disks of same norm at various points of $\Delta_3$.

\begin{figure}[h]
\begin{center}
\scalebox{1}{\parbox{13cm}{\begin{tikzpicture}  
\draw[fill=blue!10!yellow!20!white] (-3,0) -- (3,0) -- (0,5.19615) -- (-3,0);
\node at (0,-0.3) {\footnotesize $p^3=0$};
\node[rotate=60] at (-1.9,2.6) {\footnotesize $p^2=0$};
\node[rotate=-60] at (1.9,2.6) {\footnotesize $p^1=0$};
\draw[red,thick] (1.9,0.4) to [out=160,in=270] (-1,1.3)  to [out=90,in=250] (-0.2,3.6);
\draw[fill=black] (-0.2,3.6) circle (0.05cm);
\draw[fill=black] (1.9,0.4) circle (0.05cm);
\node[red] at (-0.4,2.5) {$\mathcal{C}$};
\draw[fill=black] (-1,1.3) circle (0.05cm);
\draw[black]   (-1.5,1.8) -- (-1.5,0.8) -- (-0.5,0.8) -- (-0.5,1.8) -- (-1.5,1.8);
\draw[thick,black,->] (-1.5,1.3) to [out=180,in=270] (-8,1); 
\draw[black,fill=blue!10!yellow!20!white] (-10,1) -- (-6,1) -- (-6,5) -- (-10,5) -- (-10,1); 
\draw[thick,red] (-7,1) to [out=140,in=270] (-8,3) to [out=90,in=240] (-7.5,5);
\draw[fill=green!50!black] (-8,3) circle (0.075cm);
\node[green!50!black] at (-7.5,3) {$p(t)$};
\draw[thick,rotate=80,blue!50!white] (1.6,8.4) ellipse (1cm and 0.5cm);
\draw[fill=green!50!black] (-7.9,4) circle (0.075cm);
\node[green!50!black] at (-6.9,4.25) {$p(t+dt)$};
\draw[ultra thick,blue,->] (-8,3.1) -- (-7.9,3.95);
\node[blue] at (-8.35,3.4) {$\dot{p} dt$};
\end{tikzpicture}}}
\vspace{-1.25cm}
\caption{Schematic representation of the geometry of the dynamics: an infinitesimal displacement along the curve $\mathcal{C}$ is described by $\dot{p}^\mu dt$, which is a vector of norm $\sqrt{g_{tt}}\,dt$. The blue ellipse in the detail on the left indicates all ending points of vectors with norm $\sqrt{g_{tt} dt}$ (with respect ot the Shahshahani metric), but the curve $\mathcal{C}$ passes through only one of these points at $t+dt$.}
\label{Fig:SimplexUnitTraject}
\end{center}
\end{figure}
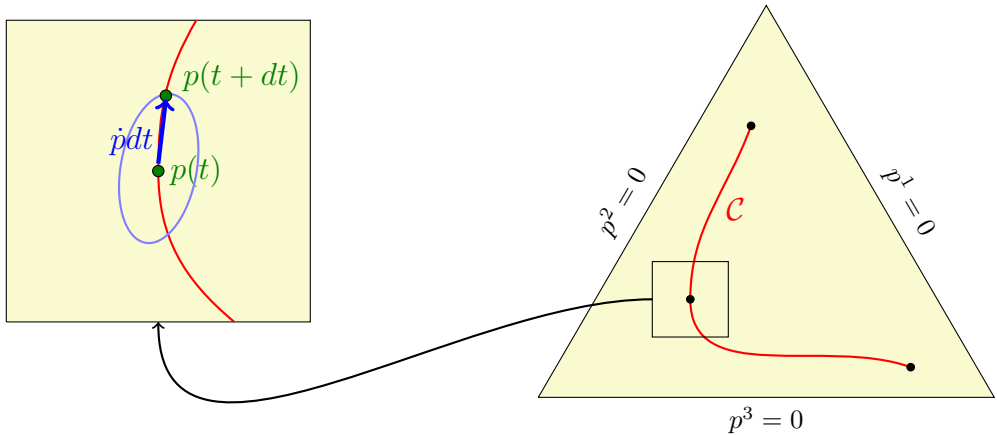

It should, however, be noted that for $N>1$ the curve $\mathcal{C}$ is not uniquely determined by $g_{tt}$. While for $N=1$ (\emph{i.e.} for systems that are described by a single degree of freedom), a universal description of the dynamics in terms of the Fisher information was elaborated in \cite{Filoche:2024xka}, for $N>1$ the dynamics also depends on the direction of $\dot{p}^\mu$ (and not only its norm). Formulated differently, for an infinitesimal time translation $t\to t+dt$, the Fisher information only provides a subspace of the simplex $\Delta_{N+1}$ on which $p^\mu(t+dt)$ is located: this subspace is a sphere (with respect to the Shahshahani metric) with radius $\sqrt{g_{tt}}$ and which is centered at $p^\mu(t)$. This is schematically shown for $N=2$ in Figure~\ref{Fig:SimplexUnitTraject}: at each point of the curve $p(t)$, $p(t+dt)=p(t)+\dot{p}\,dt+\mathcal{O}(dt^2)$ is a point located on this sphere.

\subsection{Sampling and Noise}
\subsubsection{Sampled Statistical Model}
We assume that the curve $\mathcal{C}$ is determined through an underlying theory, which furnishes equations of motion that determine $\dot{p}^\mu$ and thus allow to compute $p(t)$ for any $t\in\Xi$. In many applications, however, this model is unknown and, in fact, one tries to determine it through a number of suitable measurements. This means, that $p^\mu(t)$ is measured for a (discrete or countably infinite) number of points in time. In this work, we shall assume that these measurements are spaced-out at equal time intervals $dt$, \emph{i.e.} we are measuring $p(t)$ at times 
\begin{align}
t\in\widehat{\Xi}=\{t_0+k\,dt|k\in \mathbb{Z}\}\cap \Xi\,,\label{HatXi}
\end{align}
for some suitable $t_0\in \Xi$. Furthermore, the measuring procedure generally does not precisely reproduce $p(t)$ but rather a value close to it. Indeed, in many practical applications, the probabilities are obtained from a sampling process, in which a finite number $n\in\mathbb{N}$ of samples is taken, from which a value $\widehat{p}(t)$ close to $p(t)$ is infered. Concretely, in this work, we shall discuss \emph{multinomial} (or \emph{categorical}) sampling: at any of the $t\in\widehat{\Xi}$ we obtain empirical probabilities $\hp{\mu}(t)$ by performing $n$ independent draws with replacement (\emph{i.e.} every time an element is selected it is then put back into the original group for the next draw) from a multinomial distribution with probabilities $p^\mu(t)$. We assume this procedure to be independent at each draw and thus produces $n$ independent and identically distributed (i.i.d.) measurements. If $n_\mu\leq n$ of the draws return the outcome $\mu$, we may define the \emph{sampled probability distribution} 
\begin{align}
&\hp{\mu}(t)=\frac{n_\mu}{n}\,,&&\text{where} &&\widehat{p}=(\widehat{p}^1,\ldots,\widehat{p}^{N+1})\in \Delta_{N+1}^{(n)}\,, 
\end{align} 
where\footnote{In \cite{Csiszar_Types,Csiszar_Korner_2011} (see also \cite{CoverInformation} (Chapter 11, p.348) and Appendix~\ref{App:CentralLimit}) $\Delta_{N+1}^{(n)}$ is called the \emph{set of types with denominator }$n$, while $\widehat{p}$ is called the \emph{type class} (or composition class) of the i.i.d. draw of $n$ elements. (Part of) the framework developed in this Section is also called the method of types.}  $\Delta_{N+1}^{(n)}$ is the intersection of the simplex $\Delta_{N+1}$ with the $(N+1)$-dimensional hypercubic lattice with lattice constant $1/n$
\begin{align}
\Delta_{N+1}^{(n)}=\left\{\left(\frac{n_1}{n},\ldots,\frac{n_{N+1}}{n}\right)\bigg| n_1,\ldots,n_{N+1}\in\mathbb{N}^*\text{ and } \sum_{k=1}^{N+1}n_k=n\right)\,.
\end{align}
In this way, we obtain a new model 
\begin{align}
\widehat{\mathcal{C}}=\{\widehat{p}^\mu(t)|\mu\in\mathbb{V}\text{ and }t\in \widehat{\Xi}\}\,,\label{SampledModel}
\end{align}
which we shall call the \emph{sampled statistical model} and which corresponds to countably many ordered points on $\Delta_{N+1}^{(n)}\subset \Delta_{N+1}$. In addition to being a discrete set rather than a continuous curve (as pointed out in \cite{CoverInformation}, the number of points in $\Delta_{N+1}^{(n)}$ is $\leq (n+1)^{N+1}$), $\widehat{\mathcal{C}}$ a priori further differs from $\mathcal{C}$ in (\ref{statisticalmodel}): while for given $t\in \widehat{\Xi}$, the probability $\widehat{p}$ is close to $p$, they are generally separated by a distance (in the sense of the Shahshahani metric) $\mathfrak{a}(\widehat{p},p)$, as is schematically shown in Figure~\ref{Fig:SimplexTrajectory}. This distance depends on $\widehat{p}$ and is therefore subject to a stochastic process. To get a better understanding of the 'difference' between $\widehat{\mathcal{C}}$ and $\mathcal{C}$, it is therefore more appropriate to study the expectation value $\langle \mathfrak{a}^2(p,\widehat{p})\rangle$ as an average over multiple samplings.


\subsubsection{Distance between $\mathcal{C}$ and $\widehat{\mathcal{C}}$}\label{Sect:DistModels}

The probability $\mathbb{P}$ for $\widehat{p}$ being returned in the sampling at $t\in\widehat{\Xi}$, is given by the multinomial distribution
\begin{align}\label{eq:prob_extract}
\mathbb{P}\left(\widehat{p}(t),t\right)=\frac{n!}{n_1!n_2!\ldots n_{N+1}!}\,\prod_{\mu=1}^{N+1}\left(p^\mu(t)\right)^{n_\mu}\,.
\end{align}

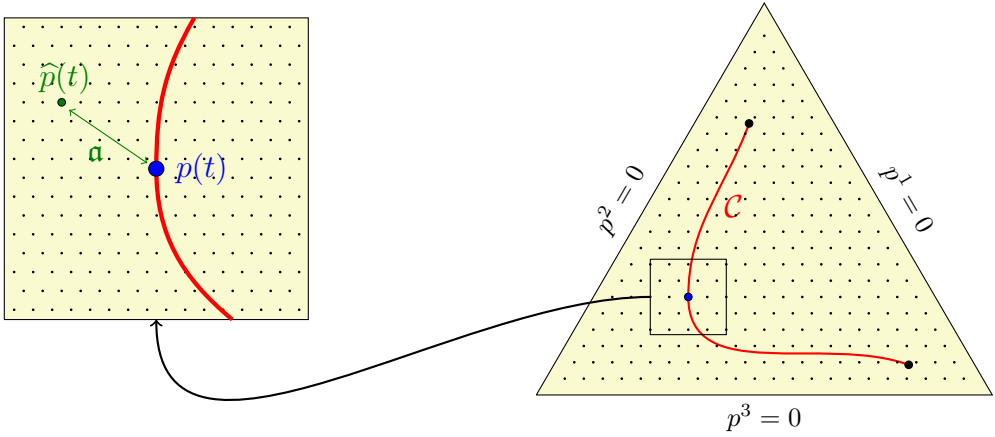
\begin{figure}[h!]
\begin{center}
\scalebox{1}{\parbox{13cm}{\begin{tikzpicture}  
\draw[fill=blue!10!yellow!20!white] (-3,0) -- (3,0) -- (0,5.19615) -- (-3,0);
\node at (0,-0.3) {\footnotesize $p^3=0$};
\node[rotate=60] at (-1.9,2.6) {\footnotesize $p^2=0$};
\node[rotate=-60] at (1.9,2.6) {\footnotesize $p^1=0$};
\draw[red,thick] (1.9,0.4) to [out=160,in=270] (-1,1.3)  to [out=90,in=250] (-0.2,3.6);
\draw[fill=black] (-0.2,3.6) circle (0.05cm);
\draw[fill=black] (1.9,0.4) circle (0.05cm);
\node[red] at (-0.4,2.5) {$\mathcal{C}$};
\foreach \i in {1,2,3,4,5,6,7,8,9,10,11,12,13,14,15,16,17,18,19,20,21,22}
                  \stepcounter{lp}
          \draw[fill=black] (-3+0.125+\i*0.25,0.2165) circle (0.01cm);
\foreach \i in {1,2,3,4,5,6,7,8,9,10,11,12,13,14,15,16,17,18,19,20,21}
                  \stepcounter{lp}
          \draw[fill=black] (-3+0.25+\i*0.25,0.433) circle (0.01cm);
\foreach \i in {1,2,3,4,5,6,7,8,9,10,11,12,13,14,15,16,17,18,19,20}
                  \stepcounter{lp}
          \draw[fill=black] (-3+0.125+0.25+\i*0.25,0.433+0.2165) circle (0.01cm);
\foreach \i in {1,2,3,4,5,6,7,8,9,10,11,12,13,14,15,16,17,18,19}
                  \stepcounter{lp}
          \draw[fill=black] (-3+2*0.25+\i*0.25,2*0.433) circle (0.01cm);
\foreach \i in {1,2,3,4,5,6,7,8,9,10,11,12,13,14,15,16,17,18}
                  \stepcounter{lp}
          \draw[fill=black] (-3+0.125+2*0.25+\i*0.25,2*0.433+0.2165) circle (0.01cm);            
\foreach \i in {1,2,3,4,5,6,7,8,9,10,11,12,13,14,15,16,17}
                  \stepcounter{lp}
          \draw[fill=black] (-3+3*0.25+\i*0.25,3*0.433) circle (0.01cm);  
\foreach \i in {1,2,3,4,5,6,7,8,9,10,11,12,13,14,15,16}
                  \stepcounter{lp}
          \draw[fill=black] (-3+0.125+3*0.25+\i*0.25,3*0.433+0.2165) circle (0.01cm);  
\foreach \i in {1,2,3,4,5,6,7,8,9,10,11,12,13,14,15}
                  \stepcounter{lp}
          \draw[fill=black] (-3+4*0.25+\i*0.25,4*0.433) circle (0.01cm);  
\foreach \i in {1,2,3,4,5,6,7,8,9,10,11,12,13,14}
                  \stepcounter{lp}
          \draw[fill=black] (-3+0.125+4*0.25+\i*0.25,4*0.433+0.2165) circle (0.01cm);  
\foreach \i in {1,2,3,4,5,6,7,8,9,10,11,12,13}
                  \stepcounter{lp}
          \draw[fill=black] (-3+5*0.25+\i*0.25,5*0.433) circle (0.01cm);  
\foreach \i in {1,2,3,4,5,6,7,8,9,10,11,12}
                  \stepcounter{lp}
          \draw[fill=black] (-3+0.125+5*0.25+\i*0.25,5*0.433+0.2165) circle (0.01cm);  
\foreach \i in {1,2,3,4,5,6,7,8,9,10,11}
                  \stepcounter{lp}
          \draw[fill=black] (-3+6*0.25+\i*0.25,6*0.433) circle (0.01cm);  
\foreach \i in {1,2,3,4,5,6,7,8,9,10}
                  \stepcounter{lp}
          \draw[fill=black] (-3+0.125+6*0.25+\i*0.25,6*0.433+0.2165) circle (0.01cm);  
\foreach \i in {1,2,3,4,5,6,7,8,9}
                  \stepcounter{lp}
          \draw[fill=black] (-3+7*0.25+\i*0.25,7*0.433) circle (0.01cm);  
\foreach \i in {1,2,3,4,5,6,7,8}
                  \stepcounter{lp}
          \draw[fill=black] (-3+0.125+7*0.25+\i*0.25,7*0.433+0.2165) circle (0.01cm);  
\foreach \i in {1,2,3,4,5,6,7}
                  \stepcounter{lp}
          \draw[fill=black] (-3+8*0.25+\i*0.25,8*0.433) circle (0.01cm);  
\foreach \i in {1,2,3,4,5,6}
                  \stepcounter{lp}
          \draw[fill=black] (-3+0.125+8*0.25+\i*0.25,8*0.433+0.2165) circle (0.01cm);  
\foreach \i in {1,2,3,4,5}
                  \stepcounter{lp}
          \draw[fill=black] (-3+9*0.25+\i*0.25,9*0.433) circle (0.01cm);  
\foreach \i in {1,2,3,4}
                  \stepcounter{lp}
          \draw[fill=black] (-3+0.125+9*0.25+\i*0.25,9*0.433+0.2165) circle (0.01cm);  
\foreach \i in {1,2,3}
                  \stepcounter{lp}
          \draw[fill=black] (-3+10*0.25+\i*0.25,10*0.433) circle (0.01cm);  
\foreach \i in {1,2}
                  \stepcounter{lp}
          \draw[fill=black] (-3+0.125+10*0.25+\i*0.25,10*0.433+0.2165) circle (0.01cm);  
\foreach \i in {1}
                  \stepcounter{lp}
          \draw[fill=black] (-3+11*0.25+\i*0.25,11*0.433) circle (0.01cm);  
\draw[fill=blue] (-1,1.3) circle (0.05cm);
\draw[black]   (-1.5,1.8) -- (-1.5,0.8) -- (-0.5,0.8) -- (-0.5,1.8) -- (-1.5,1.8);
\draw[thick,black,->] (-1.5,1.3) to [out=180,in=270] (-8,1); 
\draw[black,fill=blue!10!yellow!20!white] (-10,1) -- (-6,1) -- (-6,5) -- (-10,5) -- (-10,1); 
\foreach \j in  {0,1,2,3,4,5,6,7} 
\foreach \i in {0,1,2,3,4,5,6,7,8,9,10,11,12,13,14,15}
                  \stepcounter{lp}
          \draw[fill=black] (-9.87+\i*0.25,1.13+\j*0.5) circle (0.01cm);

\foreach \j in  {0,1,2,3,4,5,6,7} 
\foreach \i in {0,1,2,3,4,5,6,7,8,9,10,11,12,13,14}
                  \stepcounter{lp}
          \draw[fill=black] (-9.87+0.125+\i*0.25,1.13+0.25+\j*0.5) circle (0.01cm);
          
\draw[ultra thick,red] (-7,1) to [out=140,in=270] (-8,3) to [out=90,in=240] (-7.5,5);
\draw[fill=blue] (-8,3) circle (0.1cm);
\node[blue] at (-7.4,3) {$p(t)$};
\draw[fill=green!50!black] (-9.87+0.125+2*0.25,1.13+11*0.25) circle (0.05cm);
\draw[green!50!black,<->] (-9.87+0.125+2*0.25+0.1,1.13+11*0.25-0.1) -- (-8-0.12,3+0.075);
\node[green!50!black] at (-8.8,3.2) {$\mathfrak{a}$};
\node[green!50!black] at (-9.2,4.2) {$\widehat{p}(t)$};
\end{tikzpicture}}}
\vspace{-1.25cm}
\caption{Schematic representation of the probabilities $\widehat{p}$ on the simplex $\Delta_{3}^{(n)}$ for $N=2$. The curve $\mathcal{C}$ lies on $\Delta_{3}$, such that the $p(t)$ can be located anywhere on the red curve passing through the yellow triangle. The sampled probabilities $\widehat{p}(t)$ are elements of $\Delta_{3}^{(n)}$ and are thus confined to one of the black dots. The probability that a single sampling at time $t$ returns exactly $\widehat{p}(t)$ is given by the multinomial distribution $\mathbb{P}(\widehat{p}(t),t)$ in (\ref{eq:prob_extract}). For large $n$ the latter is approximated by the normal distribution in (\ref{eq:ProbHatp}), which is governed by the distance $\mathfrak{a}(\widehat{p},p)$ in (\ref{ShashaDist}).}
\label{Fig:SimplexTrajectory}
\end{center}
\end{figure}

\begin{figure}[h!]
\begin{center}
\includegraphics[width = 7.5cm]{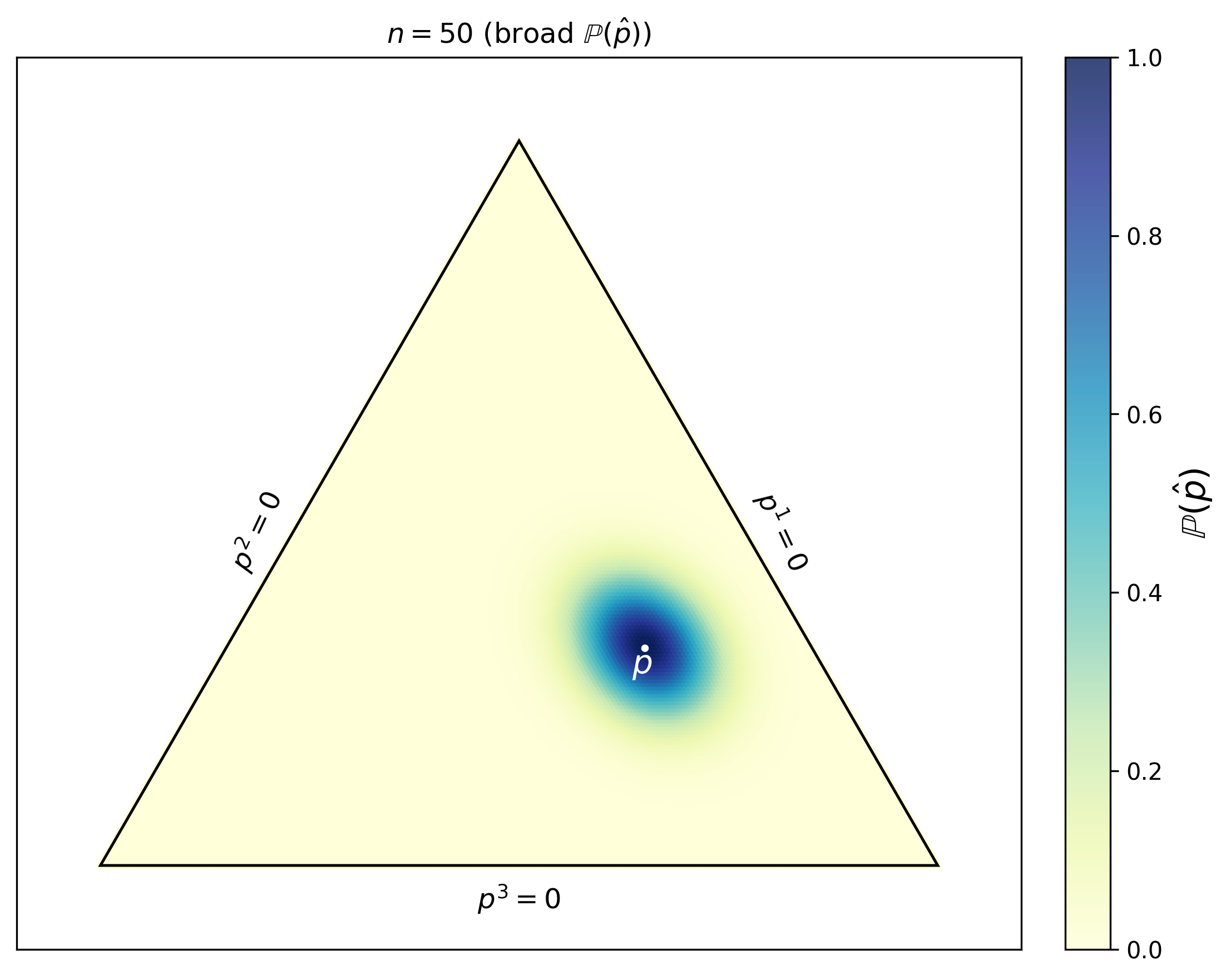}\hspace{1cm}\includegraphics[width = 7.5cm]{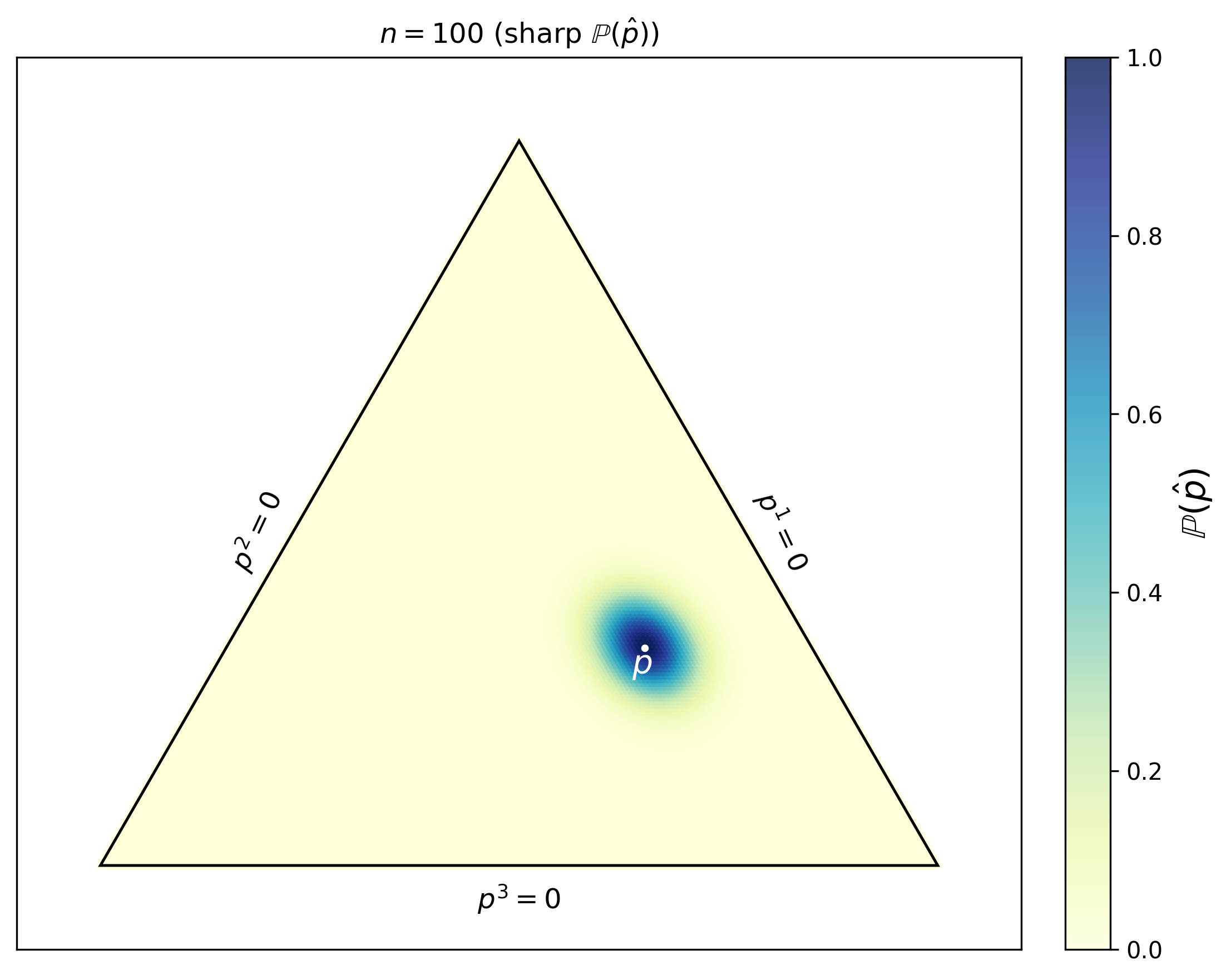}
\end{center}
\caption{Representation of the probability $\mathbb{P}$ in (\ref{eq:ProbHatp}) on the simplex $\Delta_{3}$ for $N=2$ and $p(t)=(0.2,0.5,0.3)$ for $n=50$ (left) and $n=100$ (right). Points of the same colour have the same distance (measured in the Shahshahani metric) to $p$. Comparing the two plots shows, how the probability $\mathbb{P}$ becomes more and more sharply peaked around $p$ as $n$ is increased. }
\label{Fig:simplex_larg_dev}
\end{figure}

\noindent
We remark that $\mathbb{P}(\widehat{p}(t),t)$ depends implicitly on $\mathcal{C}$ through $p^\mu(t)$ and for large $n$ can be further analysed using the \emph{method of types} \cite{Csiszar_Types,Csiszar_Korner_2011,Longo} (for reviews see also \cite{CoverInformation}), which is briefly recalled in Appendix~\ref{App:CentralLimit}: the probability $\mathbb{P}(\widehat{p}(t),t)$ can be approximated as in eq.~(\ref{DefKullbackLeibler}):
\begin{align}
&\mathbb{P}(\widehat{p}(t),t)\approx e^{-n D(\widehat{p}(t)||p(t))}\,,&&\text{with} &&D(\widehat{p}(t)||p(t))=\sum_{\mu=1}^{N+1}\widehat{p}^\mu(t)\,\log\left(\frac{\widehat{p}^\mu(t)}{p^\mu(t)}\right)\,,\label{eq:ProbHatp}
\end{align}
where $\approx$ denotes equality up to polynomial factors in $n$ and $D(\widehat{p}(t)||p(t))$ is the \emph{Kullback-Leibler divergence} \cite{KullbackLeibler} (see also \cite{Csiszar1,Csiszar2,Amari1,amari2000methods,LESNE_2014,Nielsen}). The latter is positive definite, convex and has a minimum for $\widehat{p}=p$ around which it allows an expansion of the form
\begin{align}
D(\widehat{p}||p)=\frac{1}{2}\sum_{\mu=1}^{N+1}\frac{(\widehat{p}^\mu-p^\mu)^2}{p^\mu}+\mathcal{O}((\widehat{p}-p)^3)\,.\label{ExpandKullback}
\end{align}
Using the framework of \emph{large deviation theory}  \cite{Hoefding,DemboZeitouni,BenderOrszag} (see \emph{e.g.} \cite{Touchette_2009} for a review), we shall use this form in the following Section~\ref{Sect:Sampling} to compute approximations of different expectation values for large $n$. Before, however, in light of the discussion of the previous Section~\ref{Sect:GeometryShash}, we remark

\begin{wrapfigure}{r}{0.45\textwidth}
\vspace{-0.45cm}
\centering
\includegraphics[width = 7.5cm]{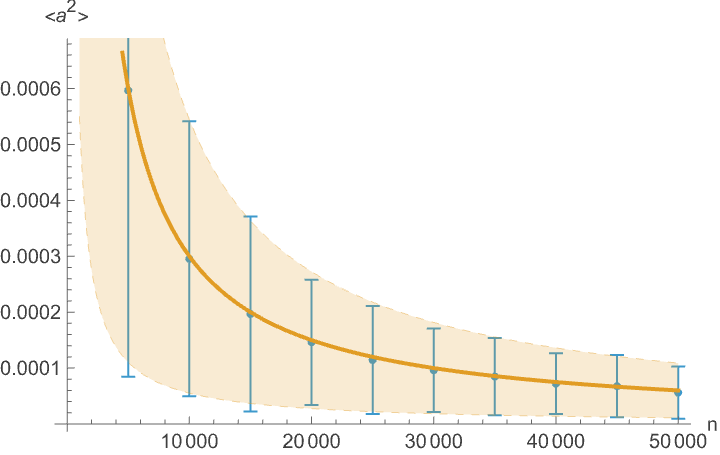}
\vspace{-0.15cm}
\caption{Expectation value $\langle\mathfrak{a}^2(\widehat{p},p)\rangle_p$ for $N=3$ and $p=(0.1,0.2,0.3,0.4)$: the blue points show the result of a numerical computation of $\langle\mathfrak{a}^2(\widehat{p},p)\rangle_p$ along with its standard deviation, while the orange curve represents the value in (\ref{ExpDistance}), with the shaded region representing the standard deviation.\\[-50pt]}
\label{Fig:ExpDistance}
${}$\\[-0.3cm]
\end{wrapfigure}

\noindent
\noindent
that the~form of the leading term in (\ref{ExpandKullback}) can be interpreted geometrically as a distance on the simplex $\Delta_{N+1}$
\begin{align}
\mathfrak{a}^2(\widehat{p},p)&=\sum_{\mu=1}^{N+1}\frac{(\hp{\mu}-p^\mu)^2}{p^\mu}\nonumber\\
&=\sum_{\mu,\nu=1}^{N+1}\mathfrak{g}_{\mu\nu}\,(\widehat{p}^\mu-p^\mu)\,(\widehat{p}^\nu-p^\nu)\,.\label{ShashaDist}
\end{align}
with $\mathfrak{g}_{\mu\nu}$ the Shahshahani metric as in (\ref{NDimShashahani}). This geometric interpretation is schematically shown in Figure~\ref{Fig:SimplexTrajectory}. With the expansion (\ref{ExpandKullback}), the probability (\ref{eq:ProbHatp}) is thus approximated by a normal distribution, with the Shahshahani distance to $p(t)$, as is numerically shown in Figure~\ref{Fig:simplex_larg_dev}. Notably, as explained in more detail in Appendix~\ref{App:DistExp} for large $n$, the expectation value $\langle\mathfrak{a}^2(\widehat{p},p)\rangle_p$ and the variance of $\mathfrak{a}^2(\widehat{p},p)$ can be approximated as

\begin{align}
&\langle\mathfrak{a}^2(\widehat{p},p)\rangle \sim_{n\to\infty}\frac{N}{n}\,,&&\text{and} &&\text{Var}(\mathfrak{a}^2(\widehat{p},p))\sim_{n\to\infty}\frac{2N}{n^2}\,.\label{ExpDistance}
\end{align}
A comparison of these results to numerical computations for $N=3$ is shown in Figure~\ref{Fig:ExpDistance}, which shows excellent agreement for sufficiently large $n$. 

The non-vanishing expectation value of the distance $\mathfrak{a}^2(\widehat{p},p)$ quantifies how 'different' the samples $\widehat{p}(t)$ are from the $p(t)$ at any given moment $t\in \widehat{X}$. Notice that the behaviour of $\mathfrak{a}^2(\widehat{p},p)$ as $\sim1/n$ is expected from the Central Limit Theorem (CLT, see for example \cite{Schervish} as reviewed in Appendix~\ref{App:CentralLimit}) and our assumption that the $n$ samples are i.i.d.: more details on this relation can be found in Appendix~\ref{App:CentralLimit}, particularly how (\ref{ExpDistance}) allows to approximate the probability $\mathbb{P}(\widehat{p}(t),t)$ by a normal distribution with respect to the Shashahani-distance of $\widehat{p}$ and $p$.

In this work, we shall mostly be interested in the impact of the difference between $\mathcal{C}$ and $\widehat{\mathcal{C}}$ on the dynamics of the system: as we mentioned before, the dynamics of the model $\mathcal{C}$ when evolving from $t_1\to t_2$ is described by a smooth curve on $\Delta_{N+1}$. However, from the point of view of $\widehat{\mathcal{C}}$, in addition to this, we also observe a purely statistical effect from the two samplings $\widehat{p}(t_1)$ and $\widehat{p}(t_2)$. In other words, even when $t_1=t_2$, two independent samplings will generally result in two different $\widehat{p}$ and thus two distinct points on $\Delta^{(n)}_{N+1}$, whose distance is governed by (\ref{ExpDistance}). In a first step (see Section~\ref{Sect:Sampling}), we shall be interested in quantifying the distance between $\widehat{p}(t_1)$ and $\widehat{p}(t_2)$ for $t_1<t_2$. To this end, we shall characterise the bias of the Fisher information (see eq.~(\ref{ExpValueFisher})) stemming from two such samplings. In a second step, we shall propose the clustering of the degrees of freedom of $\mathcal{C}$ as one means to reduce this bias: we shall also comment on how this can be used as a strategy for extracting information about dynamical systems for which limited data are available. In a follow up work \cite{Companion}, we shall discuss further strategies, related to a more efficient use of the data as a function of time.

\section{Time Series and Sampling}\label{Sect:Sampling}
In this Section we calculate the expectation value and the variance of the Fisher information for a statistical model $\widehat{\mathcal{C}}$ obtained from sampled probabilities $\widehat{p}(t)$ at times $t\in\widehat{\Xi}$ (see eq.~(\ref{HatXi})).

\subsection{Fisher Information on $\widehat{\mathcal{C}}$}\label{Sect:sampling_fish_inf}
We start by defining the Fisher information $\widehat{g}_{tt}$ for the statistical model $\widehat{\mathcal{C}}$: following a proposal in \cite{FILOCHE2025130647}, we replace (\ref{FisherInformationDef}) by the following discretised form 
\begin{align}\label{eq:metr_phat}
&\widehat{g}_{tt}(t):=\sum_{\mu=1}^{N+1}\mathfrak{s}^\mu(\widehat{p})\,\left(\frac{\hp{\mu}\left(t+\tfrac{dt}{2}\right)-\hp{\mu}\left(t-\tfrac{dt}{2}\right)}{dt}\right)^2\,,&&\forall t\in \{t_0+\left(k+\tfrac{1}{2}\right)\,dt|k\in \mathbb{Z}\}\cap \Xi\,.
\end{align}
Here $\frac{\hp{\mu}(t+dt/2)-\hp{\mu}(t-dt/2)}{dt}$ is a discretised form of the derivative $\dot{p}^\mu(t)$ and we have defined the

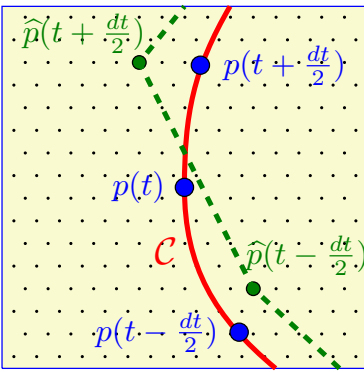
\begin{wrapfigure}[13]{l}{0.3\textwidth}
\vspace{-0.3cm}
\scalebox{1.2}{\centering\parbox{4cm}{\centering
\begin{tikzpicture}  
\draw[blue,fill=blue!10!yellow!20!white] (-10,1) -- (-6,1) -- (-6,5) -- (-10,5) -- (-10,1); 
\foreach \j in  {0,1,2,3,4,5,6,7} 
\foreach \i in {0,1,2,3,4,5,6,7,8,9,10,11,12,13,14,15}
                  \stepcounter{lp}
          \draw[fill=black] (-9.87+\i*0.25,1.13+\j*0.5) circle (0.01cm);

\foreach \j in  {0,1,2,3,4,5,6,7} 
\foreach \i in {0,1,2,3,4,5,6,7,8,9,10,11,12,13,14}
                  \stepcounter{lp}
          \draw[fill=black] (-9.87+0.125+\i*0.25,1.13+0.25+\j*0.5) circle (0.01cm);
          
\draw[ultra thick,red] (-7,1) to [out=140,in=270] (-8,3) to [out=90,in=240] (-7.5,5);
\draw[fill=blue] (-7.825,4.35) circle (0.1cm);
\node[blue] at (-6.9,4.35) {\footnotesize $p(t+\tfrac{dt}{2})$};
\draw[fill=blue] (-8,3) circle (0.1cm);
\node[blue] at (-8.5,3) {\footnotesize $p(t)$};
\node[red] at (-8.2,2.3) {$\mathcal{C}$};
\draw[fill=blue] (-7.4,1.4) circle (0.1cm);
\node[blue] at (-8.3,1.4) {\footnotesize $p(t-\tfrac{dt}{2})$};
\draw[fill=green!50!black] (-9.87+0.125+5*0.25,1.13+13*0.25) circle (0.075cm);
\node[green!50!black] at (-9.1,4.7) {\footnotesize$\widehat{p}(t+\tfrac{dt}{2})$};
\draw[fill=green!50!black] (-9.87+0.125+10*0.25,1.13+3*0.25) circle (0.075cm);
\node[green!50!black] at (-6.65,2.25) {\footnotesize$\widehat{p}(t-\tfrac{dt}{2})$};
\draw[ultra thick,dashed,green!50!black] (-6.3,1) -- (-9.87+0.125+10*0.25,1.13+3*0.25) -- (-9.87+0.125+5*0.25,1.13+13*0.25) -- (-8,5);
\end{tikzpicture}}}
\caption{Schematic representation on $\Delta_{N+1}$ of the sampling procedure to compute the Fisher information $\widehat{g}_{tt}(t)$. }
\label{Fig:SimplexSamplingArrangement}
\end{wrapfigure}


\noindent
shorthand notation
\begin{align}
\mathfrak{s}^\mu(\widehat{p})=\left\{\begin{array}{lcl}\frac{2}{\hp{\mu}\left(t+\tfrac{dt}{2}\right)+\hp{\mu}\left(t-\tfrac{dt}{2}\right)}&\text{if} & \hp{\mu}(t+\tfrac{dt}{2})+\hp{\mu}(t-\tfrac{dt}{2})\neq 0\,,\\[4pt]
0 & \text{else}\end{array}\right.\label{DefSGen}
\end{align}
For the remainder of this work we shall assume that $p(t)$ is in the interior of the simplex $\Delta_{N+1}$, such that the probability for $\mathfrak{s}^\mu$ to vanish is negligibly small (at least for sufficiently large sampling-size $n$). The latter is therefore the inverse of the average of $\widehat{p}^\mu$ between the two sampling points at $t\pm \frac{dt}{2}$. Notice that (\ref{eq:metr_phat}) is symmetric in the sampled probabilities $\widehat{p}(t\pm \tfrac{dt}{2})$. As a consequence, however, $\widehat{g}_{tt}(t)$ is defined at times $t$ that are not in $\widehat{\Xi}$, \emph{i.e.} $t\notin \widehat{\Xi}$ in (\ref{eq:metr_phat}). This is schematically shown in Figure~\ref{Fig:SimplexSamplingArrangement}: the sampled probabilities are $\widehat{p}(t+\tfrac{dt}{2})$ and $\widehat{p}(t-\tfrac{dt}{2})$, which are (according to the probability (\ref{eq:ProbHatp})) in proximity to $p(t+\tfrac{dt}{2})$ and $p(t-\tfrac{dt}{2})$ respectively. These probabilities are used to compute the discretised derivative of $\widehat{p}(t)$ at $t$, which is used to compute~$\widehat{g}_{tt}(t)$. 

The Fisher information (\ref{eq:metr_phat}) manifestly depends on $\widehat{p}^\mu$ and thus on the outcome of the sampling at both times $t\pm\tfrac{dt}{2}$. Given the probabilistic nature of the latter (\emph{i.e.} the probability (\ref{eq:prob_extract})), we are interested in its expectation value and variance. These are a priori difficult to compute, however, using the form (\ref{eq:ProbHatp}) for large $n$, we shall in the following develop an efficient way of approximating them for large sample size $n$ (and small time interval $dt$ between samplings).

\subsection{Statistics of the Dynamics}\label{Sect:CalculationExpValueVar}
\subsubsection{Normalisation}\label{Sect:CalculationPartFct}
We first calculate an approximation of the normalisation factor 
\begin{align}
&Z(p(t))=\sum_{x\in\Delta_{N+1}^{(n)}}e^{-n D(x||p(t))}\,.
\label{DefNormalisationFactor}
\end{align}
This is not only a necessary step to compute expectation values of other quantities, but also serves as a blueprint for similar approximations later on. 
For large enough $n\gg 1$ the points of $\Delta_{N+1}^{(n)}$ lie dense in $\Delta_{N+1}$, such that we can approximate the summation in $Z(p(t))$ as an integral over~$\Delta_{N+1}$
\begin{align}
Z(p(t))\sim_{n\to\infty} n^{N+1}\,\int_0^1 dx^1\ldots\int_0^1 dx^{N+1}\,\delta\left(\sum_{\nu=1}^{N+1}x^\nu-1\right)\,e^{-n D(x||p(t))}\,.\nonumber
\end{align}
Since the integrals are strongly localised around $x=p(t)$, we shall compute them with the help of a saddle-point approximation using the expansion (\ref{ExpandKullback}) of the Kullback-Leibler divergence\footnote{Concretely, we are using Laplace's method (see chapter 6.4 of \cite{BenderOrszag}) to approximate the integrals which are strongly localised around the minimum of $D(x||p(t))$.}
\begin{align}
Z(p(t))\sim_{n\to\infty} n^{N+1}\,\int_0^1 dx^1\ldots\int_0^1 dx^{N+1}\,\delta\left(\sum_{\nu=1}^{N+1}x^\nu-1\right)\,\text{exp}\left(-\frac{n}{2}\sum_{\mu=1}^{N+1}\frac{(x^\mu-p^\mu(t))^2}{p^\mu(t)}\right)\,.\label{SaddlePointExp}
\end{align}
Due to the Dirac $\delta$, the integral here is in fact $N$-dimensional, owing to the fact that $\Delta_{N+1}$ is a co-dimension 1 subspace of $\mathbb{R}^{N+1}$. We next change to the new variables  $u^\mu=\sqrt{\tfrac{n}{2p^\mu}}\,(x^\mu-p^\mu)$, where we implicitly assume that $p(t)$ is in the interior of $\Delta_{N+1}$, such that $p^\mu(t)\neq 0$
\begin{align}
Z&\sim_{n\to\infty} \sqrt{\prod_{\mu=1}^{N+1}2n\,p^\mu(t)}\int_{-\sqrt{\tfrac{n}{2p^1}} p^1}^{\sqrt{\tfrac{n}{2p^1}} (1-p^1)} du^1\ldots\int_{-\sqrt{\tfrac{n}{2p^{N+1}}} p^{N+1}}^{\sqrt{\tfrac{n}{2p^{N+1}}} (1-p^{N+1})} du^{N+1}\,\delta\left(\sum_{\nu=1}^{N+1}\sqrt{\tfrac{2p^\nu}{n}}\,u^\nu\right)\,e^{-\sum_{\mu=1}^{N+1}(u^\mu)^2}\,.\label{GeneralBoundaries}
\end{align}
Notice that due to this change of variables, higher order terms in the expansion of $D(x||p(t))$ in (\ref{ExpandKullback}), which we have neglected in (\ref{SaddlePointExp}), are in fact suppressed by powers of $n$. Furthermore, given the Gaussian nature of the integrands, extending the domain for the individual $u^\mu$ from the bounded intervals to $\mathbb{R}$ only yields corrections that are exponentially suppressed for $n\gg1$ and which we shall therefore neglect.\footnote{We recall that for $0<a,b<\infty$ we have for example
\begin{align}
\int_{-a\sqrt{n}}^{b\sqrt{n}}e^{-x^2}\,dx=\frac{\sqrt{\pi}}{2}\left(\text{Erf}(a\sqrt{n})+\text{Erf}(b\sqrt{n})\right)\sim_{n\to\infty} \sqrt{\pi}-\frac{e^{-na^2}}{2a\sqrt{n}}-\frac{e^{-nb^2}}{2b\sqrt{n}}\,,
\end{align}
where Erf denotes the error function.}
We thus obtain the approximation
{\allowdisplaybreaks
\begin{align}
Z&\sim_{n\to \infty}\sqrt{\prod_{\mu=1}^{N+1}2n\,p^\mu(t)}\,\int_{\mathbb{R}^{N+1}}d^{N+1}u\,\delta\left(\sum_{\nu=1}^{N+1}\sqrt{\tfrac{2p^\nu}{n}}\,u^\nu\right)\,e^{-\sum_{\mu=1}^{N+1}(u^\mu)^2}
=\sqrt{\frac{ n\prod_{\mu=1}^{N+1}2\pi n\, p^\mu}{2\pi\sum_{\mu=1}^{N+1} p^\mu}}\,.\label{PartitionFunction}
\end{align}
}

\subsubsection{Expectation Value}\label{Sect:ExpValuegtt}
With the definition (\ref{eq:metr_phat}) of $\widehat{g}_{tt}$ and the probability $\mathbb{P}(\widehat{p}^\mu(t),t)$ in (\ref{eq:prob_extract}) for a sampling at $t$ to produce $\widehat{p}^\mu$, we have for the expectation value of the Fisher information
\begin{align}
\langle \widehat{g}_{tt}(t)\rangle=\hspace{-0.7cm}\sum_{{\widehat{p}(t+dt/2)\in \Delta_{N+1}^{(n)}}\atop{\widehat{p}(t-dt/2)\in \Delta_{N+1}^{(n)}}}\sum_{\mu=1}^{N+1}\mathfrak{s}^\mu(\widehat{p})&\,\left(\frac{\hp{\mu}\left(t+\tfrac{dt}{2}\right)-\hp{\mu}\left(t-\tfrac{dt}{2}\right)}{dt}\right)^2 \mathbb{P}\left(\widehat{p}\left(t+\tfrac{dt}{2}\right),t+\tfrac{dt}{2}\right)\,\mathbb{P}\left(\widehat{p}\left(t-\tfrac{dt}{2}\right),t-\tfrac{dt}{2}\right)\,.\nonumber
\end{align}
Using similar approximations as for $Z(p(t))$ in the previous Subsubsection~\ref{Sect:CalculationPartFct}, we can write the expectation value in the following integral form
\begin{align}
\langle \widehat{g}_{tt}(t)\rangle\sim_{n\to \infty}& \frac{2}{dt^2}\int_{\mathbb{R}^{N+1}}\widetilde{d^{N+1}x}[p_+]\int_{\mathbb{R}^{N+1}}\widetilde{d^{N+1}y}[p_-]\,
\sum_{\mu=1}^{N+1}\frac{\left(p^\mu_++\sqrt{\frac{2 p^\mu_+}{n}}\,x^\mu-p^\mu_--\sqrt{\frac{2 p^\mu_-}{n}}\,y^\mu\right)^2}{p^\mu_++\sqrt{\frac{2 p^\mu_+}{n}}\,x^\mu+p^\mu_-+\sqrt{\frac{2 p^\mu_-}{n}}\,y^\mu}\,,\label{SampleMetricExpansion}
\end{align}
where for ease of readability, we have introduced the shorthand notation $p_\pm^\mu:=p^\mu(t\pm dt/2)$ and defined the integral measures
\begin{align}
\widetilde{d^{N+1}u}[p]:=\frac{(2n)^{\frac{N+1}{2}}}{Z(p)}\,\sqrt{\prod_{\mu=1}^{N+1}p^\mu(t)}\,\delta\left(\sum_{\lambda=1}^{N+1}\sqrt{\frac{2p^\lambda}{n}}\,u^\lambda\right)\,e^{-\sum_{\rho=1}^{N+1}(u^\rho)^2}\,d^{N+1}u\,.\label{DefGaussianIntMeasure}
\end{align}
The integrand in (\ref{SampleMetricExpansion}) is understood as the following series expansion, where we assume that $p(t)$ is in the interior of $\Delta_{N+1}$ (see Appendix~\ref{App:ExpectValue} for comments on the convergence of this series expansion)
\begin{align}
&\frac{\left(p^\mu_++\sqrt{\frac{2 p^\mu_+}{n}}\,x^\mu-p^\mu_--\sqrt{\frac{2 p^\mu_-}{n}}\,y^\mu\right)^2}{p^\mu_++\sqrt{\frac{2 p^\mu_+}{n}}\,x^\mu+p^\mu_-+\sqrt{\frac{2 p^\mu_-}{n}}\,y^\mu}=\frac{\left(p^\mu_+-p^\mu_-\right)^2}{p^\mu_++p^\mu_-}+\frac{8p_+^\mu p_-^\mu\left(\sqrt{p_-^\mu}\, x^\mu-\sqrt{p_+^\mu}\, y^\mu\right)^2}{n(p_+^\mu+p^\mu_-)^3}+\ldots+\mathfrak{o}(n^{-1})\,.\label{ExpansionTermSingle}
\end{align}
Here, for ease of readability, we have not displayed a term of order $\mathcal{O}(n^{-1/2})$, which (due to the symmetries of the integrals in (\ref{SampleMetricExpansion})) does not contribute to $\langle \widehat{g}_{tt}(t)\rangle$.\footnote{For a more complete discussion of this computation including also terms up to order $\mathfrak{o}(n^{-2}dt^{-4})$, which are needed for the calculation of the variance of $\widehat{g}_{tt}(t)$, see Appendix~\ref{App:ExpectValue}.}
Using the generalised Gaussian integrals compiled in Appendix~\ref{App:GaussInteg}, we obtain
{\allowdisplaybreaks
\begin{align}
\langle \widehat{g}_{tt}\rangle\sim_{n\to\infty}& \frac{2}{dt^2}\sum_{\mu=1}^{N+1}\bigg[\frac{\left(p^\mu_+-p^\mu_-\right)^2}{p^\mu_++p^\mu_-}+\frac{4}{n}\frac{p_+^\mu p_-^\mu(p_+^\mu+p_-^\mu-2p_+^\mu p_-^\mu)}{(p_+^\mu+p^\mu_-)^3}+\mathfrak{o}(n^{-1})\bigg]\,.
\end{align}}
Assuming also the temporal spacing $dt$ to be a small parameter, we can further expand
\begin{align}
\langle \widehat{g}_{tt}\rangle&\sim_{{n\to\infty}\atop{dt\to 0}} \frac{2}{dt^2}\sum_{\mu=1}^{N+1}\left[\frac{(\dot{p}^\mu(t))^2 dt^2}{2\,p^\mu(t)}+\frac{1-p^\mu}{n}+\mathfrak{o}(n^{-1})\right]=g_{tt}(t)+\frac{2N}{n\,dt^2}+\mathfrak{o}(n^{-1})\,.\label{ExpValueFisher}
\end{align}
Thus, to leading order (\emph{i.e.} to order $\mathcal{O}(n^{-1}dt^{-2})$) the bias $\langle \widehat{g}_{tt}(t)\rangle-g_{tt}(t)$ is universal, in the sense that it does not depend neither on the statistical model $\mathcal{C}$ (\emph{i.e.} the probabilities $p^\mu(t)$) nor the Fisher information $g_{tt}(t)$ (or on the derivatives $\dot{p}^\mu(t)$) in any other form. Instead it is only a function of basic parameters related to the sampling procedure itself (\emph{i.e.} $n$ and $dt$) as well as the number of (independent) degrees of freedom $N$. As discussed in Appendix~\ref{App:ExpectValue}, only the first subleading term $\mathcal{O}(n^{-2}dt^{-2})$ depends on $\mathcal{C}$. An illustrative example confirming this result is shown in Figures~\ref{Fig:ExpectClusterFunN} and \ref{Fig:pathogens_FisherInformationTime} in the Example Section~\ref{Sect:Examples}.

\subsubsection{Variance}\label{Sect:Vargtt}
In order to compute the variance associated with $\langle\widehat{g}_{tt}\rangle$ in (\ref{ExpValueFisher}), we first consider in the same manner as (\ref{SampleMetricExpansion})
\begin{align}
\langle(\widehat{g}_{tt})^2&\rangle\sim_{n\to\infty}\frac{4}{dt^4}\int_{\mathbb{R}^{N+1}}\widetilde{d^{N+1}x}[p_+]\int_{\mathbb{R}^{N+1}}\widetilde{d^{N+1}y}[p_-]\,\nonumber\\
&\times \sum_{\mu,\nu=1}^{N+1}\frac{\left(p^\mu_++\sqrt{\frac{2 p^\mu_+}{n}}\,x^\mu-p^\mu_--\sqrt{\frac{2 p^\mu_-}{n}}\,y^\mu\right)^2}{p^\mu_++\sqrt{\frac{2 p^\mu_+}{n}}\,x^\mu+p^\mu_-+\sqrt{\frac{2 p^\mu_-}{n}}\,y^\mu}\,\frac{\left(p^\nu_++\sqrt{\frac{2 p^\nu_+}{n}}\,x^\nu-p^\nu_--\sqrt{\frac{2 p^\nu_-}{n}}\,y^\nu\right)^2}{p^\nu_++\sqrt{\frac{2 p^\nu_+}{n}}\,x^\nu+p^\nu_-+\sqrt{\frac{2 p^\nu_-}{n}}\,y^\nu}\,,\label{VarPreExpand}
\end{align}
where the integrand is again understood as a limited series expansion in $n^{-1}$. The details of the computation up to order $\mathcal{O}(n^{-2})$ (and to leading order in $dt$) are outlined in Appendix~\ref{App:Variance} and we find to order $\mathcal{O}(n^{-2})$ and to leading order (at each order of $n$) in~$dt$
\begin{align}
\langle (g_{tt})^2\rangle\sim_{{n\to\infty}\atop{dt\to 0}}(g_{tt})^2+\frac{4(N+2)}{ndt^2}\,g_{tt}+\frac{4N(N+2)}{n^2dt^4}+\mathfrak{o}(n^{-2})\,.
\end{align}
Each term represents the leading saddle-point approximation for a term in the expansion of the integrand. Notice in particular that contributions of order $\mathcal{O}(n^{-2} dt^{-2})$ would require to go beyond the expansion (\ref{ExpandKullback}) of the Kullback-Leibler divergence in the probability $\mathbb{P}(\widehat{p}(t\pm\tfrac{dt}{2}),t\pm \tfrac{dt}{2})$, which we shall not consider here. We therefore find for the variance to order $\mathcal{O}(n^{-2})$ and to leading order (at each order of $n$) in $dt$ (see eq.~(\ref{VarianceFinalApp}))
\begin{align}
\text{Var}(\widehat{g}_{tt}(t))=\langle(\widehat{g}_{tt})^2\rangle-\langle(\widehat{g}_{tt})\rangle^2\sim_{{n\to\infty}\atop{dt\to 0}}\frac{8 }{n dt^2}\,g_{tt}(t)+\frac{8N}{n^2 dt^4}+\mathfrak{o}(n^{-2})\,.\label{VarianceFinal}
\end{align}
Notice that the leading terms in $dt$ at each order of $n$, depend on the number of degrees of freedom $N$ and on $\mathcal{C}$ only through $g_{tt}$. An illustrative example confirming this result is shown in Figures~\ref{Fig:ExpectClusterFunN} and \ref{Fig:pathogens_FisherInformationTime} in the Example Section~\ref{Sect:Examples}.

Before closing this Section, we would like to stress an important aspect of our results compared to other approaches in the literature. From a certain perspective, we are comparing the sampled statistical model $\widehat{\mathcal{C}}$ with an underlying theoretical model $\mathcal{\mathcal{C}}$. There are numerous discussions of mutual information and mutual entropy between two probability distributions (see \emph{e.g.} \cite{RiekeNeuro,PhysRevLett.80.197,PhysRevE.66.051903} for a direct application in neurology and \cite{Beirlant1997NonparametricEE} for a general review). A broader framework for capturing mutual information in this sense is the \emph{method of sieves} (see \cite{Grenander}). Such approaches establish how different $\widehat{\mathcal{C}}$ is from $\mathcal{C}$ at a given time $t$. More concretely, such approaches discuss estimators that provide certain \emph{static} distances between these models (similar to our discussion in Section~\ref{Sect:DistModels}). In the current Section, we have provided a way to estimate the difference of the \emph{dynamics} described by the statistical models. That is, rather than analysing a (Riemannian) distance between to points of the curves $\widehat{\mathcal{C}}$ and $\mathcal{C}$, we have analysed the Fisher information as a way to describe how far one is moving along the respective curves after an infinitesimal shift in time $t\to t+dt$. This requires to compare the two curves at (at least) two different points and therefore average over multiple samplings. This also explains, why our results are not a trivial application of the Central Limit Theorem (see Appendix~\ref{App:CentralLimit}). Notice, however, that our final results (\ref{ExpValueFisher}) and (\ref{VarianceFinal}) show the same order of magnitude, since in many examples the bias and variance of (mutual) entropies is found to be of order $\mathcal{O}(1/n)$ (see \emph{e.g.}~\cite{Basharin,Paninski}).



\section{Clustering}\label{Sect:Clustering}
Given the bias (\ref{ExpValueFisher}) for the Fisher information as a measure for the dynamical change along the model $\mathcal{C}$, reconstructing the latter from a time-series of samples can be difficult depending on the available data. Indeed (\ref{ExpValueFisher}) (along with (\ref{VarianceFinal})) allows to quantify how much information that is recovered from the sampling is due to statistical noise of the data-taking process and unrelated to the actual dynamics of the underlying system. Geometrically, since $g_{tt}(t)$ measures how far the system moves along the curve $\mathcal{C}$ during the time evolution $t\to t+dt$, the result (\ref{ExpValueFisher}) determines how much this motion is superposed by statistical noise from the measuring process. In situations where dynamical processes are reconstructed from analysing a finite amount of data, it is important to understand how much information can actually be recovered: how much of the model $\mathcal{C}$ can be reconstructed from a given discrete model $\widehat{\mathcal{C}}$?

In this Section, we shall analyse this question from the perspective of \emph{effective degrees of freedom} that can (reliably) be described from $\widehat{\mathcal{C}}$. Indeed, if the bias (\ref{ExpValueFisher}) is too large, it is not possible to correctly capture the temporal evolution of all $N+1$ degrees of freedom. However, the fact that the leading term in (\ref{ExpValueFisher}) is proportional to $N$, suggests that this bias can be lowered if the dynamics of a smaller number of effective degrees of freedom is considered. From an information theoretical perspective, this corresponds to the \emph{clustering} of the probability distribution. In the following, we shall describe such a 'coarse-graining' approach from an information geometric perspective.

\subsection{Notation and General Notions}
Our starting point is a statistical model $\mathcal{C}$ on the discrete set $\mathbb{V}$, as outlined in Section~\ref{Sect:ProbabilityDistribution}. We consider a \emph{clustering} of $\mathbb{V}$ into a disjoint union of $\ell\leq N+1$ non-empty subsets $\{\mathbb{A}_1,\ldots,\mathbb{A}_\ell\}$ (which we shall call clusters in the following), \emph{i.e.} 
\begin{align}
&\mathbb{V}=\bigcup\limits_{a=1}^\ell \mathbb{A}_a&&\text{with} && \begin{cases}
        \mathbb{A}_a \neq \emptyset & \forall a = 1,...,\ell\,,\\
        \mathbb{A}_a \cap \mathbb{A}_b = \emptyset & \text{if }a \neq b. 
    \end{cases}  
\end{align}
We then define the surjective clustering map
\begin{align}
&f:\,\mathbb{V}\longrightarrow \mathbb{W}\cong\{1,\ldots,\ell\}\,,\label{ClusteringFunction}
\end{align}
which assigns any element of $\mathbb{V}$ to exactly one of the $\mathbb{A}_a$. We note that we do not assume $f$ in (\ref{ClusteringFunction}) to depend on $t$: indeed, we assume the clustering to be the same for all $t\in\Xi$. The function $f$ allows to define a statistical model as a curve on the simplex $\Delta_\ell$
\begin{align}
\mathcal{C}_f=\{q(t)|t\in\Xi\}\,,\label{StatModelCluster}
\end{align}
with the clustered probabilities
\begin{align}\label{eq:clust_probs}
    q^a(t) :=\sum_{\mu\in\mathbb{A}_a}p^\mu(t), && \forall a=1,...,\ell\,,
\end{align}
which satisfy $\sum_{a=1}^\ell q^a(t)=1$ $\forall t\in\Xi$. Mimicking (\ref{eq:information}), the information associated with the $a$-th cluster is then defined as
\begin{align}
&\mathbb{I}^a(t) := \log(q^a(t))\,,&& \text{such that} &&\dot{\mathbb{I}}^a(t) = \frac{\dot{q}^a}{q^a}(t)\,,&&\forall a=1,...,\ell.\label{eq:clust_inform}
\end{align}
This allows us to define a Fisher information $g_{tt}^f$ for $\mathcal{C}_f$
\begin{align}\label{eq:clust_metric}
&g_{tt}^f(t) = \sum_{a=1}^\ell \frac{(\dot{q}^a(t))^2}{q^a(t)}=\sum_{a=1}^\ell q^a(t)\,\left(\dot{\mathbb{I}}^a(t)\right)^2\,,&&\text{with} &&\Delta g_{tt}(t) = g_{tt}(t)-g_{tt}^f(t)\geq 0\hspace{0.5cm} \forall t\in\Xi\,.
\end{align}
The difference $\Delta g_{tt}$ (which is always non-negative \cite{amari2000methods}) is a measure for the coarse-graining of information due to the clustering. Indeed, $\Delta g_{tt}(t)=0$ if and only if the clustering is a \textit{sufficient statistic}, in which case any observable computed on $\mathcal{C}$ and $\mathcal{C}_f$ are identical as functions of time \cite{Fisher,amari2000methods} (see also \cite{FILOCHE2025130647}). A sufficient statistic is also characterised by the condition
\begin{align}\label{eq:suf_stat}
&\frac{d r^\mu}{dt}(t)=0\hspace{0.5cm}\begin{array}{l}\forall t\in\Xi\,,\\\forall \mu=1,\ldots, N+1\,,\end{array}&& \text{where} && r^\mu(t) = \frac{p^\mu(t)}{q^{f(\mu)}(t)}\,.
\end{align}
This condition is equivalent to $\dot{\mathcal{I}}^\mu=\dot{\mathbb{I}}^{f(\mu)}$ $\forall \mu=1,\ldots,N+1$, such that indeed $\Delta g_{tt}=0$.

\subsection{Information Clustering}\label{Sec:infclust}
As is manifest from (\ref{eq:clust_metric}), any clustering that is not a sufficient statistics reduces the Fisher information and therefore our understanding of the dynamics of the model $\mathcal{C}$. Intuitively, by clustering, we have coarse-grained the degrees of freedom of the system, which (outside of very particular circumstances) reduces the information we have to describe it. In order to minimise this loss of information, we need to cluster degrees of freedom together that 'behave similarly'. This idea was applied in \cite{FILOCHE2025130647} not only to theoretical models that describe the spread of a pathogen through a population, but also real-world epidemiological data of over 500.000 infected individuals during the Covid-pandemic in France: it was demonstrated that a clustering of different variants of SARS-CoV-2 (which constitute the different degrees of freedom of the epidemiological system) allows to monitor and highlight competitive advantages of variants that spread faster and more efficiently than others and link them to particular genetic mutations.  

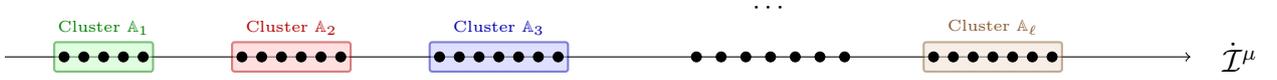
\begin{figure}[h]
\begin{center}
\begin{tikzpicture}[scale=1.3]
\draw[->] (0,0) -- (12,0);
\node at (12.5,0) {$\dot{\mathcal{I}}^\mu$};
\fill[green!25!white,opacity = 0.5,draw=green!50!black,thick,rounded corners=1pt] (0.5,-0.15) rectangle (1.5,0.15);
\node[green!50!black] at (1,0.30) {\tiny $\text{Cluster }\mathbb{A}_1$};
\fill[red!25!white,opacity = 0.5,draw=red!70!black,thick,rounded corners=1pt] (2.3,-0.15) rectangle (3.5,0.15);
\node[red!70!black] at (2.9,0.3) {\tiny $\text{Cluster }\mathbb{A}_2$};
\fill[blue!25!white,opacity  = 0.5,draw=blue!70!black,thick,rounded corners=1pt] (4.3,-0.15) rectangle (5.7,0.15);
\node[blue!70!black] at (5.0,0.3) {\tiny $\text{Cluster }\mathbb{A}_3$};
\fill[brown!25!white,opacity =0.5,draw=brown!70!black,thick,rounded corners=1pt] (9.3,-0.15) rectangle (10.7,0.15);
\node[brown!70!black] at (10.0,0.3) {\tiny $\text{Cluster }\mathbb{A}_\ell$};
\node at (7.75,0.5) {\footnotesize $\cdots$};
\foreach \x in {2.4,2.6,2.8,3.0,3.2,3.4}{
  \draw[fill=black] (\x,0) circle (0.05cm);
}
\foreach \x in {4.4,4.6,4.8,5.0,5.2,5.4,5.6}{
  \draw[fill=black] (\x,0) circle (0.05cm);
}
\foreach \x in {7.0,7.25,7.5,7.75,8.0,8.25,8.5}{
  \draw[fill=black] (\x,0) circle (0.05cm);
}
\foreach \x in {0.6,0.8,1.0,1.2,1.4}{
  \draw[fill=black] (\x,0) circle (0.05cm);
}
\foreach \x in {9.4,9.6,9.8,10.0,10.2,10.4,10.6}{
  \draw[fill=black] (\x,0) circle (0.05cm);
}
\end{tikzpicture}
\end{center}
\caption{Illustration of the clustering according to the change of information $\dot{\mathcal{I}}^\mu$: for fixed $t$, the black dots represent the values $\dot{\mathcal{I}}^\mu(t)$ for $\mu=1,\ldots,N+1$. The clustering groups degrees of freedom with 'similar' $\dot{\mathcal{I}}^\mu(t)$ into the clusters $\mathbb{A}_{a}$ for $a=1,\ldots,\ell$.}
\label{Fig:SchemClustering}
\end{figure}

At the core of the approach in \cite{FILOCHE2025130647} is a clustering that combines degrees of freedom with similar change of information $\dot{\mathcal{I}}^\mu$, as schematically shown in Figure~\ref{Fig:SchemClustering}. 

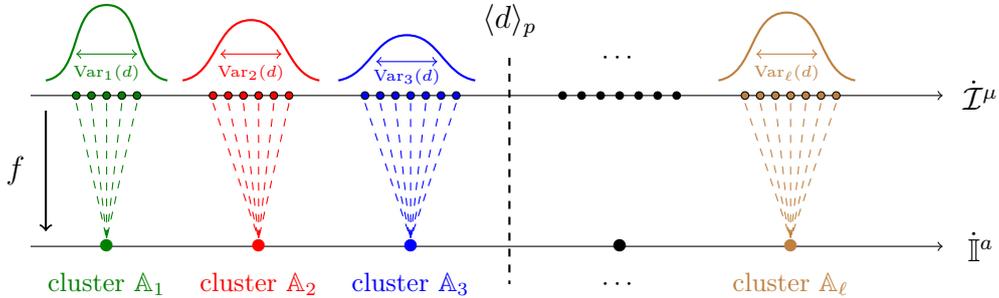
\begin{figure}[h]
\begin{center}
\begin{tikzpicture}
\draw[->] (0,0) -- (12,0);
\node at (12.5,0) {$\dot{\mathcal{I}}^\mu$};
\draw[fill=green!50!black] (0.6,0) circle (0.05cm);
\draw[thin,dashed,green!50!black] (0.6,-0.1) -- (1,-1.9);
\draw[fill=green!50!black] (0.8,0) circle (0.05cm);
\draw[thin,dashed,green!50!black] (0.8,-0.1) -- (1,-1.9);
\draw[fill=green!50!black] (1,0) circle (0.05cm);
\draw[thin,dashed,green!50!black] (1,-0.1) -- (1,-1.9);
\draw[fill=green!50!black] (1.2,0) circle (0.05cm);
\draw[thin,dashed,green!50!black] (1.2,-0.1) -- (1,-1.9);
\draw[fill=green!50!black] (1.4,0) circle (0.05cm);
\draw[thin,dashed,green!50!black] (1.4,-0.1) -- (1,-1.9);
\draw[thick,green!50!black] (0.2,0.2) to [out=10,in=180] (1,1.2) to [out=0,in=170] (1.8,0.2);
\draw[green!50!black,<->] (0.6,0.55) -- (1.4,0.55);
\node[green!50!black]  at (1,0.3) {\tiny $\text{Var}_1(d)$};
\draw[fill=red] (2.6,0) circle (0.05cm);
\draw[thin,dashed,red] (2.6,-0.1) -- (3,-1.9);
\draw[fill=red] (2.8,0) circle (0.05cm);
\draw[thin,dashed,red] (2.8,-0.1) -- (3,-1.9);
\draw[fill=red] (3,0) circle (0.05cm);
\draw[thin,dashed,red] (3,-0.1) -- (3,-1.9);
\draw[fill=red] (3.2,0) circle (0.05cm);
\draw[thin,dashed,red] (3.2,-0.1) -- (3,-1.9);
\draw[fill=red] (3.4,0) circle (0.05cm);
\draw[thin,dashed,red] (3.4,-0.1) -- (3,-1.9);
\draw[fill=red] (2.4,0) circle (0.05cm);
\draw[thin,dashed,red] (2.4,-0.1) -- (3,-1.9);
\draw[thick,red] (2,0.2) to [out=10,in=180] (2.9,1) to [out=0,in=170] (3.8,0.2);
\draw[red,<->] (2.5,0.55) -- (3.3,0.55);
\node[red]  at (2.9,0.3) {\tiny $\text{Var}_2(d)$};
\draw[fill=blue] (4.4,0) circle (0.05cm);
\draw[thin,dashed,blue] (4.4,-0.1) -- (5,-1.9);
\draw[fill=blue] (4.6,0) circle (0.05cm);
\draw[thin,dashed,blue] (4.6,-0.1) -- (5,-1.9);
\draw[fill=blue] (4.8,0) circle (0.05cm);
\draw[thin,dashed,blue] (4.8,-0.1) -- (5,-1.9);
\draw[fill=blue] (5,0) circle (0.05cm);
\draw[thin,dashed,blue] (5,-0.1) -- (5,-1.9);
\draw[fill=blue] (5.2,0) circle (0.05cm);
\draw[thin,dashed,blue] (5.2,-0.1) -- (5,-1.9);
\draw[fill=blue] (5.4,0) circle (0.05cm);
\draw[thin,dashed,blue] (5.4,-0.1) -- (5,-1.9);
\draw[fill=blue] (5.6,0) circle (0.05cm);
\draw[thin,dashed,blue] (5.6,-0.1) -- (5,-1.9);
\draw[thick,blue] (4.05,0.2) to [out=10,in=180] (4.95,0.8) to [out=0,in=170] (5.85,0.2);
\draw[blue,<->] (4.55,0.45) -- (5.35,0.45);
\node[blue]  at (4.95,0.25) {\tiny $\text{Var}_3(d)$};
\draw[fill=brown] (9.4,0) circle (0.05cm);
\draw[thin,dashed,brown] (9.4,-0.1) -- (10,-1.9);
\draw[fill=brown] (9.6,0) circle (0.05cm);
\draw[thin,dashed,brown] (9.6,-0.1) -- (10,-1.9);
\draw[fill=brown] (9.8,0) circle (0.05cm);
\draw[thin,dashed,brown] (9.8,-0.1) -- (10,-1.9);
\draw[fill=brown] (10,0) circle (0.05cm);
\draw[thin,dashed,brown] (10,-0.1) -- (10,-1.9);
\draw[fill=brown] (10.2,0) circle (0.05cm);
\draw[thin,dashed,brown] (10.2,-0.1) -- (10,-1.9);
\draw[fill=brown] (10.4,0) circle (0.05cm);
\draw[thin,dashed,brown] (10.4,-0.1) -- (10,-1.9);
\draw[fill=brown] (10.6,0) circle (0.05cm);
\draw[thin,dashed,brown] (10.6,-0.1) -- (10,-1.9);
\draw[thick,brown] (9.05,0.2) to [out=10,in=180] (9.95,1.1) to [out=0,in=170] (10.85,0.2);
\draw[brown,<->] (9.55,0.55) -- (10.35,0.55);
\node[brown]  at (9.95,0.3) {\tiny $\text{Var}_\ell(d)$};
\draw[thick,dashed] (6.3,0.5) -- (6.3,-2.5);
\node at (6.3,1) {$\av$};
\draw[->] (0,-2) -- (12,-2);
\node at (12.5,-2) {$\dot{\mathbb{I}}^a$};
\node[green!50!black] at (1,-2) {$\bullet$};
\node[green!50!black] at (1,-2.5) {\footnotesize \text{cluster }$\mathbb{A}_1$};
\node[red] at (3,-2) {$\bullet$};
\node[red] at (3,-2.5) {\footnotesize \text{cluster }$\mathbb{A}_2$};
\node[blue] at (5,-2) {$\bullet$};
\node[blue] at (5,-2.5) {\footnotesize \text{cluster }$\mathbb{A}_3$};
\draw[fill=black] (7.75,0) circle (0.05cm);
\draw[fill=black] (7.5,0) circle (0.05cm);
\draw[fill=black] (7.25,0) circle (0.05cm);
\draw[fill=black] (7,0) circle (0.05cm);
\draw[fill=black] (8,0) circle (0.05cm);
\draw[fill=black] (8.25,0) circle (0.05cm);
\draw[fill=black] (8.5,0) circle (0.05cm);
\node at (7.75,-2) {$\bullet$};
\node at (7.75,-2.5) {\footnotesize $\cdots$};
\node at (7.75,0.5) {\footnotesize $\cdots$};
\node[brown] at (10,-2) {$\bullet$};
\node[brown] at (10,-2.5) {\footnotesize \text{cluster }$\mathbb{A}_\ell$};
\node at (-0.2,-1) {$f$};
\draw[thick,->] (0.2,-0.2) -- (0.2,-1.8);
\end{tikzpicture}
\end{center}
\caption{Illustration of the variance of the couplings over all clusters. The change in the Fisher information is the expectation value over the variances (\ref{VarClusterDefForm}) of the couplings $d^\mu$ over all clusters.}
\label{Fig:SchemClusteringVariance}
\end{figure}

\noindent
Indeed, the change of the Fisher information (\ref{eq:clust_metric}) can equivalently be written in the form
\begin{align}
 \Delta g_{tt} = \sum_{\mu=1}^{N+1} p^\mu \left(\dot{\mathcal{I}}^\mu-\dot{\mathbb{I}}^{f(\mu)}\right)^2\,,\label{DeltaGClusterInit}
\end{align}
which is interpreted as an average over the (squares of the) differences $\dot{\mathcal{I}}^\mu-\dot{\mathbb{I}}^{f(\mu)}$. In order to provide a more direct interpretation of $ \Delta g_{tt} $, we re-write it using the effective couplings $d^\mu$ introduced in (\ref{Couplings}). To this end, we start from
\begin{align}
&\dot{\mathbb{I}}^a=\frac{\sum_{\mu\in\mathbb{A}_a}\dot{p}^\mu}{\sum_{\mu\in\mathbb{A}_a}p^\mu}\,,&&\text{with} &&\begin{array}{l}\dot{q}^a=\sum_{\mu\in\mathbb{A}_a}\dot{p}^\mu=\sum_{\mu\in\mathbb{A}_a}p^\mu\,d^\mu-q^a\,\av\,,\\[8pt] 
q^a=\sum_{\mu\in\mathbb{A}_a}p^\mu\,,
\end{array}\label{ClusterInfoDotIa}
\end{align}
which we re-write as
\begin{align}
&\dot{\mathbb{I}}^a=\tfrac{1}{q^a}\,\sum_{\mu\in\mathbb{A}_a}p^\mu d^\mu-\av=:\mathfrak{d}^a-\av\,,&&\text{with} &&\efc{a}:=\frac{\sum_{\mu\in\mathbb{A}_a}p^\mu d^\mu}{\sum_{\mu\in\mathbb{A}_a}p^\mu}\,,
\end{align}
where $\efc{a}$ can be interpreted as the effective coupling of the $a$th cluster and $\av$ is defined in (\ref{Couplings}). With this notation, we find for the clustered Fisher information (\ref{eq:clust_metric})
\begin{align}
g_{tt}^{f}&=\sum_{a=1}^\ell q^a\left(\efc{a}-\av\right)^2
&=\sum_{a=1}^\ell q^a\,(\efc{a})^2-2\,\av\,\sum_{a=1}^\ell \sum_{\mu\in\mathbb{A}_a}p^\mu d^\mu+\av^2=\sum_{a=1}^\ell q^a\,(\efc{a})^2-\av^2\,.
\end{align}
We thus have
\begin{align}
\Delta g_{tt}=g_{tt}-g_{tt}^f=\sum_{a=1}^\ell\left(\sum_{\mu\in\mathbb{A}_a}p^\mu\,(d^\mu)^2-\frac{\left(\sum_{\nu\in\mathbb{A}_a}p^\nu d^\nu\right)^2}{\sum_{\nu\in\mathbb{A}_a}p^\nu}\right)\,.
\end{align}
Using furthermore the definition of the $r^\mu$ in (\ref{eq:suf_stat}) (such that $p^\mu=q^{f(\mu)}\,r^\mu$), 
we obtain\begin{align}
g_{tt}-g_{tt}^f&=\sum_{a=1}^\ell \left(q^a\sum_{\mu\in\mathbb{A}_a}r^\mu\,(d^\mu)^2-\frac{(q^a)^2\left(\sum_{\nu\in\mathbb{A}_a}r^\nu\,d^\nu\right)^2}{q^a}\right)\nonumber\\
&=\sum_{a=1}^\ell q^a\left(\left(\sum_{\mu\in\mathbb{A}_a}r^\mu\,(d^\mu)^2\right)-\left(\sum_{\mu\in\mathbb{A}_a}r^\mu d^\mu\right)^2\right)=:\sum_{a=1}^\ell q^a\,\text{Var}_a(d)\,.\label{VarianceCluster}
\end{align}
Here 
\begin{align}
&\text{Var}_a(d):=\left(\sum_{\mu\in\mathbb{A}_a}r^\mu\,(d^\mu)^2\right)-\left(\sum_{\mu\in\mathbb{A}_a}r^\mu\, d^\mu\right)^2\,,&&\forall a\in\{1,\ldots,\ell\}\,,\label{VarClusterDefForm}
\end{align}
is interpreted as the the variance of the coupling $d^\mu$ within the cluster $a$. Notice that
\begin{align}
\efc{a}=\frac{\sum_{\mu\in\mathbb{A}_a}p^\mu d^\mu}{q^a}=\sum_{\mu\in\mathbb{A}_a}r^\mu\,d^\mu\,,\label{VarianceRelatedCoupling}
\end{align}
is the expectation value of the coupling $d^\mu$ within the cluster $a$. The result (\ref{VarianceCluster}) therefore implies that $g_{tt}-g_{tt}^f$ is the average of the variance of the couplings over all clusters. This is schematically depicted in Figure~\ref{Fig:SchemClusteringVariance}.

\noindent

Before closing this Subsection, we remark that (\ref{VarianceCluster}) (with (\ref{VarClusterDefForm})) depends on the couplings, which is primarily of conceptual interest, since it links the loss of information due to clustering to the couplings $d^\mu$. However, it renders it somewhat difficult to determine $\text{Var}_a(d)$ directly from the $\mathcal{C}$, \emph{i.e.} in terms of the $p^\mu$ and their derivatives. To this end, it is more advantageous to return to (\ref{DeltaGClusterInit}) and write
\begin{align}
&\Delta g_{tt}=\sum_{a=1}^\ell q^a\left(\sum_{\mu\in \mathbb{A}_a} r^\mu(\dot{\mathcal{I}}^\mu)^2-(\dot{\mathbb{I}}^a)^2\right)\,,&&\text{and} &&
\text{Var}_a(d)=\sum_{\mu\in\mathbb{A}_a}\frac{p^\mu}{q^a}(\dot{\mathcal{I}}^\mu)^2-(\dot{\mathbb{I}}^a)^2\,,\label{FormDeltaGFromC}
\end{align}
which is indeed completely determined by $\mathcal{C}$.

\subsection{Information Coarse Graining}
The fact that the clustering (\ref{ClusteringFunction}) reduces the degrees of freedom that enter into the Fisher information suggests that the bias of $g_{tt}^f$ is proportional to the number of clusters rather than $N$. To see that this intuition is indeed correct, following (\ref{eq:metr_phat}), we first define the sampled clustered Fisher information
\begin{align}
\widehat{g}^f_{tt}(t):=\sum_{a=1}^{\ell}\mathfrak{s}^a(\widehat{p})\,\left(\sum_{\mu\in\mathbb{A}_a}\frac{\hp{\mu}\left(t+\tfrac{dt}{2}\right)-\hp{\mu}\left(t-\tfrac{dt}{2}\right)}{dt}\right)^2\,,&&\forall t\in \{t_0+\left(k+\tfrac{1}{2}\right)\,dt|k\in \mathbb{Z}\}\cap \Xi\,,\label{ClusteredFisherInformation}
\end{align}
where we have defined the following generalisation of (\ref{DefSGen})
\begin{align}
\mathfrak{s}^a(\widehat{p})=\left\{\begin{array}{lcl}\frac{2}{\sum_{\mu\in\mathbb{A}_a}\left(\hp{\mu}(t+dt/2)+\hp{\mu}(t-dt/2)\right)}&\text{if} & \sum_{\mu\in\mathbb{A}_a}\left(\hp{\mu}(t+dt/2)+\hp{\mu}(t-dt/2)\right)\neq 0\,,\\[4pt]
0 & \text{else}\end{array}\right.\label{DefSaCluster}
\end{align}
For $p(t)$ in the interior of $\Delta_{N+1}$, we can approximate the expectation value of $\widehat{g}^f_{tt}$ in the same manner as in Section~\ref{Sect:ExpValuegtt}
\begin{align}
\langle\widehat{g}^f_{tt}&\rangle\sim_{n\to\infty}\frac{2}{dt^2}\int_{\mathbb{R}^{N+1}}\widetilde{d^{N+1}x}[p_+]\int_{\mathbb{R}^{N+1}}\widetilde{d^{N+1}y}[p_-]\, \sum_{a=1}^{\ell}\frac{\left(\sum_{\mu\in\mathbb{A}_a}\left(p^\mu_++\sqrt{\frac{2 p^\mu_+}{n}}\,x^\mu-p^\mu_--\sqrt{\frac{2 p^\mu_-}{n}}\,y^\mu\right)\right)^2}{\sum_{\nu\in\mathbb{A}_a}\left(p^\nu_++\sqrt{\frac{2 p^\nu_+}{n}}\,x^\nu+p^\nu_-+\sqrt{\frac{2 p^\nu_-}{n}}\,y^\nu\right)}\,,\nonumber
\end{align}
where the integrand is understood as a series expansion in inverse powers of $n$ and as before $p_\pm^\mu:=p^\mu(t\pm dt/2)$. Performing the Gaussian integrals as in Section~\ref{Sect:ExpValuegtt} (see Apendix~\ref{App:GaussInteg}), we find 
\begin{align}
\langle\widehat{g}^f_{tt}\rangle&\sim_{n\to\infty}\frac{2}{dt^2}\sum_{a=1}^\ell\bigg[\left(\frac{dt^2\sum_{\mu,\nu\in\mathbb{A}_a}\dot{p}^\mu\dot{p}^\mu}{2\sum_{\rho\in\mathbb{A}_a}p^\rho}+\mathcal{O}(dt^4)\right)\nonumber\\
&\hspace{1cm}+\frac{1}{n}\,\left(\frac{\sum_{\mu,\nu\in\mathbb{A}_a}\sqrt{p^\mu p^\mu}(\delta^{\mu\nu}-\sqrt{p^\mu p^\nu})}{\sum_{\rho\in\mathbb{A}_a}p^\rho}+\mathcal{O}(dt^2)\right)+\mathfrak{o}(n^{-1})\bigg]\,.
\end{align}
We therefore have to leading order
\begin{align}
\langle\widehat{g}^f_{tt}\rangle\sim_{{n\to\infty}\atop{dt\to 0}}g_{tt}^f+\frac{2(\ell-1)}{n dt^2}+\mathfrak{o}(n^{-1})\,,\label{ExpValueClusterFin}
\end{align}
which indeed corresponds to a reduction of the leading term of the bias by a factor of $\frac{N}{\ell-1}\geq 1$ relative to (\ref{ExpValueFisher}). We note in passing that we find for the variance
\begin{align}
\text{Var}(\widehat{g}^f_{tt}(t))
&\sim_{{n\to \infty}\atop{dt\to 0}} \frac{8 }{n dt^2}\,g^f_{tt}(t)+\left(\frac{8(\ell-1)}{n^2 dt^4}+\mathcal{O}(dt^{-2})\right)\,,\label{VarianceClusterApp}
\end{align}
which reduces to (\ref{VarianceFinal}) for $\ell=N+1$, that is, however, smaller for $\ell<N+1$.

The results (\ref{VarianceCluster}) (or equivalently (\ref{VarClusterDefForm})) on one hand and (\ref{ExpValueClusterFin}) (as well as (\ref{VarianceClusterApp})) on the other hand, show that clustering can indeed be used as a meaningful strategy to extract non-trivial information of a system, when only limited data is available. Indeed, the former result allows the quantify the amount of information that is lost by clustering (\emph{i.e.} coarse graining) the degrees of freedom that enter into the model $\mathcal{C}$.  The result (\ref{ExpValueClusterFin}) in turn quantifies how much the 'noise' in measuring the change of information in the system is reduced by keeping track of fewer degrees of freedom (for which relative statistical errors) can be reduced. Notice also, while this latter part (\emph{i.e.} (\ref{ExpValueClusterFin})) is primarily determined by the sampling procedure (typically the amount of available data), the change (\ref{VarianceCluster}) depends on how the degrees of freedom are clustered together. It can therefore be minimised independently of the available amount of data, by combining degrees of freedom that behave 'similarly'. This was indeed the strategy proposed in \cite{FILOCHE2025130647}. We shall discuss an example of the balance between (\ref{VarianceCluster}) and (\ref{ExpValueClusterFin}) for a concrete model in Section~\ref{Sect:Examples}.


\subsection{Clustering from Sampled Data}\label{Sect:ClusteringSampledData}
As mentioned above, the change in the Fisher information (\ref{VarianceCluster}) due to clustering can be minimised by a judicious choice of the number and type of clusters.  However, for practical applications, an important question is to what extent the clustering can be optimised based on the sampled probabilities $\widehat{p}$, rather than the $p$ of the (potentially unknown) underlying model. In \cite{FILOCHE2025130647}, it was proposed to cluster the degrees of freedom according to $\dot{I}^\mu$. Thus, in order to better understand this question, we first study the expectation value of the change of information $\widehat{\dot{\mathcal{I}}}{}^\mu$ related to the discrete model $\widehat{\mathcal{C}}$, which we define as
\begin{align}
&\widehat{\dot{\mathcal{I}}}{}^\mu(t):=\mathfrak{s}^\mu(\widehat{p})\,\frac{\widehat{p}^\mu(t+dt/2)-\widehat{p}^\mu(t-dt/2)}{dt}\,,&&\forall t\in \{t_0+\left(k+\tfrac{1}{2}\right)\,dt|k\in \mathbb{Z}\}\cap \Xi\,,\label{DefSampledInfo}
\end{align}
with $\mathfrak{s}^\mu(\widehat{p})$ defined in (\ref{DefSGen}). We therefore have for the expectation value
\begin{align}
\langle \widehat{\dot{\mathcal{I}}}{}^\mu(t)\rangle=\hspace{-0.4cm}\sum_{{\widehat{p}(t+dt/2)\in \Delta_{N+1}^{(n)}}\atop{\widehat{p}(t-dt/2)\in \Delta_{N+1}^{(n)}}}\hspace{-0.2cm}\mathfrak{s}^\mu(\widehat{p})\,\frac{\widehat{p}^\mu(t+dt/2)-\widehat{p}^\mu(t-dt/2)}{dt}\,\mathbb{P}\left(\widehat{p}\left(t+\tfrac{dt}{2}\right),t+\tfrac{dt}{2}\right)\,\mathbb{P}\left(\widehat{p}\left(t-\tfrac{dt}{2}\right),t-\tfrac{dt}{2}\right)\,.\nonumber
\end{align}
Using similar approximations as in Section~\ref{Sect:CalculationExpValueVar}, we find to leading orders in $n^{-1}$ and $dt$
\begin{align}
\langle \widehat{\dot{\mathcal{I}}}{}^\mu(t)\rangle\sim_{{n\to \infty}\atop{dt\to 0}}\left[\left(\dot{\mathcal{I}}^\mu+\mathcal{O}(dt)\right)+\frac{1}{n}\left(\dot{\mathcal{I}}^\mu\,\frac{1}{2}+\mathcal{O}(dt)\right)+\frac{1}{n^2}\left(\dot{\mathcal{I}}^\mu\,\frac{1-p^\mu}{p^\mu}\,\frac{3}{4}+\mathcal{O}(dt)\right)+\mathcal{O}(n^{-3})\right]\,,\nonumber
\end{align}
such that the leading contribution can in fact be written as
\begin{align}
\langle \widehat{\dot{\mathcal{I}}}{}^\mu(t)\rangle\sim_{n\to \infty\atop dt\to 0}\dot{\mathcal{I}}^\mu(t)\,\left(1+\frac{1}{2n}\right)\,.\label{ExpInformationForm}
\end{align}
Thus, in leading approximation, the expectation value $\langle \widehat{\dot{\mathcal{I}}}{}^\mu(t)\rangle$ is proportional to $\dot{\mathcal{I}}^\mu(t)$, with a $\mu$-independent factor. Thus, since such a factor does not enter into the clustering strategies employed in \cite{FILOCHE2025130647}, the clustering of the expectation values $\langle \widehat{\dot{\mathcal{I}}}{}^\mu(t)\rangle$ yields exactly the same result as the clustering of the $\dot{\mathcal{I}}^\mu(t)$. However, we note that the Variance of $\widehat{\dot{\mathcal{I}}}{}^\mu(t)$ is approximated as
\begin{align}
\text{Var}(\widehat{\dot{\mathcal{I}}}{}^\mu(t))\sim_{n\to \infty\atop dt\to 0}\frac{1}{n}\left(\frac{2}{dt^2}\,\frac{1-p^\mu}{p^\mu}-(\dot{\mathcal{I}}^\mu)^2\right)\,,\label{VarInformationForm}
\end{align} 
which depends on $p$ and can become large for small $p^\mu$, close to the boundary of the simplex $\Delta_{N+1}$. We also note that for specific situations, it may become possible to reduce this variance by making better use of multiple (data) points of $\widehat{\mathcal{C}}$: indeed, in Section~\ref{Sect:Examples}, we provide some numerical evidence for a concrete model that the variance can be reduced by combining data from more than 2 points in $\widehat{\mathcal{C}}$ (see also \cite{Companion}). We also note that, after clustering $\widehat{\mathcal{C}}$, we define the change of information of a cluster $a$
\begin{align}
\widehat{\dot{\mathbb{I}}}{}^a(t):=\mathfrak{s}^a(\widehat{p})\,\sum_{\mu\in\mathbb{A}_a}\frac{\hp{\mu}\left(t+\tfrac{dt}{2}\right)-\hp{\mu}\left(t-\tfrac{dt}{2}\right)}{dt}\,,
\end{align}
where $\mathfrak{s}^a$ is as in (\ref{DefSaCluster}). We then obtain for its expectation value and variance
\begin{align}
&\langle \widehat{\dot{\mathbb{I}}}{}^a(t)\rangle\sim_{n\to \infty\atop dt\to 0}\dot{\mathbb{I}}^a(t)\,\left(1+\frac{1}{2n}\right)\,,&&\text{and} &&\text{Var}(\widehat{\dot{\mathbb{I}}}{}^a(t))\sim_{n\to \infty\atop dt\to 0}\frac{1}{n}\left(\frac{2}{dt^2}\,\frac{1-q^a}{q^a}-(\dot{\mathbb{I}}^a)^2\right)\,.\label{ExpClusterInfoForm}
\end{align}
Since $q^a=\sum_{\mu\in\mathbb{A}_a}p^\mu$, generally the variance $\text{Var}(\widehat{\dot{\mathbb{I}}}{}^a(t))$ can be smaller than $\text{Var}(\widehat{\dot{\mathcal{I}}}{}^\mu(t))$. Indeed, in Figure~\ref{Fig:ExpectClusterInf} (see Section~\ref{Sect:Examples}), we demonstrate this at the example of a particular model.

\section{Illustrative Example: S$\text{I}^{N+1}$R Model}\label{Sect:Examples}
In this section, we illustrate the theoretical results presented above through an example. To this end, we choose a simple \emph{SIR-model} that describes the spread of $N+1$ variants of a pathogen in a host population. These variants, endowed with slightly different properties and initial conditions, represent the degrees of freedom of the system, allowing us to exhibit the effects of clustering. 

The dynamical system we are studying is a simple \emph{compartmental model}\footnote{Such models have a long standing history: the earliest incarnations go back over a century \cite{Hamer,HamerLect1,HamerLect2,HamerLect3,Ross1911,Ross1916,RossHudson1916II,RossHudson1916III,McKendrick1912,McKendrick1914,McKendrick1926,Kermack:1927} and they have been studied and extended ever since. For a number of excellent reviews see for example \cite{BookAnderson,BookBauer,BookBrauerWu,Capasso1993MathematicalSO,BookDiekmann,BookKeeling,BookMartcheva,cacciapaglia2021field}. In this work, we only deal with a simple model, whose basic form (\emph{i.e.} for one pathogen) was proposed in \cite{Kermack:1927}.},  which divides the population into a number of different classes depending on their epidemiological status. The rate at which individuals transition from one compartment into another (\emph{i.e.} the rate at which they become infected or recover) are described by the following coupled differential equations 
\begin{align}
&\frac{dS}{dt} (t)= -\sum_{\nu=1}^{N+1} \gamma^\nu I^\nu(t) S(t)\,,&& \frac{dR}{dt}(t) = \sum_{\nu=1}^{N+1}\epsilon^\nu I^\nu(t)\,,&& \frac{dI^\mu}{dt}(t) = \gamma^\mu I^\mu(t) S(t) -\epsilon^\mu I^\mu(t)\,,\label{eq:model_equations}
\end{align}
where $S(t)\in[0,1]$ and $R(t)\in[0,1]$ represent the fraction of the whole population of \textit{susceptible} and \textit{recovered} individuals respectively, while $I^\mu(t)\in[0,1]$ (for $\mu=1,\ldots, N+1$) is the fraction of the population infected with the $\mu$-th variant of the pathogen. Furthermore, $\gamma^\mu$ (infection rate of the $\mu$-th variant) and $\epsilon^\mu$ (recovery rate of the $\mu$-th variant) are fixed numerical coefficients, which represent epidemiological properties of the $\mu$-th variant. The equations (\ref{eq:model_equations}) need to be supplemented by suitable initial conditions at $t_0=0$, such that $S(t_0)+\sum_{\mu=1}^{N+1}I^\mu(t_0)+R(t_0)=1$.\footnote{Notice, since (\ref{eq:model_equations}) implies 
$\tfrac{d}{dt}\left(S+\sum_{\mu=1}^{N+1}I^\mu+R\right)=0$, this fixes the normalisation of the total size of the population for all $t>t_0$.} From an epidemiological perspective, (\ref{eq:model_equations}) describes a fairly simple dynamics and is thus only understood as an illustrative example to exhibit some of the concepts outlined above.

\subsection{Statistical Model and Sampling}

\begin{figure}[h!]
\begin{center}
\includegraphics[width = 7.5cm]{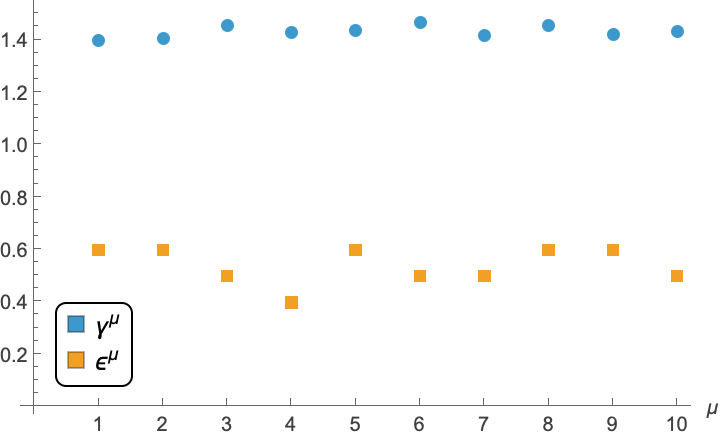}\hspace{1cm}\includegraphics[width = 7.5cm]{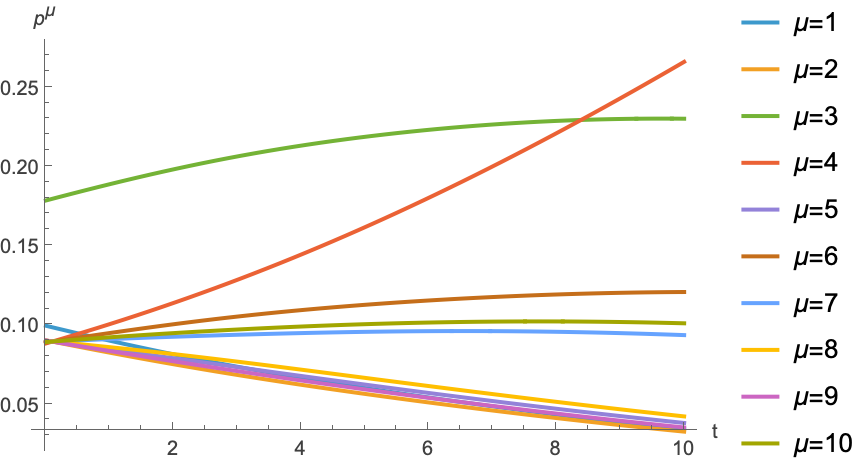}
\end{center}
\caption{Left panel: numerical choice of the transmission and recovery rates for $N+1=10$ pathogens. We shall use this choice for all numerical results in the remainder of this Section. Right panel: the probabilities $p^\mu(t)$ that constitute the statistical model $\mathcal{C}$ in (\ref{eq:probs}) as functions of time. They have been obtained as numerical solutions from (\ref{eq:model_equations}) using a Runge-Kutta formalism, with $\gamma^\mu$ and $\epsilon^\mu$ as shown in the left panel and with initial conditions such that $S(0)=0.9445$ and $R(0)=0$.}
\label{Fig:pathogens_probs}
\end{figure}

Based on (\ref{eq:model_equations}), we define a statistical model $\mathcal{C}$, by defining the probability distribution $p^\mu(t)$
\begin{align}
\mathcal{C}=\left\{p^\mu(t)=\frac{I^\mu(t)}{\sum_{\nu=1}^{N+1}I^\nu(t)}\bigg| \mu\in\{1,\ldots,N+1\}\text{ and }t\geq 0\right\}\,.\label{eq:probs}
\end{align}
From an epidemiological perspective, $p^\mu(t)$ describes the fraction of individuals infected with the $\mu$-th variant of the pathogen at time $t$. Figure~\ref{Fig:pathogens_probs} shows a numerical solution for the probabilities $p^\mu$ as functions of time, for a given choice of the infection and recovery rates for a system with $N+1=10$ variants. We shall discuss this particular $\mathcal{C}$ for the remainder of this Section (\emph{i.e.} in particular all numerical plots in this Section are based on this numerical solution of the $\widehat{p}^\mu$).\footnote{We also remark that we shall neglect the numerical error stemming from the Runge-Kutta procedure used to obtain the numerical solutions shown in the right panel of Figure~\ref{Fig:pathogens_probs}, as they are negligible compared to the statistical errors from the sampling process we shall discuss below.}

The couplings introduced in (\ref{Couplings}) that govern the dynamics of $\mathcal{C}$ can be determined from

\begin{wrapfigure}[7]{r}{0.45\textwidth}
\parbox{7cm}{\centering
\includegraphics[width = 6.5cm]{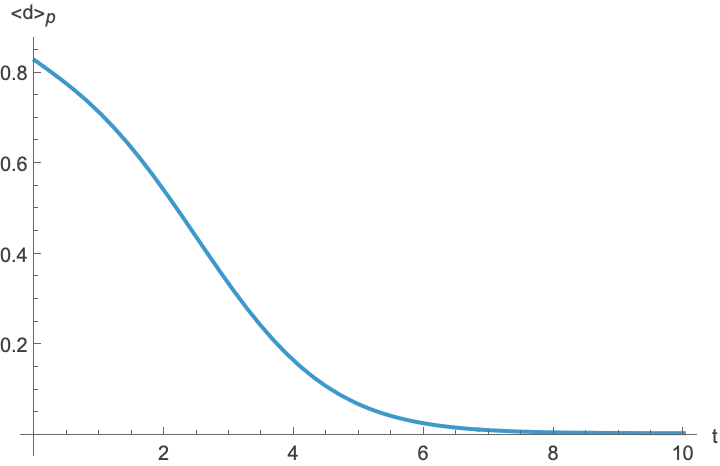}
\caption{The expectation value $\av$ (\ref{CouplingsSIR}) as a function of time for the numerical solutions found in Figure~\ref{Fig:pathogens_probs}.}
\label{Fig:pathogens_average}
${}$\\[-20pt]}
\end{wrapfigure}
\begin{figure}[h!]
\begin{center}
\includegraphics[width = 7.5cm]{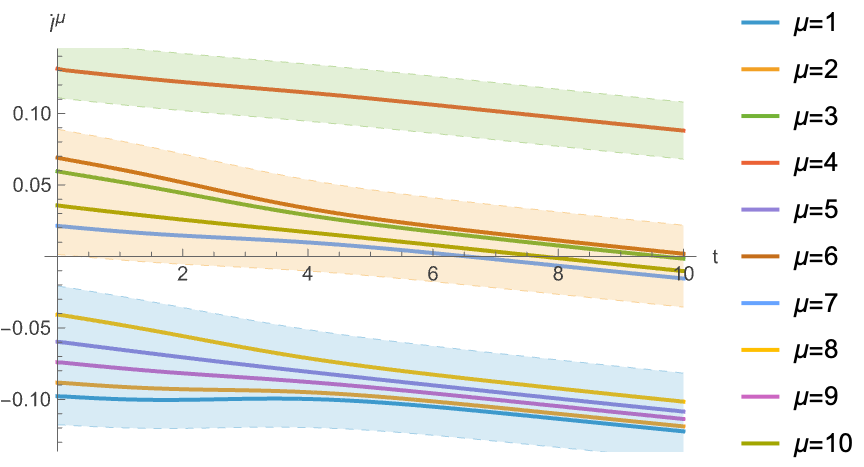}\hspace{1cm}\includegraphics[width = 7.5cm]{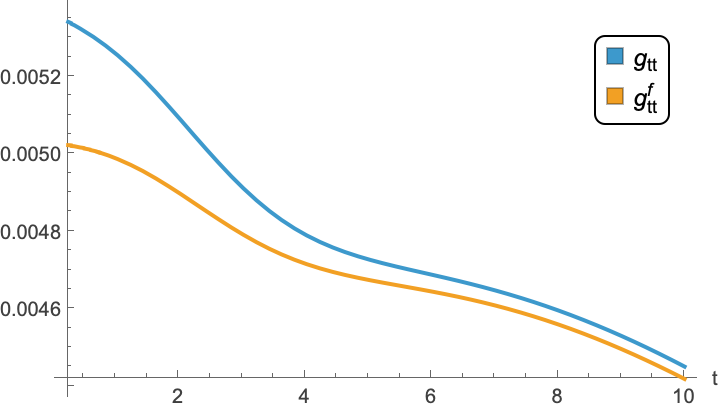}
\end{center}
\caption{Left panel: time derivative of the self-information $\dot{\mathcal{I}}^\mu$ for each of the $N+1=10$ variants as functions of time. Variants showing a similar $\dot{\mathcal{I}}^\mu$ are combined into a cluster, which is indicated by the shaded regions: cluster 1 (blue): $\mu\in\mathbb{A}_1=\{1, 2, 5, 8, 9\}$, cluster 2 (orange): $\mu\in\mathbb{A}_2=\{3, 6, 7, 10\}$, cluster 3 (green): $\mu\in\mathbb{A}_3=\{4\}$. Right panel: Fisher information $g_{tt}(t)$ (blue curve) (\ref{FisherInformationIs}) for the model $\mathcal{C}$ and $g_{tt}^f(t)$ (orange curve) (\ref{eq:clust_metric}) for the clustered model $\mathcal{C}_f$.}
\label{Fig:pathogens_Infos}
\end{figure}


\noindent
the differential equations (\ref{eq:model_equations})
\begin{align}
d^\mu(t)&=\frac{\dot{I}^\mu(t)}{I^\mu(t)}=\gamma^\mu \,S(t)-\epsilon^\mu\,,\nonumber\\
\av(t)&=\frac{\sum_{\nu=1}^{N+1}\dot{I}^\nu(t)}{\sum_{\nu=1}^{N+1}I^\nu(t)}\nonumber\\
&=\frac{\sum_{\nu=1}^{N+1}(\gamma^\nu S(t)-\epsilon^\nu)I^\nu(t)}{\sum_{\nu=1}^{N+1}I^\nu(t)}\,.\label{CouplingsSIR}
\end{align}
Notice that both $d^\mu(t)$ and $\av(t)$ not only depend on time, but also on quantities that are not immediately functions of the $p^\mu(t)$: notably $d^\mu(t)$ depends on $S(t)$ (along with the rates $\gamma^\mu$ and $\epsilon^\mu$). As was argued in \cite{FILOCHE2025130647}, it determines how well the $\mu$-th variant is adapted to the host population (represented by the number of susceptible) or, in other words, how it couples to the larger dynamical system represented by (\ref{eq:model_equations}). We have plotted $\av$ as a function of time in Figure~\ref{Fig:pathogens_average}, which generically requires the knowledge of the underlying equations (\ref{eq:model_equations}). The left panel of Figure~\ref{Fig:pathogens_Infos} shows the time derivative of the self-information (\ref{eq:information}) for each of the 10 variants, which can be obtained through (\ref{Couplings}) from (\ref{CouplingsSIR}). The $\dot{\mathcal{I}}^\mu$ (along with the $p^\mu$) allow to compute the Fisher information (see (\ref{FisherInformationIs})), which is shown as the blue curve in the right panel of Figure~\ref{Fig:pathogens_Infos}.

\begin{figure}[h!]
\begin{center}
\includegraphics[width = 7.5cm]{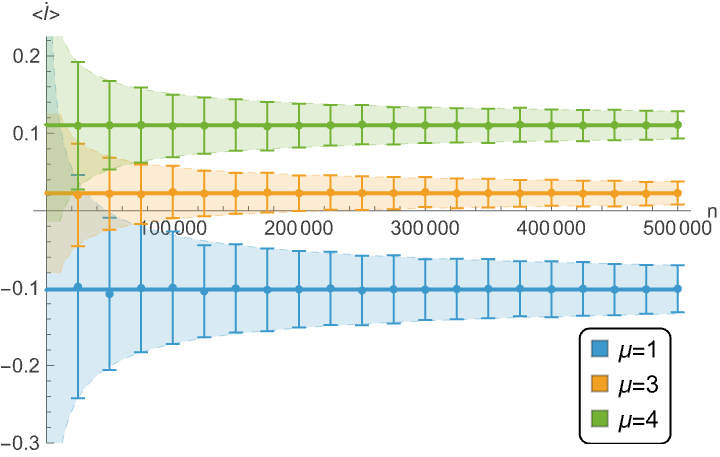}\hspace{1cm}\includegraphics[width = 7.5cm]{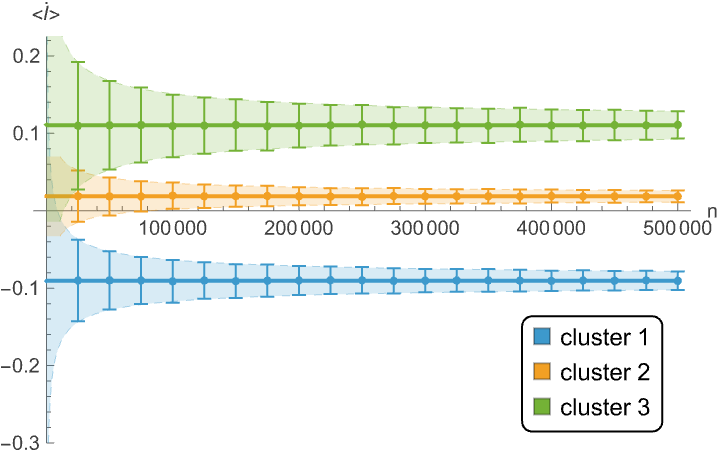}
\end{center}
\caption{Left panel: expectation value of $\langle \widehat{\dot{\mathcal{I}}}{}^\mu(t)\rangle$ at $t=5$ for the variants $\mu=1,3,4$ as a function of $n$. Right panel: expectation value of $\langle \widehat{\dot{\mathbb{I}}}{}^a(t)\rangle$ at $t=5$ for the three clusters $a=1,2,3$ as a function of $n$. In both plots, the discrete points denote the expectation values of 1000 samplings of $\widehat{p}^\mu(t\pm dt)$ (for $dt=0.25$ and given $n$) and the error bars represent the associated standard deviation. The solid line represents the analytically determined values (\ref{ExpInformationForm}) (left) and (\ref{ExpClusterInfoForm}) (right), with the shaded region representing the standard deviation determined from (\ref{VarInformationForm}) (left) and (\ref{ExpClusterInfoForm}) (right).}
\label{Fig:ExpectClusterInf}
\end{figure}

As explained in Section~\ref{Sect:Sampling}, from $\mathcal{C}$ we can obtain a (discrete) model $\widehat{\mathcal{C}}$ on $\Delta_{N+1}^{(n)}$ by sampling the $p^\mu(t)$ at periodic time intervals. From the latter, we can compute $\widehat{\dot{\mathcal{I}}}{}^\mu$ (see (\ref{DefSampledInfo})), whose expectation value was discussed in Section~\ref{Sect:ClusteringSampledData}. The left panel of Figure~\ref{Fig:ExpectClusterInf} shows a numerical evaluation of these expectation values $\langle\widehat{\dot{\mathcal{I}}}{}^\mu(t)\rangle$ (for three variants) at a given $t$ for different values of $n$ and compares it to the analytically computed value in (\ref{ExpInformationForm}), showing excellent agreement.

\begin{figure}[h!]
\begin{center}
\includegraphics[width = 7.5cm]{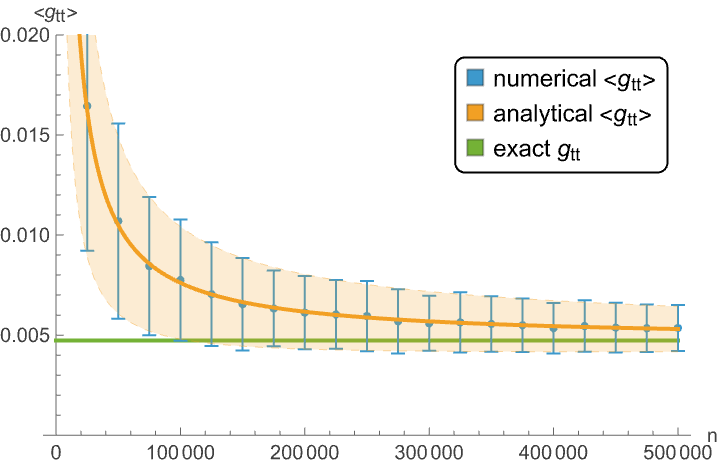}\hspace{1cm}\includegraphics[width = 7.5cm]{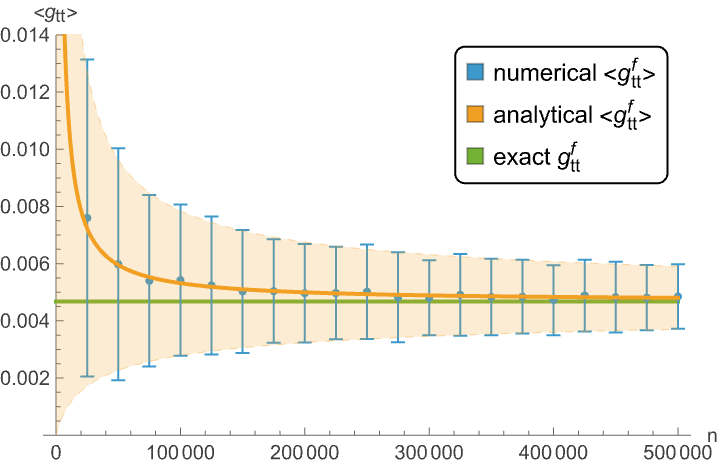}
\end{center}
\caption{Left panel: expectation value of the Fisher information $\langle \widehat{g}_{tt}(t)\rangle$ in (\ref{eq:metr_phat}) at $t=5$ as a function of $n$. Right panel: expectation value of the Fisher information $\langle \widehat{g}^f_{tt}(t)\rangle$ in (\ref{ClusteredFisherInformation}) for the clustered system at $t=5$ as a function of $n$. In both plots, the discrete points denote the expectation values of 500 samplings of $\widehat{p}^\mu(t\pm dt)$ (for $dt=0.25$ and given $n$) and the error bars represent the associated standard deviation. The solid orange line represents the analytically determined values (\ref{ExpValueFisher}) (left) and (\ref{ExpValueClusterFin}) (right), with the shaded region representing the standard deviation determined from (\ref{VarianceFinal}) (left) and (\ref{VarianceClusterApp}) (right). Notice also that $\Delta g_{tt}(t=5)\sim 0.000053$, such that the difference between the green line in the left panel and the one in the right panel is negligible compared to the statistical effects shown in these plots. For comparison, the green solid line represents the exact value of $g_{tt}$ (left) in (\ref{eq:metr_phat}) and $g_{tt}^f$ (right) in (\ref{eq:clust_metric}) obtained from the dynamical equations (\ref{eq:model_equations}).}
\label{Fig:ExpectClusterFunN}
\end{figure}

\begin{figure}[h!]
\begin{center}
\includegraphics[width = 7.5cm]{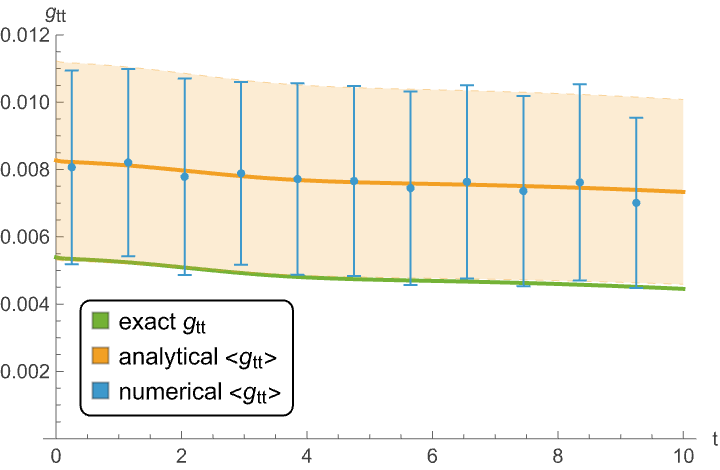}\hspace{1cm}\includegraphics[width = 7.5cm]{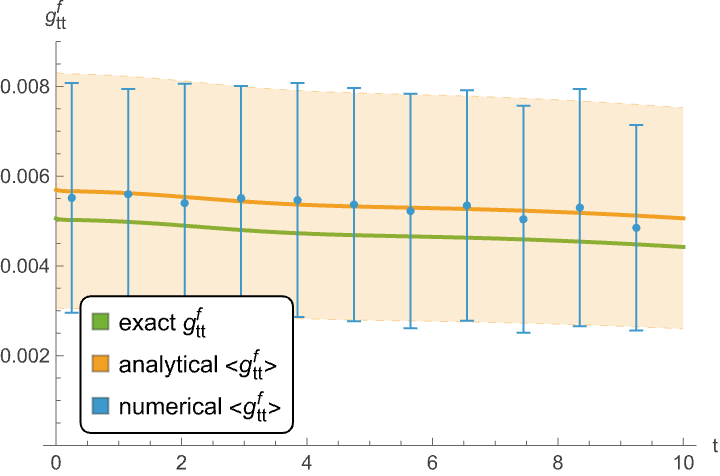}
\end{center}
\caption{Left panel: expectation value of the Fisher information $\langle \widehat{g}_{tt}(t)\rangle$ in (\ref{eq:metr_phat}) for $n=100.000$ as a function of $t$. Right panel: expectation value of the Fisher information $\langle \widehat{g}^f_{tt}(t)\rangle$ in (\ref{ClusteredFisherInformation}) for the clustered system for $n=100.000$ as a function of $t$. In both plots, the discrete points denote the expectation values of 500 samplings of $\widehat{p}^\mu(t\pm dt)$ (for $dt=0.25$) and the error bars represent the associated standard deviation. The solid orange line represents the analytically determined values (\ref{ExpValueFisher}) (left) and (\ref{ExpValueClusterFin}) (right), with the shaded region representing the standard deviation determined from (\ref{VarianceFinal}) (left) and (\ref{VarianceClusterApp}) (right). For comparison, the green solid line represents the exact value of $g_{tt}$ (left) in (\ref{eq:metr_phat}) and $g_{tt}^f$ (right) in (\ref{eq:clust_metric}) obtained from the dynamical equations (\ref{eq:model_equations}).}
\label{Fig:pathogens_FisherInformationTime}
\end{figure}

With the help of the sampled $\widehat{p}^\mu (t)$, we can calculate the Fisher information for the model $\widehat{\mathcal{C}}$, as defined in (\ref{eq:metr_phat}). We have plotted the expectation value $\langle \widehat{g}_{tt}\rangle$ as a function of $n$ for fixed $t$ (left panel of Figure~\ref{Fig:ExpectClusterFunN}) and as a function of $t$ for fixed $n$ (left panel of Figure~\ref{Fig:pathogens_FisherInformationTime}), showing excellent agreement between the analytically determined values for large $n$ and small $dt$, with the numerical expressions.

\subsection{Clustering}
We next consider clustering the $N+1$ variants. following the proposal in \cite{FILOCHE2025130647}, we group together degrees of freedom with similar $\dot{\mathcal{I}}^\mu(t)$. The latter are plotted in the left panel of Figure~\ref{Fig:pathogens_Infos}, which naturally suggest the following clustering function for $\ell = 3$ (as in (\ref{ClusteringFunction}))
\allowdisplaybreaks
\begin{align}
f:\,\mathbb{V}=\{1,\ldots,10\}&\longrightarrow \mathbb{W}=\{1,\ldots,\ell\}\nonumber\\
\mu &\longmapsto f(\mu)=\left\{\begin{array}{lcl}1 & \text{if} & \mu\in\mathbb{A}_1=\{1, 2, 5, 8, 9\}\,,\\
2 & \text{if} & \mu\in\mathbb{A}_2=\{3,6,7,10\}\,, \\ 3 & \text{if} & \mu\in\mathbb{A}_3=\{4\}\,, \end{array}\right.\label{ClusterFunctionExample}
\end{align}
where we have $\mathbb{V}=\mathbb{A}_1\cup \mathbb{A}_2\cup\mathbb{A}_3$. This clustering is also schematically indicated in the left panel of Figure~\ref{Fig:pathogens_Infos}.\footnote{In \cite{FILOCHE2025130647}, dedicated clustering algorithms (like K-means \cite{MR90073,MR214227,Lloyd,Forgy}) have been used to cluster real world epidemiological data. In the current simple example, we shall not require such methods.} We can thus define a clustered statistical model $\mathcal{C}_f$ as in (\ref{StatModelCluster}). The Fisher information $g_{tt}^f$ (defined in (\ref{eq:clust_metric})) associated with this model is plotted as the orange curve in Figure~\ref{Fig:pathogens_Infos}: from this Figure it is clear that (\ref{ClusterFunctionExample}) is not a sufficient statistics, since $\Delta g_{tt}=g_{tt}(t)-g_{tt}^f(t)> 0$ $\forall t\geq 0$. However, as is evident in the left panel of Figure~\ref{Fig:ExpectClusterFunN}, for the entire range of $n$ considered, $\Delta g_{tt}\sim 0.000053$ is (much) smaller than the bias of the metric as computed in (\ref{ExpValueFisher}). Therefore, clustering the degrees of freedom leads to a significant net reduction in the bias, thus allowing the metric $g_{tt}^f$ to be determined with higher accuracy and thus allowing a better description of the dynamics of $\mathcal{C}_f$. 

These results highlight the reduction due to clustering in the bias of the Fisher information obtained from sampling of the system at fixed time intervals, which thus leads to a more accurate description of the temporal evolution of the system based on these data. However, while the choice of the clusters is fairly straight-forward to determine from the underlying model $\mathcal{C}$ (\emph{i.e.} a fairly natural grouping appears from the time evolution of the $\dot{\mathcal{I}}^\mu$, as is evident from the left panel of Figure~\ref{Fig:pathogens_Infos}), it is not evident to which degree this clustering can be obtained from the sampled data alone: the expectation value $\langle \widehat{\dot{\mathbb{I}}}{}^a(t)\rangle$ for a given $t$ is shown in the right panel of Figure~\ref{Fig:ExpectClusterInf}. Comparing with the left panel (which shows the expectation value $\langle \widehat{\dot{\mathcal{I}}}{}^\mu(t)\rangle$ for one variant in each of the three clusters), we see that notably the standard deviation is much smaller in the clustered case.

\begin{figure}[h!]
\begin{center}
\includegraphics[width = 7.5cm]{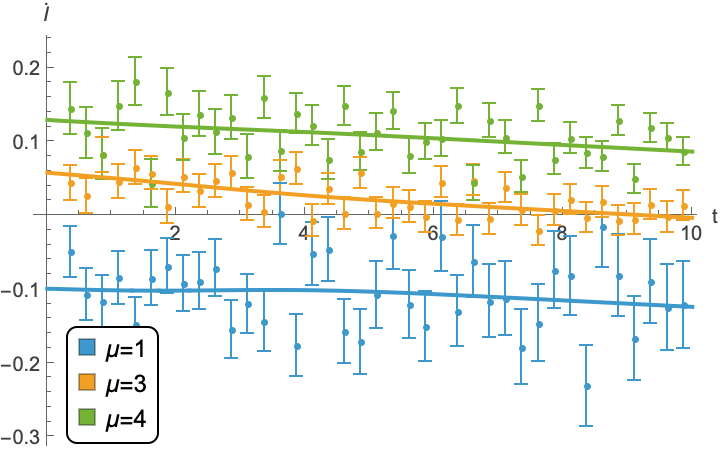}\hspace{1cm}\includegraphics[width = 7.5cm]{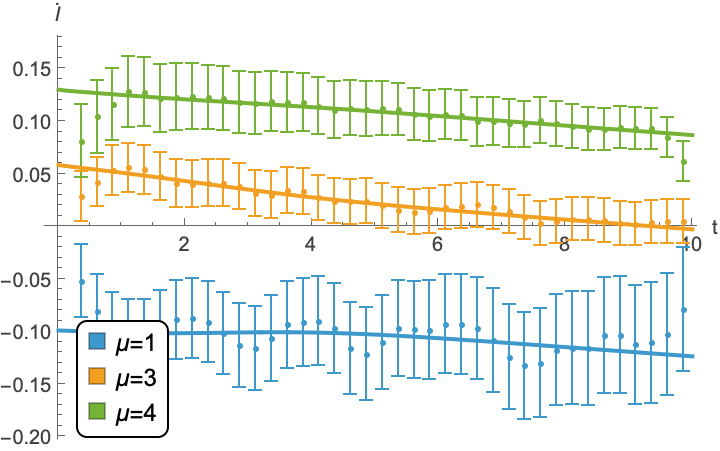}
\end{center}
\caption{ Left panel: comparison of the change of information $\widehat{\dot{\mathcal{I}}}{}^\mu$ (solid points) obtained from a single sampling of $\mathcal{C}$ with the exact value $\dot{\mathcal{I}}^\mu$ (solid line) for three variants. The error bars indicate the standard deviation as in (\ref{DefSampledInfo}). As is evident, for certain times $t$, the theoretical value of $\dot{\mathcal{I}}^\mu$ lies outside of the range set by the latter. Right panel: same comparison after Gaussian filtering of $\widehat{C}$ with the function $\varphi(k)=\frac{e^{-4k^2/9}}{\sum_{m=-3}^3 e^{-4m^2/9}}$. Both plots use $n=250.000$ and $dt=0.25$.}
\label{Fig:pathogens_FisherInformationTimeFilter}
\end{figure}

As mentioned in the previous Section, while the expectation value (\ref{ExpInformationForm}) leads to the same clustering as the (exact) $\dot{\mathcal{I}}{}^\mu$, the large standard deviation of this result may pose problems in practice. In order to appreciate this point, the left panel of Figure~\ref{Fig:pathogens_FisherInformationTimeFilter} shows examples of $\widehat{\dot{\mathcal{I}}}^\mu$ based on a model $\widehat{\mathcal{C}}$ obtained from a sampling with $n=250.000$, in relation to $\dot{\mathcal{I}}^{\mu}$ calculated from $\mathcal{C}$.\footnote{The standard deviation for an individual data point has been imposed from (\ref{DefSampledInfo}). Alternatively, it can be obtained by re-sampling techniques such as the jackknife \cite{jackknife} or the bootstrap \cite{Young,Schervish}, leading to comparable results.} While the three clusters are globally still separated, it is generally difficult at a fixed $t$ to associate the $\widehat{{\dot{\mathcal{I}}}^\mu}$ with one of the three clusters. This observation suggests that it is more advantageous to also take into account other data points of $\widehat{\mathcal{C}}$ for the clustering procedure. As an example, the right panel of Figure~\ref{Fig:pathogens_FisherInformationTimeFilter} shows the change of information for different variants of $\widehat{\mathcal{C}}$ after the $\widehat{p}$ have been filtered with a Gauss weight.\footnote{Concretely, we have replaced the $\widehat{p}$ in the definition (\ref{DefSampledInfo}) by $\widetilde{p}^\mu(t)=\sum_{k=-\ell}^\ell\varphi(k)\,\widehat{p}^\mu(t+k dt)$, where $\varphi$ is specified in the caption of Figure~\ref{Fig:pathogens_FisherInformationTimeFilter}.} As is evident, the individual data points are much closer to the theoretical values of $\dot{\mathcal{I}}^\mu$ which allows to unambiguously cluster them. The importance of Gaussian filtering has already been noted in \cite{FILOCHE2025130647} and we shall discuss it (along with other techniques related to the time-dependence of $\mathcal{C}$) in more detail in \cite{Companion}.

\begin{figure}[h]
\begin{center}
\includegraphics[width = 7.5cm]{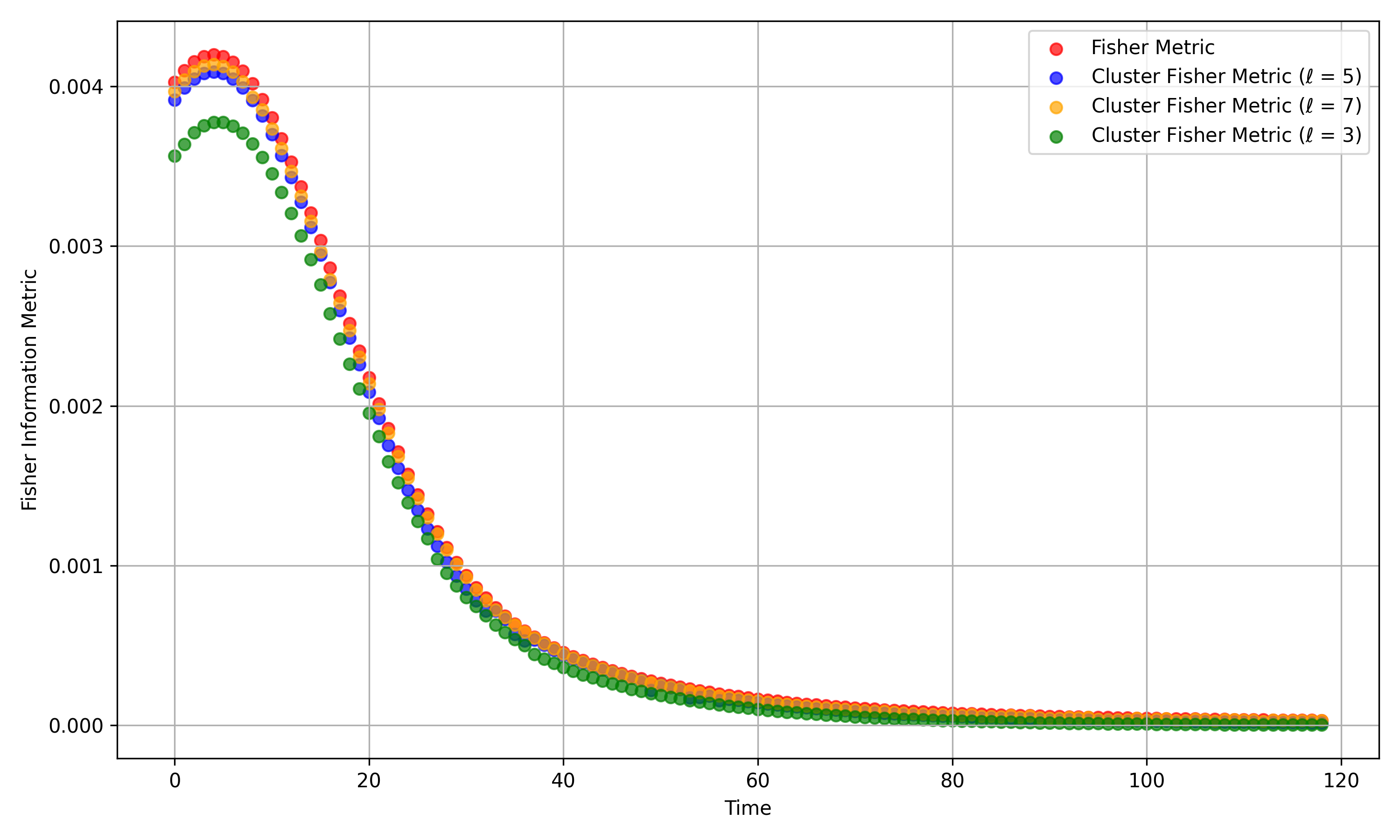}\hspace{1cm}\includegraphics[width = 7.5cm]{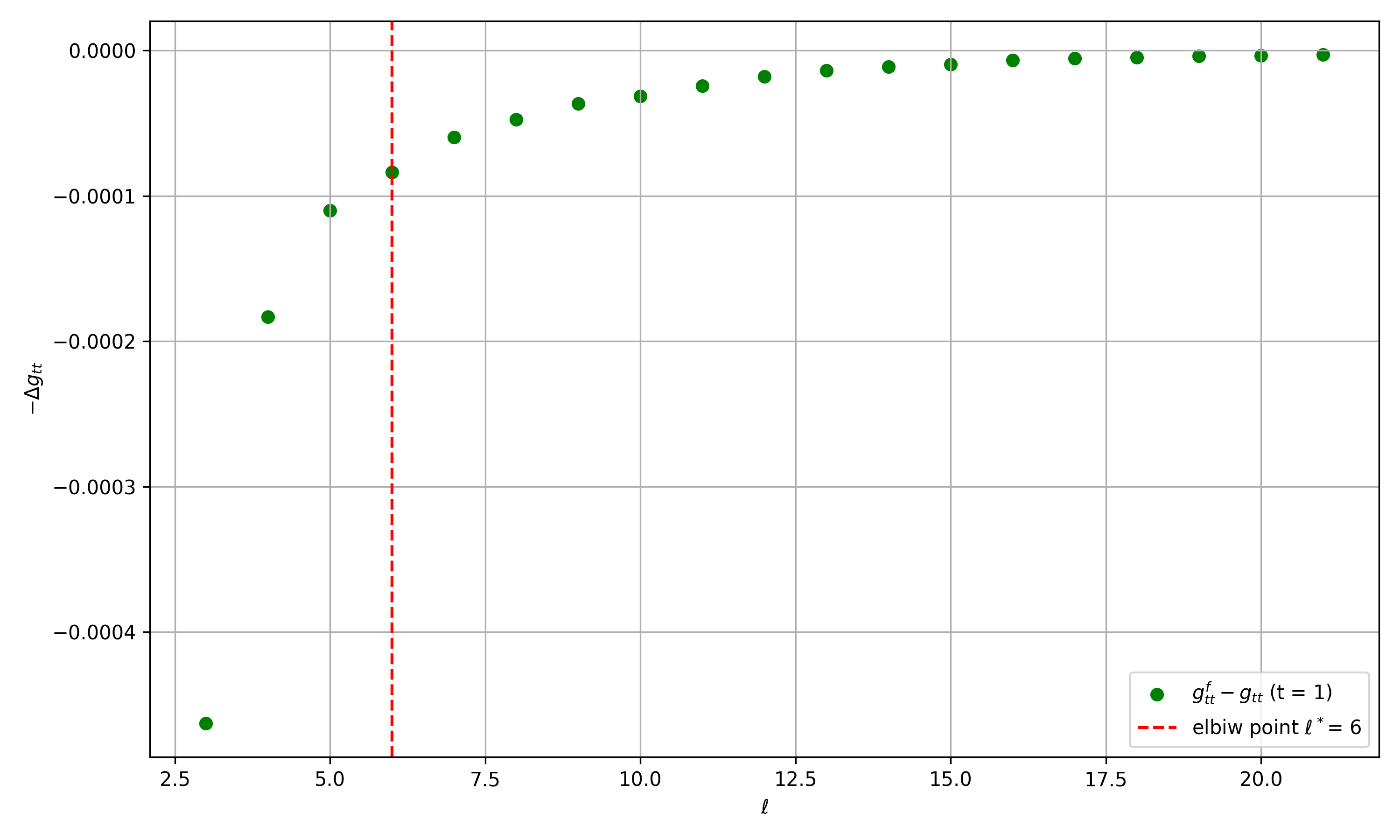}
\end{center}
\caption{Left panel: comparison between $g_{tt}(t)$ and $g_{tt}^f(t)$ (for several $\ell$) obtained from a SIR model with $N+1=50$. Increasing the number of clusters $\ell$ generally decreases $\Delta g_{tt}$. Right panel: difference $-\Delta g_{tt}=g_{tt}^f-g_{tt}$ over $\ell$ obtained from the Fisher information in the left panel for $t=1$. We find an elbow for this curve at $\ell^*=6$, thus providing a strategy to find the value for $\ell$ that maximises the performances of the clustering: indeed, increasing the number of $\ell$ beyond $\ell^*$ leads to a diminishing decrease of $-\Delta g_{tt}$.}
\label{Fig:delta_comp}
\end{figure}

Finally, before closing this Section, we also comment on the method in which the clustering $\mathcal{C}_f$ is obtained. As mentioned before, in the example described in this Section, the form of the $\dot{\mathcal{I}}$ (see Figure~\ref{Fig:pathogens_Infos}) suggests a natural clustering. In other examples this might be less obvious and more sophisticated methods need to be applied to find a clustering that minimises the loss of information in (\ref{eq:clust_metric}) (see also \cite{FILOCHE2025130647}). For a SIR model (\ref{eq:model_equations}) with $N+1=50$, Figure~\ref{Fig:delta_comp} shows the loss of information $\Delta g_{tt}$ as a function of the number of clusters. Here a K-means algorithm \cite{MR90073,MR214227,Lloyd,Forgy} was used to cluster the degrees of freedom. As is shown by the right panel of this Figure, an optimal number of clusters can be obtained by studying the shape of the curve of $\Delta g_{tt}$ as a function of $\ell$.

\section{Conclusions}\label{Sect:Conclusions}
In this paper, we study the impact of statistical noise on the sampled time-series of dynamical systems. Using the framework of information geometry, we interpret statistical models (as defined in (\ref{statisticalmodel})) geometrically, in the form of (smooth) curves on a hyperplane $\Delta_{N+1}$ of $\mathbb{R}^{N+1}$, where $N$ denotes the number of degrees of freedom of the system. A key quantity in describing the dynamics of the model is the Fisher information, which captures the norm of an infinitesimal displacement in time along $\mathcal{C}$. This norm is calculated with respect to the Shahshahani metric (\ref{NDimShashahani}), which more generally allows to define a distance between probability distributions on the hyperplane $\Delta_{N+1}$. 

While this framework allows to describe dynamical systems (more precisely statistical models) in a unified fashion, in many applications we are interested in the somewhat different problem of reconstructing $\mathcal{C}$ from a given time-series of data. Concretely, here we consider that at periodic time intervals $dt$, i.i.d. samples of size $n$ are taken, leading to a discrete statistical model $\widehat{\mathcal{C}}$ (see (\ref{SampledModel})), which geometrically is represented by an ordered sequence of discrete points in $\Delta_{N+1}$. In this work, we are interested in the question to what extent the dynamics of $\mathcal{C}$ can be inferred from $\widehat{\mathcal{C}}$, or more concretely to gauge the impact of the statistical noise (in the form of the sample size $n$ and the sampling interval $dt$). While (a variant of) the central limit theorem (see Appendix~\ref{App:CentralLimit}) implies that for given time $t$ and for large $n$, the expectation value of the (Shahshahani) distance between the probabilities $p(t)$ appearing in $\mathcal{C}$ and the sampled probabilities $\widehat{p}$ behaves like $ \frac{N}{n}$ (see (\ref{ExpDistance})), gauging the impact on the dynamics is more subtle: indeed, to this end, the probabilities cannot simply be compared at fixed time $t$, but instead we need to consider the rate of change of the probabilities with time, \emph{i.e.} we need to consider (discretised versions of) the derivatives of $\widehat{p}$, which require the input from samplings at multiple times: in this work we us a proposal in \cite{FILOCHE2025130647} for the Fisher information on $\widehat{\mathcal{C}}$ (see (\ref{eq:metr_phat})), which replaces the time derivatives of $p(t)$ in (\ref{FisherInformationDef}) by a discretised version of the time series $\widehat{p}$. In this way, $\widehat{g}_{tt}(t)$ depends on the sampled probabilities $\widehat{p}^\mu$ at $t\pm dt/2$, which allows us to compute its expectation value (\ref{ExpValueFisher}) and variance (\ref{VarianceFinal}) to leading order in $n$ and $dt$. We illustrate (and validate) these theoretical results by comparing with the numerical computations of a simple theoretical model (see the left panels of Figures~\ref{Fig:ExpectClusterFunN} and \ref{Fig:pathogens_FisherInformationTime}). Remarkably, the leading term in the bias of $\widehat{g}_{tt}$ is $\sim\tfrac{2N}{n\,dt^2}$, and thus independent of the details of the dynamics and only a function of the quality of the sampling (\emph{i.e.} $n$ and $dt$) as well as the number of degrees of freedom. In cases, where the sampling cannot be changed (which is indeed a limiting factor in many applications), our results suggest two strategies of reducing the bias:
\begin{itemize}
\item[\emph{(i)}] coarse grain the information in $\widehat{\mathcal{C}}$ to reduce the (effective) number of degrees of freedom~$N$ 
\item[\emph{(ii)}] change the definition of  estimator to make better use of the estimator $\widehat{g}_{tt}$ in (\ref{eq:metr_phat}) to make better use of more data of the time series $\widehat{\mathcal{C}}$
\end{itemize}
In this work, we have discussed in detail strategy \emph{(i)}, while \emph{(ii)} shall be discussed in the follow-up work \cite{Companion}. Indeed, following a proposal in \cite{FILOCHE2025130647}, we have considered grouping the $(N+1)$ probabilities comprising $\mathcal{C}$ into $\ell$ clusters thus defining a new clustered statistical model $\mathcal{C}_f$ (see (\ref{StatModelCluster})), which represents a curve on the (lower-dimensional) simplex $\Delta_{\ell}$. This has two effects:
\begin{itemize}
\item The Fisher information $g^f_{tt}$ associated with $\mathcal{C}_f$ is generally smaller than $g_{tt}$: the clustering leads to a loss of information on the dynamics of the system. Indeed, we have quantified this loss (see (\ref{VarianceCluster})) in terms of the variances of the couplings $d^\mu$ within each cluster. As defined in \cite{FILOCHE2025130647}, these couplings are a universal form of describing the interaction of an individual degree of freedom with the remaining dynamical system. Notably, these variances (and thus also the loss of information) can be minimised by grouping together degrees of freedom that 'behave similarly': our result (\ref{VarianceCluster}) quantifies this statement and links it to the particularities of the dynamics. Indeed, the couplings are specific to the model under consideration and in the SIR example studied in Section~\ref{Sect:Examples} represent a combination of infection- and recovery rate at a given epidemiological situation (see~(\ref{CouplingsSIR})).
\item By keeping track of a reduced number of degrees of freedom (\emph{i.e.} for $\ell<N+1$), the bias (\ref{ExpValueClusterFin}) and variance (\ref{VarianceClusterApp}) of the Fisher information $\widehat{g}_{tt}^f$ obtained from the sampled data (see (\ref{ClusteredFisherInformation}) for the definition) is generally smaller. Indeed, these results are numerically illustrated in the right panels of Figures~\ref{Fig:ExpectClusterFunN} and \ref{Fig:pathogens_FisherInformationTime}.
\end{itemize}
Our results allow to explicitly quantify the trade off between the loss of information and the reduction in the bias due to the clustering of the dynamical system. Moreover, we are able to link the changes in the (sampled) Fisher information to quantities pertaining either to the underlying physical system or the sampling process respectively. This allows to gauge in concrete situations what level of coarse graining of the system is required to describe the dynamics adequately in view of a finite amount of data.

A question remaining in practice is how to find a clustering, which minimises the loss of (Fisher) information: a concrete proposal to this end was made in \cite{FILOCHE2025130647}, by clustering the degrees of freedom according to the change of their self-information (see (\ref{eq:information})). The expectation value of the individual self-informations (see (\ref{ExpInformationForm})) is to leading order only multiplied by a universal factor (which only depends on $n$ and $dt$). Therefore,  the clustering of these expectation values yields the same result as for the $\dot{\mathcal{I}}^\mu$ stemming from $\mathcal{C}$ itself. However, our theoretical computations (\ref{VarInformationForm}) and numerical examples (see left panel of Figure~\ref{Fig:pathogens_FisherInformationTimeFilter}) show that the associated variance can become significantly large, thus leading to probelms on how to identify in practice the clusters from the sampled data. However, in line with strategy \emph{(ii)} mentioned above, preliminary results also indicate that these results can drastically be improved by first applying a Gaussian filtering to the $\widehat{p}^\mu(t)$, \emph{i.e.} averaging the sampled data as functions of time first. 

This is indeed a first indication that the combination of the two strategies \emph{(i)} and \emph{(ii)} provides powerful tools to analyse numerical systems when only finite amounts of data are available. We shall provide further evidence to this idea in \cite{Companion}, where we shall discuss different ways to extract the Fisher information from the sampled data: we show that replacing the proposed $\widehat{g}_{tt}$ in (\ref{eq:metr_phat}) by a different estimator, the same time series sampling gives a much better representation of the underlying $g_{tt}$ and provides a reduced bias. We shall discuss different proposals for such estimators and describe their properties.

While a major part of the inspiration in this paper stems from epidemiological models (notably the paper \cite{FILOCHE2025130647}\footnote{Indeed, in this work it was demonstrated that clustering variants of SARS-CoV-2 (\emph{i.e.} the degrees of freedom of the system) according to the change of their self information, groups them according to their adaptation to a larger dynamical system that is constituted by the entire population. It was demonstrated, that this clustering reveals genetic specificities and mutations of the different variants that provided competitive and evolutionary advantages over others. Systematically studying these advantages over a longer period also exhibits, how these advantaged shift in the context of the larger dynamical system and countless parameters (biological, epidemiological, socio-demographic, etc.) that influence the reproductive capabilities of the virus.}), our results hold very generically and can be applied to a number of different situations. Within an epidemiological context, the sampled data stem from the (infected) patients that are tested in periodic intervals (\emph{e.g.} each week in the case of SARS-CoV-2) and reveal microbiological and genetic properties of the pathogen that spreads through the population. These data are naturally limited and one usually tries to describe (and predict) the spread of the disease with this finite amount of data as best as possible. Such limitations arise in countless other situations and the tools developed in this work provide quantitative insights into how much of the corresponding dynamics can be captured by the available data. As we have outlined in Section~\ref{Sect:Sampling}, while other works have addressed similar questions, to our knowledge, the approach based on the Fisher information presented in Section~\ref{sec:info_theory} (pioneered in \cite{FILOCHE2025130647}) is novel and we foresee numerous applications in the future.


\section*{Acknowledgements}
We would like to thank Andrea Cimarelli, Baptiste Filoche, Marta Nunes and Francesco Sannino for numerous enlightening discussions.
\textit{This research was supported by EU funding within the NextGenerationEU-MUR PNRR Extended Partnership initiative on Emerging Infectious Diseases (Project no. PE00000007, INF-ACT)}

\appendix

\section{Tools in Information Theory and Statistics}\label{App:CentralLimit}
In this Appendix, we review a number of concepts in statistics and probability theory, which are relevant for the bulk of this paper. For general reviews on this topics, we refer the reader to a number of excellent books \cite{CoverInformation,Schervish,Goldmann,Feller,Feller2,wasserman2010statistics,DemboZeitouni,Nielsen}.

\noindent
We first review the \emph{Central Limit Theorem (CLT)}, which can be stated formally as follows
\begin{center}
\parbox{15.5cm}{{\bf Theorem B.97, p.642 in \cite{Schervish}}: {\it Suppose that $\{X_i\}_{i=1}^\infty$ is a sequence that is i.i.d. with finite mean $\mu$ and finite variance $\sigma^2$. Let $\overline{X}_n$ be the average of the first $n$ $X_i$'s. Then $\sqrt{n}(\overline{X}_n-\mu)\stackrel{\mathcal{D}}{\longrightarrow} N(0,\sigma^2)$, the normal distribution with mean $0$ and variance $\sigma^2$.}}
\end{center}
Here the arrow represents convergence in distribution\footnote{As stated in \cite{Schervish} for an estimator $E$, the sequence $\{X_i\}_{n=1}^\infty$ converges in distribution to $X$ if $\lim_{n\to\infty} E(f(X_n))=E(f(X))$ for every bounded continuous function $f$, which is denoted $X_n \stackrel{\mathcal{D}}{\longrightarrow} X$.} A formulation closer to our discussion in Section~\ref{Sect:DistModels} can be stated as follows (see \emph{e.g.}~\cite{Paninski}, multinomial CLT): the empirical measure of $\widehat{p}$ is asymptotically normal, concentrated on an ellipse of size $\sim n^{-1/2}$ around the true discrete measure $p$. This is indeed what we have found in (\ref{ExpDistance}).

The \emph{method of types} \cite{Csiszar_Types,Csiszar_Korner_2011,Longo} (see also \cite{Wolfowitz} and \cite{Boltzmann,Schroedinger} for similar earlier ideas in the context of physics) studies sequences $X=(X_1\,,X_2\,,\ldots,X_n)$ of $n$ symbols from a given alphabet $\mathfrak{X}=(\alpha_1,\ldots,\alpha_{k})$. Following \cite{CoverInformation}, the \emph{type} of a sequence is $P_X=(\mathcal{N}(\alpha_1,X),\ldots,\mathcal{N}(\alpha_k,X))$, where $\mathcal{N}(\alpha_i,X)$ is the number of occurrences of $\alpha_i$ within $X$. In the context of the multinomial distribution (\ref{eq:prob_extract}), we in particular consider the \emph{set of types with denominator} $n$
\begin{align}
\mathcal{P}_n=\left\{\left(\frac{n_1}{n}\,,\frac{n_2}{n}\,,\ldots\,,\frac{n_k}{n}\right)\bigg|0\leq n_i\leq n\text{ and }\sum_{i=1}^k n_i=n\right\}\,,
\end{align}
and define the \emph{type class} (also called composition class)
\begin{align}
&T(Q)=\left\{(X_1,\ldots,X_n)\big|P_X=Q\right\}\,,&&\text{for} &&Q\in\mathcal{P}_n\,.
\end{align}
If $|\mathcal{P}_n|$ denotes the number of elements in $\mathcal{P}_n$, the number of elements in $T(Q)$ satisfies \cite{Csiszar_Types,CoverInformation}
\begin{align}
&|\mathcal{P}_n|^{-1} e^{n H(Q)}\leq |T(Q)|\leq e^{n H(Q)}\,,&&\text{with} &&H(Q)=-\sum_{i=1}^k Q_i\,\log Q_i\,,
\end{align}
where $H$ is called the (Shannon) \emph{entropy} \cite{ShannonEntropy,WeaverShannon}.\footnote{We also remind that $\log$ denotes the logaritm to base $e$.} In \cite{Csiszar_Types}, this relation is abbreviated as $|T(Q)|\approx e^{n H(Q)}$, where $\approx$ denotes equality up to a polynomial factor in $n$. For any $Q\in\mathcal{P}_n$ the probability $\mathbb{P}(P_X,Q)$ for a sequence $X$ of $n$ elements drawn i.i.d. from $\mathcal{X}$ according to $Q$ is
\begin{align}
&\mathbb{P}(P_X,Q)= e^{-n (D(P_X||Q)+H(P_X))}\,,&&\text{with} &&D(P_X||Q)=\sum_{i=1}^k P_{X,i}\,\log\frac{P_{X,i}}{Q_i}\,,\label{DefKullbackLeibler}
\end{align}
where $D(P_X||Q)$ is called the \textit{Kullback-Leibler divergence}~\cite{KullbackLeibler} (see also \cite{Csiszar1,Csiszar2,Amari1,amari2000methods,LESNE_2014,Nielsen}). This result can also be stated as $\mathbb{P}(P_X,Q)\approx  e^{-n D(P_X||Q)}$ (see \cite{Sanov,Csiszar1}), where $\approx$ denotes equality up to polynomial factors in $n$. 

The Kullback-Leibler divergence is always positive, possesses a minimum equal to $0$ for $P_X = Q$, and is convex. These properties imply that the probability $\mathbb{P}(P_X,Q)$ is strongly concentrated for $P_X$ close to $Q$ and quickly decreases for large deviations of the two probabilities. We can therefore use \emph{large deviation theory} \cite{Hoefding,DemboZeitouni} (see \cite{Touchette_2009} for a review) to approximate notably integrals involving the probability $\mathbb{P}(P_X,Q)$ in the form of saddle-point approximations (see \emph{e.g.} \cite{BenderOrszag}). To this end, we concretely use Laplace's method by expanding the Kullback-Leibler divergence
\begin{align}
&D(P_X||Q)=\sum_{r=2}^\infty \sum_{i=1}^k\frac{(-1)^r}{r(r-1)}\frac{(P_{X,i}-Q_i)^r}{Q_i^{r-1}}\,,&&\text{for} &&\left|\frac{P_{X,i}-Q_i}{Q_i}\right|< 1\,,
\end{align}
such that to leading order
\begin{align}
&D(P_X||Q)= \frac{1}{2}\sum_{i=1}^k\frac{(P_{X,i}-Q_i)^2}{Q_i}+\mathcal{O}((P_{X,i}-Q_i)^3)\,.
\end{align}
The Gaussian approximation $\mathbb{P}(P_X,Q)\sim \text{exp}\left(-\frac{n}{2}\sum_{i=1}^k\frac{(P_{X,i}-Q_i)^2}{Q_i}\right)$ is accurate for $P_X$ around $Q$ to order $\mathcal{O}(n^{-1/2})$, but starts to significantly deviate for larger distances \cite{Touchette_2009,BRYC1993253} (for which, however, it is strongly suppressed). For concrete saddle-point approximations of specific integrals, we refer to the bulk of the paper, notably Section~\ref{Sect:Sampling} (see also Appendix~\ref{Sect:ExpectationValues}, as well as~\cite{Daniels,MartinLof}).


\section{Expectation Values}\label{Sect:ExpectationValues}
In this Appendix, we compile details for a number of computations that appear in the main part of this paper. We present approximations of certain expectation values, by re-writing them in the form of Gaussian integrals.

\subsection{Gaussian Integrals}\label{App:GaussInteg}
Before starting the actual computations, we first tabulate a number of useful integrals of Gauss-type: let $(p^\mu)_{\mu=1,\ldots,N+1}$ be a probability distribution, \emph{i.e.} $p^\mu\in[0,1]$ and $\sum_{\mu=1}^{N+1}p^\mu=1$. Using the integral measure defined in (\ref{DefGaussianIntMeasure}) we then have
{\allowdisplaybreaks
\begin{align}
&\int_{\mathbb{R}^{N+1}}\widetilde{d^{N+1}u}[p]=1\,,\hspace{7.15cm}\int_{\mathbb{R}^{N+1}}\widetilde{d^{N+1}u}[p]\,u^\mu=0\,,\nonumber\\
&\int_{\mathbb{R}^{N+1}}\widetilde{d^{N+1}u}[p]\,u^\mu\,u^\nu=\tfrac{1}{2}(\delta^{\mu\nu}-\sqrt{p^\mu p^\nu})\,,\hspace{3.5cm}\int_{\mathbb{R}^{N+1}}\widetilde{d^{N+1}u}[p]\,u^\mu\,u^\nu\,u^\rho=0\,,\nonumber\\
&\int_{\mathbb{R}^{N+1}}\widetilde{d^{N+1}u}[p]\,u^\mu\,u^\nu\,u^\rho\,u^\sigma=\tfrac{3}{4}\,\sqrt{p^\mu p^\nu p^\rho p^\sigma}-\tfrac{1}{4}\big(\delta^{\mu\nu}\sqrt{p^\rho p^\sigma}+\delta^{\mu\rho}\sqrt{p^\nu p^\sigma}+\delta^{\mu\sigma}\sqrt{p^\nu p^\rho}\nonumber\\
&\hspace{2cm}+\delta^{\nu\rho}\sqrt{p^\mu p^\sigma}+\delta^{\nu\sigma}\sqrt{p^\mu p^\rho}+\delta^{\rho\sigma}\sqrt{p^\mu p^\nu}\big)+\tfrac{1}{4}(\delta^{\mu\nu}\delta^{\rho\sigma}+\delta^{\mu\rho}\delta^{\nu\sigma}+\delta^{\mu\sigma}\delta^{\nu\rho})\,.
\end{align}}
\subsection{$\langle\mathfrak{a}^2(\widehat{p},p)\rangle$ and $\text{Var}_{p}(\mathfrak{a}^2(\widehat{p},p))$}\label{App:DistExp}
We first compute the expectation value of the distance $\langle\mathfrak{a}^2(\widehat{p},p)\rangle$ in (\ref{ExpDistance}): here $p(t)$ denotes an arbitrary point in the interior of $\Delta_{N+1}$ (\emph{i.e.} a point on the curve described by the model $\mathcal{C}$) and $\widehat{p}\in\Delta_{N+1}^{(n)}$ has been obtained from $n\gg 1$ samples of the latter. By definition, we then find for the expectation value 
\begin{align}
\langle\mathfrak{a}^2(\widehat{p},p)\rangle&=\sum_{\widehat{p}\in \Delta_{N+1}^{(n)}}\,\sum_{\mu=1}^{N+1}\frac{\left(\widehat{p}^\mu-p^\mu\right)^2}{p^\mu}\,\mathbb{P}(\widehat{p},t)\sim_{n\to\infty}\sum_{\widehat{p}\in \Delta_{N+1}^{(n)}}\,\sum_{\mu=1}^{N+1}\frac{\left(\widehat{p}^\mu-p^\mu\right)^2}{p^\mu}\,\frac{e^{-n D(\widehat{p}||p)}}{Z(p(t))}\,,\nonumber
\end{align}
where $\mathbb{P}(\widehat{p},t)$ is defined in (\ref{eq:prob_extract}) and we have used the approximation (\ref{eq:ProbHatp}): indeed, for $n\gg 1$, we have used the method of types (see Appendix~\ref{App:CentralLimit}) with $Z(p(t))$ defined in (\ref{DefNormalisationFactor}) to approximate the probability. Following a similar argument as in Section~\ref{Sect:CalculationExpValueVar}, for large $n$, we can replace the summation over discrete points of $\Delta_{N+1}^{(n)}$ by integrals over $\mathbb{R}^{N+1}$. For the latter, we use a similar saddle-point approximation (see \cite{BenderOrszag}) as in Section~\ref{Sect:CalculationExpValueVar}, in particular the expansion (\ref{ExpandKullback}). With  the integration measure given in (\ref{DefGaussianIntMeasure}) we can thus write\footnote{As explained in Section~\ref{Sect:CalculationExpValueVar}, we use the fact that for $n\gg 1$, the points of $\Delta_{N+1}^{(n)}$ lie dense in $\Delta_{N+1}$ and that the probability density for $\widehat{p}$ is strongly localised around $p$.} 
\begin{align}
\langle\mathfrak{a}^2(\widehat{p},p)\rangle&\sim_{n\to\infty} \frac{2}{n}\int_{\mathbb{R}^{N+1}}\widetilde{d^{N+1}u}[p]\,\sum_{\mu=1}^{N+1}(u^\mu)^2=\frac{2}{n}\,\sum_{\mu=1}^{N+1}\frac{1-p^\mu}{2}=\frac{N}{n}\,,
\end{align}
where we have used the Gaussian integrals compiled in Appendix~\ref{App:GaussInteg} as well as the fact that $\sum_{\mu=1}^{N+1}p^\mu=1$.

In order to determine the variance $\text{Var}(\mathfrak{a}^2(\widehat{p},p))$, we first compute
\begin{align}
\langle(\mathfrak{a}^2(\widehat{p},p))^2\rangle&=\sum_{\widehat{p}\in \Delta_{N+1}^{(n)}}\,\left(\sum_{\mu=1}^{N+1}\frac{\left(\widehat{p}^\mu-p^\mu\right)^2}{p^\mu}\right)\left(\sum_{\nu=1}^{N+1}\frac{\left(\widehat{p}^\nu-p^\nu\right)^2}{p^\nu}\right)\,\mathbb{P}(\widehat{p},t)\nonumber\\
&\sim_{n\to\infty}\sum_{\widehat{p}\in \Delta_{N+1}^{(n)}}\,\sum_{\mu,\nu=1}^{N+1}\frac{\left(\widehat{p}^\mu-p^\mu\right)^2}{p^\mu}\,\frac{\left(\widehat{p}^\nu-p^\nu\right)^2}{p^\nu}\,\frac{e^{-\frac{n}{2}\,\sum_{\rho=1}^{N+1}\frac{(\hp{\rho}(t)-p^\rho(t))^2}{p^\rho(t)}}}{Z(p(t))}\,.
\end{align}
Replacing the summation over $\Delta_{N+1}^{(n)}$ as before by integrals, we find 
\begin{align}
\langle(\mathfrak{a}^2(\widehat{p},p))^2\rangle&\sim_{n\to\infty} \frac{4}{n^2}\int_{\mathbb{R}^{N+1}}\widetilde{d^{N+1}u}[p]\,\sum_{\mu,\nu=1}^{N+1}(u^\mu)^2(u^\nu)^2\nonumber\\
&=\frac{4}{n^2}\,\sum_{\mu,\nu=1}^{N+1}\left(\frac{1}{4}\left(1-p^\mu-p^\nu+3p^\mu p^\nu\right)+\frac{1}{2}\,\delta^{\mu\nu}(1-2 p^\mu)\right)=\frac{N(N+2)}{n^2}\,,
\end{align}
For the variance we therefore find 
\begin{align}
\text{Var}(\mathfrak{a}^2(\widehat{p},p))=\langle(\mathfrak{a}^2(\widehat{p},p))^2\rangle-(\langle\mathfrak{a}^2(\widehat{p},p)\rangle)^2\sim_{n\to\infty}\frac{2N}{n^2}\,.
\end{align}


\subsection{Higher Order Expectation Values}
In this Appendix, we shall present more details for the computation of the expectation value and variance of $\widehat{g}_{tt}$ in Sections~\ref{Sect:ExpValuegtt} and \ref{Sect:Vargtt} respectively.
\subsubsection{Expectation Value}\label{App:ExpectValue}
We start from the series expansion (\ref{ExpansionTermSingle}) of the integrand in (\ref{SampleMetricExpansion})
{\allowdisplaybreaks
\begin{align}
&\frac{\left(p^\mu_++\sqrt{\frac{2 p^\mu_+}{n}}\,x^\mu-p^\mu_--\sqrt{\frac{2 p^\mu_-}{n}}\,y^\mu\right)^2}{p^\mu_++\sqrt{\frac{2 p^\mu_+}{n}}\,x^\mu+p^\mu_-+\sqrt{\frac{2 p^\mu_-}{n}}\,y^\mu}\nonumber\\
&=\frac{\left(p^\mu_+-p^\mu_-\right)^2}{p^\mu_++p^\mu_-}+\frac{\sqrt{2}(p_+^\mu-p_-^\mu)\left(3p_-^\mu\sqrt{p_+^\mu}\,x^\mu+(p_+^\mu)^{3/2}\, x^\mu-(p_-^\mu)^{3/2}\, y^\mu -3p_+^\mu \sqrt{p_-^\mu}\,y^\mu\right)}{\sqrt{n}(p_+^\mu+p_-^\mu)^2}\nonumber\\
&\hspace{0.5cm}+\frac{8p_+^\mu p_-^\mu\left(\sqrt{p_-^\mu}\, x^\mu-\sqrt{p_+^\mu}\, y^\mu\right)^2}{n(p_+^\mu+p^\mu_-)^3}-\frac{8\sqrt{2}p_+^\mu p_-^\mu\left(\sqrt{p_+^\mu}x^\mu+\sqrt{p_-^\mu}y^\mu\right)\left(\sqrt{p_-^\mu}x^\mu-\sqrt{p_+^\mu}y^\mu\right)^2}{n^{3/2}(p_+^\mu+p_-^\mu)^4}\nonumber\\
&\hspace{0.5cm}+\frac{16p_+^\mu p_-^\mu\left(\sqrt{p_+^\mu}x^\mu+\sqrt{p_-^\mu}y^\mu\right)^2\left(\sqrt{p_-^\mu}x^\mu-\sqrt{p_+^\mu}y^\mu\right)^2}{n^{2}(p_+^\mu+p_-^\mu)^5}+\mathfrak{o}(n^{-2})\,.\label{ExpansionTermSingleApp}
\end{align}}
A few comments are in order concerning the convergence of this series: First of all, the left-hand side of (\ref{ExpansionTermSingleApp}) is ill-defined when $p^\mu_++\sqrt{\tfrac{2 p^\mu_+}{n}}\,x^\mu+p^\mu_-+\sqrt{\tfrac{2 p^\mu_-}{n}}\,y^\mu=0$. However, this is precisely the region in the parameter space of $(x^\mu,y^\mu)$ where $\mathfrak{s}^\mu(\widehat{p})$ in (\ref{eq:metr_phat}) vanishes. Secondly, outside of this region, the radius of convergence of (\ref{ExpansionTermSingleApp}) is given through
\begin{align}
\frac{1}{\sqrt{n}}<\left|\frac{p^\mu_++ p^\mu_-}{\sqrt{2p_+^\mu} x^\mu+\sqrt{2p_-^\mu} y^\mu}\right|\,.\label{RadiusConvergence}
\end{align}
The right hand side vanishes if $p^\mu_++ p^\mu_-=0$, \emph{i.e.} if $\mathcal{C}$ is on the boundary of the simplex $\Delta_{N+1}$, which we therefore implicitly exclude in our analysis. Still, for fixed $n$, (\ref{RadiusConvergence}) further limits the parameter space (notably for $(x^\mu,y^\mu)$) in which (\ref{ExpansionTermSingleApp}) is applicable. Notice, however, that in the excluded region (namely for large $(x^\mu,y^\mu)$) the integral measure (\ref{DefGaussianIntMeasure}) is exponentially suppressed, and therefore leads to corrections of (\ref{SampleMetricExpansion}) that are exponentially suppressed for large $n$: indeed the boundaries which restrict the integral to the simplex $\Delta_{N+1}$ (see (\ref{GeneralBoundaries})) in fact render $(x^\mu,y^\mu)$ compatible with (\ref{RadiusConvergence}), while the inclusion of the problematic region in the integrals in (\ref{SampleMetricExpansion}) changes the result by terms exponentially suppressed in $n$ (see also the comment in the footnote in Section~\ref{Sect:CalculationPartFct}). Thus, in Section~\ref{Sect:CalculationExpValueVar} (and elsewhere in this work), when we replace the integrand in (\ref{ExpansionTermSingle}) by (\ref{ExpansionTermSingleApp}), we do not understand the latter to be a(n everywhere) convergent series expansion, but rather we understand it as an approximation that renders the final integral (\ref{ExpansionTermSingle}) correct up to subleading terms in $n$. We also point to a similar discussion in \cite{Paninski} concerning the series expansions of bias estimates of entropy and mutual information, completing previous work in the literature \cite{Miller1955NoteOT,Carlton,TrevesPanzeri,Victor} (see also \cite{Blyth})). 

In (\ref{ExpansionTermSingleApp}) we have included terms up to order $\mathcal{O}(n^{-2})$ (which shall become relevant in the computation of the variance), which, however, are only correct to leading order in $dt$, as we shall see below: indeed, terms of order $\mathcal{O}(n^{-2})$ also arise by including higher orders in the expansion of the Kullback-Leibler divergence in (\ref{ExpandKullback}), which, however, only provide terms of order $\mathcal{O}(n^{-2} dt^{-2})$ instead of $\mathcal{O}(n^{-2} dt^{-4})$. In (\ref{ExpansionTermSingleApp}), we have also displayed explicitly terms of half-integer orders in $n^{-1}$. As is evident, these contain odd powers in $x^\mu$ or $y^\mu$ and thus yield a vanishing result when substituted into (\ref{SampleMetricExpansion}). Indeed, using the generalised Gaussian integrals compiled in Appendix~\ref{App:GaussInteg}, we find for the latter up to order $\mathcal{O}(n^{-2})$
{\allowdisplaybreaks
\begin{align}
\langle \widehat{g}_{tt}\rangle\sim_{n\to\infty}& \frac{2}{dt^2}\sum_{\mu=1}^{N+1}\bigg[\frac{\left(p^\mu_+-p^\mu_-\right)^2}{p^\mu_++p^\mu_-}+\frac{4}{n}\frac{p_+^\mu p_-^\mu(p_+^\mu+p_-^\mu-2p_+^\mu p_-^\mu)}{(p_+^\mu+p^\mu_-)^3}\nonumber\\
&+\frac{4p_+^\mu p_-^\mu}{n^2(p_+^\mu+p_-^\mu)^5} \bigg((p_+^\mu+p_-^\mu)(p_+^\mu+p_-^\mu-2p_+^\mu p_-^\mu)(1-p_+^\mu -p_-^\mu+2p_+^\mu p_-^\mu)\nonumber\\
&\hspace{3.2cm}+2p_+^\mu p_-^\mu(p_+^\mu p_-^\mu-2p_+^\mu p_-^\mu (3-p_+^\mu -p_-^\mu))\bigg)+\mathfrak{o}(n^{-2})\bigg]\,.
\end{align}}
Finally, expanding to leading order in $dt$ we obtain
\begin{align}
\langle \widehat{g}_{tt}\rangle&\sim_{{n\to\infty}\atop{dt\to 0}} \frac{2}{dt^2}\sum_{\mu=1}^{N+1}\left[\frac{(\dot{p}^\mu(t))^2 dt^2}{2\,p^\mu(t)}+\frac{1-p^\mu}{n}+\frac{(1-p^\mu)^2}{2n^2p^\mu(t)}\right]\nonumber\\
&=g_{tt}(t)+\frac{2N}{n\,dt^2}-\frac{1}{2n^2 dt^2}\,\left(2N+1-\sum_{\mu=1}^{N+1}\frac{1}{p^\mu(t)}\right)\,.\label{ExpValueFisherApp}
\end{align}
Here we have used that $\sum_{\mu=1}^{N+1}p^\mu=1$.

\subsubsection{Variance}\label{App:Variance}
The integrand of (\ref{VarPreExpand}) can be expanded in the following manner
{\allowdisplaybreaks
\begin{align}
&\frac{\left(p^\mu_++\sqrt{\frac{2 p^\mu_+}{n}}\,x^\mu-p^\mu_--\sqrt{\frac{2 p^\mu_-}{n}}\,y^\mu\right)^2}{p^\mu_++\sqrt{\frac{2 p^\mu_+}{n}}\,x^\mu+p^\mu_-+\sqrt{\frac{2 p^\mu_-}{n}}\,y^\mu}\,\frac{\left(p^\nu_++\sqrt{\frac{2 p^\nu_+}{n}}\,x^\nu-p^\nu_--\sqrt{\frac{2 p^\nu_-}{n}}\,y^\nu\right)^2}{p^\nu_++\sqrt{\frac{2 p^\nu_+}{n}}\,x^\nu+p^\nu_-+\sqrt{\frac{2 p^\nu_-}{n}}\,y^\nu}\nonumber\\
&=\frac{\left(p^\mu_+-p^\mu_-\right)^2}{p^\mu_++p^\mu_-}\,\frac{\left(p^\nu_+-p^\nu_-\right)^2}{p^\nu_++p^\nu_-}\nonumber\\
&\hspace{0.2cm}+\frac{\sqrt{2}(p_+^\mu-p_-^\mu)(p_+^\nu-p_-^\nu)}{\sqrt{n}(p_+^\mu+p_-^\mu)(p_+^\nu+p_-^\nu)}\left[\frac{(p_+^\nu-p_-^\nu)}{p_+^\mu+p_-^\mu}\left(\sqrt{p_+^\mu}(p_+^\mu+3 p_-^\mu)x^\mu-\sqrt{p_-^\mu}(p_-^\mu+3 p_+^\mu)y^\mu\right)+(\mu\leftrightarrow\nu)\right]\nonumber\\
&\hspace{0.2cm}+\frac{2}{n}\bigg[\left(4\,\frac{\left(p^\mu_+-p^\mu_-\right)^2}{\left(p^\mu_++p^\mu_-\right)}\,\frac{p_+^\nu p_-^\nu\left(\sqrt{p_-^\nu}\, x^\nu-\sqrt{p_+^\nu}\, y^\nu\right)^2}{(p_+^\nu+p^\nu_-)^3}+(\mu\leftrightarrow \nu)\right)\nonumber\\
&\hspace{1cm}+\left(\frac{(p_+^\mu-p_-^\mu)\left(3p_-^\mu\sqrt{p_+^\mu}\,x^\mu+(p_+^\mu)^{3/2}\, x^\mu-(p_-^\mu)^{3/2}\, y^\mu -3p_+^\mu \sqrt{p_-^\mu}\,y^\mu\right)}{(p_+^\mu+p_-^\mu)^2}\times (\mu\leftrightarrow\nu)\right)\bigg]+\mathfrak{o}(n^{-1})\,,
\end{align}}
where for brevity, we refrain from writing explicitly the terms of order higher than $\mathcal{O}(n^{-1})$. Using the general Gaussian integrals compiled in Appendix~\ref{App:GaussInteg} we find for (\ref{VarPreExpand})
{\allowdisplaybreaks
\begin{align}
\langle (&g_{tt})^2\rangle\sim_{{n\to\infty}\atop{dt\to 0}}\sum_{\mu,\nu=1}^{N+1}\bigg[\left(\frac{(\dot{p}^\mu(t))^2}{p^\mu(t)}\,\frac{(\dot{p}^\nu(t))^2}{p^\nu(t)}+\mathfrak{o}(dt)\right)\nonumber\\
&+\frac{2}{ndt^2}\left(\frac{(\dot{p}^\mu)^2}{p^\mu}(1-p^\nu)-4\dot{p}^\mu\dot{p}^\nu+\frac{(\dot{p}^\nu)^2}{p^\nu}(1-p^\mu)+4\delta^{\mu\nu}\frac{\dot{p}^\mu \dot{p}^\nu}{\sqrt{p^\mu p^\nu}}+\mathfrak{o}(dt)\right)\nonumber\\
&+\frac{4}{n^2dt^4}\left(1-p^\mu-p^\nu+3 p^\mu p^\nu+2\delta^{\mu\nu}(1-2p^\mu)+\mathfrak{o}(dt)\right)+\mathfrak{o}(n^{-2})\bigg]\,.
\end{align}}
After performing the sums over $\mu,\nu$ (and taking into account that $\sum_{\mu=1}^{N+1}p^\mu=1$ and $\sum_{\mu=1}^{N+1}\dot{p}^\mu=0$), we thus obtain
\begin{align}
\langle (g_{tt})^2\rangle\sim_{{n\to\infty}\atop{dt\to 0}}\left((g_{tt})^2+\mathcal{O}(dt)\right)+\left(\frac{4(N+2)}{ndt^2}\,g_{tt}+\mathcal{O}(dt)\right)+\left(\frac{4N(N+2)}{n^2dt^4}+\mathcal{O}(dt)\right)+\mathfrak{o}(n^{-2})
\end{align}
Up to order $\mathcal{O}(n^{-2})$ and to leading order (at each individual order of $n$) in $dt$ we thus find with (\ref{ExpValueFisherApp}) for the variance of $\widehat{g}_{tt}(t)$
\begin{align}
&\text{Var}(\widehat{g}_{tt}(t))=\langle(\widehat{g}_{tt})^2\rangle-\langle(\widehat{g}_{tt})\rangle^2\nonumber\\
&\sim_{{n\to \infty}\atop{dt\to 0}}(g_{tt}(t))^2+\frac{4(N+2)}{n dt^2}\,g_{tt}+\frac{4N(N+2)}{n^2dt^4}-\left(g_{tt}(t)+\frac{2N}{n\,dt^2}-\frac{1}{2n^2 dt^2}\,\left(2N+1-\sum_{\mu=1}^{N+1}\frac{1}{p^\mu(t)}\right)\right)^2\nonumber\\
&\sim_{{n\to \infty}\atop{dt\to 0}} \frac{8 }{n dt^2}\,g_{tt}(t)+\left(\frac{8N}{n^2 dt^4}+\mathcal{O}(dt^{-2})\right)\,.\label{VarianceFinalApp}
\end{align}
Here we have not provided the term of order $\mathcal{O}(n^{-2} dt^{-2})$ which is subleading (either by one order of $n$ or by two orders of $dt$) relative to the terms shown in (\ref{VarianceFinalApp}). Moreover, as already previously remarked, terms of order $\mathcal{O}(n^{-2} dt^{-2})$ also arise when including subleading terms in the saddle-point approximation, notably the expansion of the Kullback-Leibler divergence in (\ref{ExpandKullback}), which we did not take into account.

\bibliographystyle{ieeetr}
\bibliography{references}

\end{document}